%% file: tremmel2016_mnras_v4_complete.tex
\documentclass[useAMS,usenatbib,twocolumn]{mn2e}
\usepackage{natbib,graphics,epsfig}
\usepackage{amsmath}
\usepackage{amssymb}
\usepackage{textcomp}
\usepackage{graphicx}
\usepackage{color}
\usepackage{hyperref}
\usepackage{txfonts}
\input{rgb}

\usepackage{graphicx}
\usepackage{color}
\input{rgb}
\usepackage[T1]{fontenc}
\usepackage{aecompl}

\newenvironment{tight_itemize}{
\begin{itemize}
  \setlength{\leftmargin}{1.5em}
  \setlength{\itemsep}{1pt}
  \setlength{\parskip}{1pt}
}{\end{itemize}}

\begin{document}

\title[The Romulus Simulations]{The Romulus Cosmological Simulations: A Physical Approach to the Formation, Dynamics and Accretion Models of SMBHs}
\author[M. Tremmel et al.]{M. ~Tremmel$^{1}$\thanks{email: mjt29@uw.edu}, M. ~Karcher$^{2}$, F. ~Governato$^{1}$, M. 
~Volonteri $^{3}$, T. ~R. ~Quinn$^{1}$, \newauthor A. ~Pontzen$^{4}$, L. ~Anderson$^{1}$, J.~Bellovary$^{5}$\\
$^1$Astronomy Department, University of Washington, Box 351580, Seattle, WA, 98195-1580\\
$^2$Statistics Department, University of Washington Seattle, WA, 98195-1580\\
$^3$Sorbonne Universit\`{e}s, UPMC Univ Paris 6 et CNRS, UMR 7095, Institut d`Astrophysique de Paris, 98 bis bd Arago, 75014 Paris, France\\
$^4$ Department of Physics and Astronomy, University College London, 132 Hampstead Road, London, NW1 2PS, 
United Kingdom\\
$^5$ Department of Physics, Queensborough Community College, 222-05, 56th Avenue, Bayside, NY 11364}

\pagerange{\pageref{firstpage}--\pageref{lastpage}} \pubyear{2015}

\maketitle

\label{firstpage}

\begin{abstract}

  We present a novel implementation of supermassive black hole (SMBH) formation, dynamics, and accretion in the massively parallel
  tree+SPH code, {\sc ChaNGa}. This approach improves the modeling of SMBHs in fully cosmological simulations, 
  allowing for a more detailed analysis of SMBH-galaxy co-evolution throughout cosmic time. Our scheme includes novel, physically motivated
  models for SMBH formation, dynamics and sinking timescales within galaxies, and SMBH accretion of rotationally supported gas. The sub-grid
  parameters that regulate star formation (SF) 
  and feedback from SMBHs and SNe are optimized against a comprehensive set of $z = 0$ galaxy scaling relations using a novel, 
  multi-dimensional parameter search.  We have incorporated our new SMBH implementation and parameter optimization 
  into a new set of high resolution, large-scale cosmological simulations called {\sc Romulus}. 
  We present initial results from our flagship simulation, {\sc Romulus25}, showing that our SMBH model results in SF efficiency, SMBH masses,
  and global SF and SMBH accretion histories at high redshift that are consistent with observations. We discuss the 
  importance of SMBH physics in shaping the evolution of massive galaxies and show how SMBH feedback is much more effective at
   regulating star formation compared to SNe feedback in this regime. Further, we show how each aspect of our SMBH model impacts this evolution 
  compared to more common approaches. Finally, we present a science application of this 
  scheme studying the properties and time evolution of an example dual AGN system, highlighting how our approach allows
  simulations to better study galaxy interactions and SMBH mergers in the context of galaxy-BH co-evolution.

\end{abstract}


\maketitle

\section{Introduction}

Supermassive Black Holes (SMBHs) are ubiquitous in galaxies across a
wide range of masses. SMBHs are observed not only in massive galaxies
\citep[e.g.][]{Gehren84,kormendy95,kormendy2013} but also in small,
bulge-less disk galaxies \citep{shields08,Filippenko03} as well as
dwarfs \citep{reines11,reines12,reines13,moran14}. Accreting SMBHs
lead to extremely energetic events throughout cosmic time, including luminous $z > 6$ quasars
powered by SMBHs with masses as high as $10^9$ M$_{\odot}$
\citep{fan01, mortlock11}. 

Despite their importance to galaxy evolution theory, understanding how
these black holes form, the mechanisms that regulate their growth, and
in what ways they affect their host galaxies are still open areas of
study.  Empirical scaling relations between SMBH mass and the stellar
mass and velocity dispersion of their host galaxies are indicative of
co-eval growth \citep{haring04,gultekin09, schramm2013,
  kormendy2013,volonteribellovary12}. While there have been attempts
  to quantify the evolution of this relationship
  \citep[e.g.][]{alexander08, bennert11, Bongiorno14,sun15}, these high redshift observations can
  be highly biased \citep{lauer07} and cannot effectively probe lower mass
  galaxies and black holes. There is also
evidence that the relationships break down at low redshift for lower mass, star
forming galaxies \citep{ReinesVolonteri15}. Thus, understanding the genesis of these
empirical relations and their mass dependency requires predictions from simulations
that accurately follow SMBH growth in low mass
(M$_{vir}<10^{10}$M$_{\odot}$) halos at both high and low redshift.



Previous works have been
fundamental in showing how energy from SMBH feedback is necessary to
shape the bright end of the galaxy luminosity
function, quench the formation of bulges in
field galaxies, and support a close causal connection between early
rapid growth, galaxy mergers, and QSO and AGN activity
\citep{dimatteoBH05, Teyssier11,IllustrisBH15,eagle15,bonoli16,volonteri16}.
With spatial resolutions
of the order of 10s to 100s of pc, physical processes involved in SMBH
accretion, feedback, and dynamics are necessarily implemented in
cosmological simulations via sub-grid prescriptions, under the broad
assumption that conditions at the smallest resolved scale drive the SMBH
evolution at much smaller scales. 

However, the simplifications inherent in these sub-grid models can hinder our understanding of the
connection between galaxy and SMBH evolution and growth, in
particular our ability to predict the merging rate of binary SMBHs and
how the inflow of gas onto the host galaxy \citep{bellovary13} feeds their
growth.  For example, a common treatment of SMBH dynamics is to assume they are always stable at the center of
their host galaxies, often obtained by shifting a SMBH towards the nearest potential
minimum \citep{IllustrisBH15}, a process we refer to as `advection'. This approach fails to capture the Gyr
timescale of sinking orbits for black holes during satellite
accretions or galaxy mergers \citep{G94,taffoni03, tremmel15} resulting in an
unrealistic coupling of SMBH and galaxy mergers, as well as artificially
high accretion rates during these perturbing events. Furthermore, SMBH
accretion is commonly calculated via a boosted Bondi-Hoyle
prescription \citep[e.g.][]{BoothBH2009}, but the assumptions of
this approach break down for gas supported by rotation rather than
internal pressure \citep{hopkins2010}.  Finally, SMBH `seeds' have often been placed based on the host
halo mass irrespective of the local gas properties
\citep{bonoli16}. This approach leads to a protracted epoch of SMBH
formation and an occupation probability artificially connected to the
observed population of active SMBHs, rather than to physically motivated
models of SMBH formation, which predict seeds that form at very high redshift \citep{begelman2006,volonteri12}.



The main goal of this Paper is to present a set of novel
implementations of SMBH physics, improving on the
way SMBH formation, dynamics, and accretion are handled with sub-grid
models in cosmological simulations. Specifically we:
 
\begin{itemize}
\setlength\itemsep{2em}
\item Connect SMBH seed formation to dense, very low
metallicity gas which allows us to predict the SMBH population in
both high mass galaxies and dwarf galaxies. 

\item Incorporate the sub-grid model
for dynamical friction presented in \citet{tremmel15} so that SMBHs experience
realistic dynamical evolution, allowing us to predict SMBH dynamics,
the frequency and mass ratio distribution of SMBH mergers, and SMBH growth
during galaxy interactions.

\item Introduce a new sub-grid model
for SMBH accretion that naturally accounts for the angular momentum
support of nearby gas at resolved scales.
 This creates a more physical picture of how, when, and where SMBHs grow compared to the more common Bondi-Hoyle prescription, while avoiding
   the additional assumptions and free parameters required by other current methods \citep[e.g.][]{rosasGuevara2015, anglesalcazar17}.

\end{itemize}

We also outline a novel approach to constrain `free' parameters within simulations, specifically
those that govern star formation and stellar feedback as well as SMBH growth and feedback. Due
to the computational cost, simulation studies have often relied on rerunning a small number of
large volumes while only changing one parameter at a time \citep[e.g.][]{eagle15}, only rarely running
grids of simplified models \citep{G07}. In this work we use a quantitative and efficient strategy, based
 on a large number of `zoomed-in' cosmological simulations, to decide
the optimal combination of sub-grid SF and SMBH related parameters
for a given set of physical modules and resolution. This general strategy is not specific to our simulations and can
be easily applied to any set of free parameters that govern any relevant physical processes.

In \S2 we describe the simulations and in \S3 we discuss the sub-grid
parameter optimization technique. We describe the sub-grid models for
star formation and feedback in \S4 and our novel approach to SMBH
physics in \S5. In \S6 we present results from our flagship 25 Mpc
volume, in \S7 we discuss the role of SMBH feedback in limiting star
formation compared to SN feedback alone, and in \S8 we show how more
common implementations of SMBH physics result in appreciably different
galaxies compared to our implementation. Finally, in \S9 we present an
example that illustrates how our SMBH implementation allows us to
study dual AGN in unprecedented detail and in \S10 we summarize our
results.  In Appendix A and B we discuss the rationale behind our
  sub-grid parameter optimization approach and explain our model for
post processing dust absorption.
\section{The {\sc Romulus} Simulations}

\subsection{{\sc ChaNGa}}

The simulations are run using the new Tree + SPH code {\sc ChaNGa}
\citep{changa15}, which includes standard physics modules previously
used in { \sc GASOLINE} \citep{wadsley04,wadsley08,Stinson06,shen10}
such as a cosmic UV background, star formation,`blastwave' SN feedback
and low temperature metal cooling.  The `blastwave' implementation of SN feedback is a well tested
approach that has been shown to reliably reproduce observable properties of galaxies, including
cored dark matter profiles in dwarf galaxies \citep{G10}. This is distinct from `super bubbles'  \citep{keller14}, a newer
approach to SN feedback that will be implemented in future simulations.. The SPH implementation includes
thermal diffusion \citep{shen10} and eliminates artificial gas surface
tension through the use of a geometric mean density in the SPH force
expression \citep{ritchie01,changa15,governato15}. This update  
accurately simulates shearing flows with Kelvin-Helmholtz instabilities. Our flagship simulation, {\sc Romulus25}, 
used up to 100,000 cores with good scaling.  {\sc ChaNGa} \citep{changa15} is part of
 the AGORA \citep{AGORA}  code comparison collaboration.

\subsection{Simulation Properties}

In addition to our flagship 25~Mpc per side uniform, periodic volume simulation ({\sc Romulus25}), we are currently running
a set of three zoom-in cluster simulations ({\sc RomulusC}) comprising halos of mass $10^{14} - 10^{15}$ M$_{\odot} $
as well as a 50~Mpc per side uniform volume ({\sc Romulus50}). Both
{\sc Romulus25} and {\sc RomulusC} will be run to $z = 0$ and {\sc Romulus50} will be run to $z = 2$. 
See Table 1 for a list of simulations and parameters. We use {\sc Romulus25} for 
the analysis in this paper because it provides a large, uniform sample of low redshift galaxies. Smaller 8 Mpc uniform volume 
simulations are also used for our comparative studies (see \S 7 and \S8).



The simulations are run assuming a $\Lambda$CDM cosmology following the most recent results from Planck
\citep[$\Omega_0=0.3086$, $\Lambda=0.6914$, h$=0.67$, $\sigma_8=0.77$;][]{plank15} and
at the same resolution, with a Plummer equivalent force softening of $250$ pc. Unlike many similar cosmological
runs, the dark matter distribution is {\it oversampled}, such that we
simulate $3.375$ times more dark matter particles than gas particles,
resulting in a dark matter particle mass of $3.39 \times 10^5$
M$_{\odot}$ and gas particle mass of $2.12 \times 10^5$
M$_{\odot}$. This is an important shift from the standard approach of
simulating the same number of gas and dark matter particles, as it
allows us to decrease numerical noise and allow for more accurate
black hole dynamics \citep{tremmel15}. Our mass
resolution is better than recent large volume simulations
\citep{IllustrisBH15,volonteri16} and our force resolution is comparable
to the highest resolution runs of the EAGLES series
\citep{eagle15}. Spline force softening converges to a Newtonian force
at scales twice the gravitational softening, $\epsilon_g$.

In order to showcase the results of our model and compare with other
common SMBH implementations, we also run a series of 8~Mpc per side
uniform volume simulations with different realizations of SMBH
physics, as well as a simulation with no SMBHs and an enhanced SN
feedback efficiency.  These smaller simulations (e.g. {\sc Romulus8}) have
the same cosmology and resolution as our main {\sc Romulus}
dataset. {\sc Romulus25}, along with the 8 Mpc runs, make up the
data used in the analysis presented in this Paper. For a complete list
of simulations, see Table 1.

\subsection{Halo and Galaxy Extraction}

For all simulations referred to in this work, 
we use the Amiga Halo Finder \citep{knollmann09} to extract individual halos.
 We calculate galaxy properties based on all of the particles within a given halo. However,
 for a better `apples-to-apples' comparison with observational results,
 we utilize the corrections from \citet{munshi13} to account for the mass of stars
 missed in observations and the baryonic effects on halo mass not accounted for in dark matter only (DMO) simulations.
 These corrections have been calibrated for
halos with virial mass $10^8 - 10^{12} $M$_{\odot}$ and are shown to be
roughly constant across this range. Specifically, M$_{\star, obs} = 0.6 $ M$_{\star, sim}$
and M$_{vir, sim} = $0.8 M$_{vir, \text{DMO}}$. We apply these corrections to halos with M$_{vir}$ as
large as $10^{13} $M$_{\odot}$. In these halos, such corrections are particularly necessary, 
as $\sim$40\% of stars exist far from halo center, either
in an extended stellar halo or in satellite galaxies, and would not be included 
in observational estimates for stellar mass.

\section{Sub-grid Parameters Optimization}

In {\sc ChaNGa}, SF and SMBH physics are regulated through a series
of sub-grid prescriptions that parameterize unresolved physics into
several free parameters. In order to set these parameters to
  their optimal values we employ a quantitative optimization
technique to map out the suitability of the parameter space and
near-converge on the `best' parameters. The idea of this approach is
similar to that of \citet{Bower2010}, but tailored specifically for more complicated
simulations where only a few galaxies can be run with 10s of different parameter combinations.
A summary of the procedure is the following (see Appendix A for a more detailed description):


\begin{enumerate}

\item We simulate a large number of sets of 4 `zoomed-in' galaxies
\citep{G07,G09} at the same resolution as {\sc Romulus}, with halo
masses ranging from 10$^{10.5}$ to 10$^{12}$ M$_{\odot}$, with dozens
of different sub-grid parameter realizations.

\item  We compare the properties of the resulting galaxies to
local empirical scaling relations, grading each parameter set
accordingly based on the logarithmic distance of each galaxy from the
relation.The score of each parameter realization is 
then the sum of the distance (in log space) of each halo from each empirical relation. 

 \item  The procedure is repeated, each time sampling in more
 detail around the best graded models until a best set of parameters is converged upon. The Kriging algorithm (see Appendix A)
 is used to efficiently explore parameter space and determine convergence.
 
\end{enumerate}

A first set of simulations was run with only SF physics and with higher weight placed on reproducing the
observed properties of lower mass galaxies, where the effect of SMBH physics should be less important.
The parameters searched were the local SF efficiency, the density
threshold for SF, and the fraction of SN energy coupled to the
surrounding gas (see \S4).  Once the best SF parameters were
identified (with the SN efficiency being the most important overall),
a second set of galaxies was run including SMBHs physics, leaving the
SF parameters unchanged but varying 1) the SMBH accretion and 2) energy
coupling efficiencies (see \S5.3). For results from the SF parameter search,
we point the reader to Appendix A and \citet{anderson17}.

The z$=$0  relations used to grade the galaxy sets were: 1) The stellar mass - halo mass relation, 2) The HI gas fraction as a function of stellar mass\footnote{ALFALFA data from private correspondence with Jessica Rosenberg.}, 3) The galaxy specific angular momentum vs stellar mass, and 4) The SMBH mass vs stellar mass (SMBHs only). The first two scaling relations  \citep{moster13,cannon11,AlfAlfaHaynes11} allow us to
respectively constrain the SF efficiency over the whole Hubble time,
and the low redshift gas depletion time (i.e the recent SF rates). Our simulations follow the HI abundance of gas
so M$_{HI}$ is derived explicitly from the total gas content of each halo.
The relationship between stellar mass, angular momentum, and morphology \citep{obreschkow14} is a
useful proxy of galaxy sizes as well as the removal of low angular momentum gas through feedback processes.
The M$_{BH}$-M$_\star$ relation \citep{schramm2013} is a final test specific for SMBH physics. 
These four scaling relations
control several fundamental aspects of galaxy formation connected to
the regulation of SF, angular momentum evolution, and the
growth of SMBHs. Taken together they provide useful, low-z constraints
to our model without {\it unconsciously} biasing our effort to
reproduce one specific scaling relation. For the sake of simplicity and to avoid biasing the analysis, 
we use just the raw logarithmic distance from each relation to determine the plausibility of each parameter set,
 implementing no weighting between different relations. However, we do exclude the dwarf galaxy from
 the morphological and SMBH relations, as explained in Appendix A.

When applied to setting three star formation parameters, the technique was able to 
converge with little user input after 27 realizations (a total of 80 simulations; see Appendix A).
For two SMBH parameters, we were able to find a suitable parameter set
after 12 realizations (a total of 48 simulations; see \S5.5).

This `zoomed-in' approach to parameter optimization allows us to
efficiently explore the parameter space without having to simulate as
many parameter realizations as would be required for a standard
random-walk Markov-chain. It presents several advantages over shutting
off or including individual physics modules \citep{genel14} or to
running a small cosmological volume multiple times
\citep{eagle15,schaye10}, the main issue being that running large
simulations, particularly those at high resolution, is computationally expensive and
will result in only a very limited parameter space exploration. Using this approach, the non linear effect of changing
more than one parameter at a time can now be followed and the search for best parameters can cover the mass
range of the final, large scale simulation (which tend to have more
massive halos than small test volumes). Finally the set of zoomed-in
runs provides a useful post main run framework to understand
significant deviations from observed properties of  galaxies or SMBHs should
they emerge from the production runs.

\section{Star Formation Physics}

As in our standard implementation \citep{Stinson06} for runs at this
resolution, star formation (SF) is regulated by:

\begin{enumerate}
\item the normalization of the SF efficiency, c$_{\star}$, used to calculate the probability
of creating a star particle from gas with dynamical time $t_{dyn}$ and characteristic star formation time, $\Delta t$, assumed to be $10^6$ yr
\begin{equation}
p =\frac{m_{gas}}{m_{star}}(1 - e^{-c_{\star} \Delta t /t_{dyn}}), 
\end{equation}
 
\item The fraction of SNe energy that is coupled to the ISM

\item the minimum density (n$_\star$) and
 maximum temperature (T$_\star$) thresholds beyond which cold gas is allowed to form stars.
 
 \end{enumerate}
\noindent The final values adopted for these three sub-grid parameters are:
\begin{tight_itemize}
\item SF efficiency c$_{\star} = 0.15$

\item Gas temperature threshold, T$_\star =10^4$ K

\item Gas density threshold, n$_\star = 0.2$ m$_{p}/$cc)

\item SNe energy coupling efficiency, $\epsilon_{SN}$, of 75\%
\end{tight_itemize}

SN feedback adopts the `blastwave' implementation \citep{Stinson06}.
Gas cooling is regulated by metal abundance as in \citet{eris11} and
SPH hydrodynamics and thermal and metal diffusion are described in
\citet{shen10} and \citet{governato15}. Our simulations do not include H$_2$ cooling as their resolution is not sufficient to model individual star forming
regions. 
We use a Kroupa IMF \citep{kroupa2001}, with the associated metal yields. 

It is important to note that without SMBH feedback, parameters
that work the best for dwarf galaxies based on our grading criteria
(see \S3) are different from those that work best for higher mass galaxies. The 
parameters used here represent those that grade the highest when dwarf galaxy
results are more heavily weighted. The idea is to start with a SF model that
performs very well at low masses and allow SMBH physics to create better
results for high mass galaxies.

One 8 Mpc cosmological simulation was also run, with $\epsilon_{SN}$=2 (see
Table 1). Note that $\epsilon_{SN}$=2 can be justified by implying a  top
heavy IMF or contribution from `early feedback' \citep{governato15}.
During our parameter search we found that this run produced galaxies in $10^{12}$ M$_{\odot}$ halos that
better matched observed relations, at the expense of dwarf galaxies. However,
strong SNe feedback alone still results in too much star formation at late times (see \S7).
We find the inclusion of SMBH feedback
as described below is necessary to reproduce the `bend' in the
M$_\star$- M$_{halo}$ relation at high halo masses while maintaining realistic
dwarf galaxy properties.

\begin{table}
\caption{Physics implementations in different simulations presented in this Paper.}
\label{symbols}
\begin{tabular}{@{}cccccc}
\hline
Name & Box Size & Accretion$^{a}$ & SMBH & $\epsilon_{SN}^{c}$  & Run to\\
           & (Mpc) &  & Dynamics$^{b}$ &  & z = \\
\hline
{\bf {\it Romulus8}} & 8 & Bondi+AM &Dyn. Frict. & 0.75  & 0.5 \\
{\bf {\it Romulus25}} & 25 & Bondi+AM & Dyn. Frict. & 0.75 & 0\\
{\bf {\it RomulusC}} & N/A & Bondi+AM & Dyn. Frict. & 0.75 & 0\\
{\bf {\it Romulus50}} & 50 & Bondi+AM & Dyn. Frict. & 0.75  & 2\\
Advect & 8 & Bondi+AM & Advection &  0.75 & 0.5\\
Bondi & 8 & Bondi & Dyn. Frict. & 0.75 & 0.5\\
highSN & 8 & N/A & N/A & 2.0 & 0.5\\
\hline
\end{tabular}
\medskip
$a$ Bondi+AM denotes the implementation described in this work\\
$b$ `Advection' denotes method utilized in \citet{sijacki07} and `Dyn. Frict' is that from \citet{tremmel15}\\
$c$ how much energy per SN is coupled to gas (in units of $10^{51} $ergs)  All the runs have identical  particle mass, force resolution and numerical parameters.
\end{table}

\section{Modeling Black Hole Physics}

\subsection{Seed Formation}

Unlike SMBH seeding methods {\it directly} tied to halo mass
thresholds that are often utilized in other large cosmological volume
simulations \citep[e.g.][]{dimatteoBH03,IllustrisBH15,eagle15}, our approach allows for a more
realistic seeding at high redshift without any \textit{a priori}
assumptions regarding halo occupation fraction.

Instead, SMBH seed formation is connected to the physical state of the gas by converting a
gas particle already selected to form a star (see \S4) into a SMBH seed instead if it has:

\begin{tight_itemize}
\item Low mass fraction of metals (Z $< 3\times 10^{-4}$)

\item  Density 15 times that of the SF threshold ($3$ m$_{p}/$cc)

\item Temperature between $9500$ K and $10000$ K
\end{tight_itemize}

\noindent These criteria ensure that black holes form only from gas that a)~is
collapsing quickly (i.e. faster than the star formation timescale as
it has not been turned into a star already) while b)~cooling relatively
slowly, approximating formation cites predicted for SMBH seed formation
\cite{begelman2006,volonteri12}

The criteria above were not chosen via an extensive parameter
search. Rather, they were empirically derived via analysis of star forming gas
particles in high redshift volume simulations. The model limits SMBH growth
to the highest density peaks in the early Universe with high Jeans mass. This 
is a marked improvement over stochastic formation from star forming gas, resulting in seed formation that occurs in environments that are different
than the average unenriched star forming region, 
as seen in higher resolution tests of SMBH formation sites \citep{agarwal14,habouzit17}.
 Because we are following conditions of gas at resolved scales
(i.e. hundreds of pc), these criteria are designed to capture the
regions where SMBH seeds should exist and then be able to grow quickly
to large masses, regardless of the specifics of the true formation
mechanism at unresolved scales.

\begin{figure}
\centering
\includegraphics[trim=15mm 0mm -40mm 25mm, clip, width=105mm]{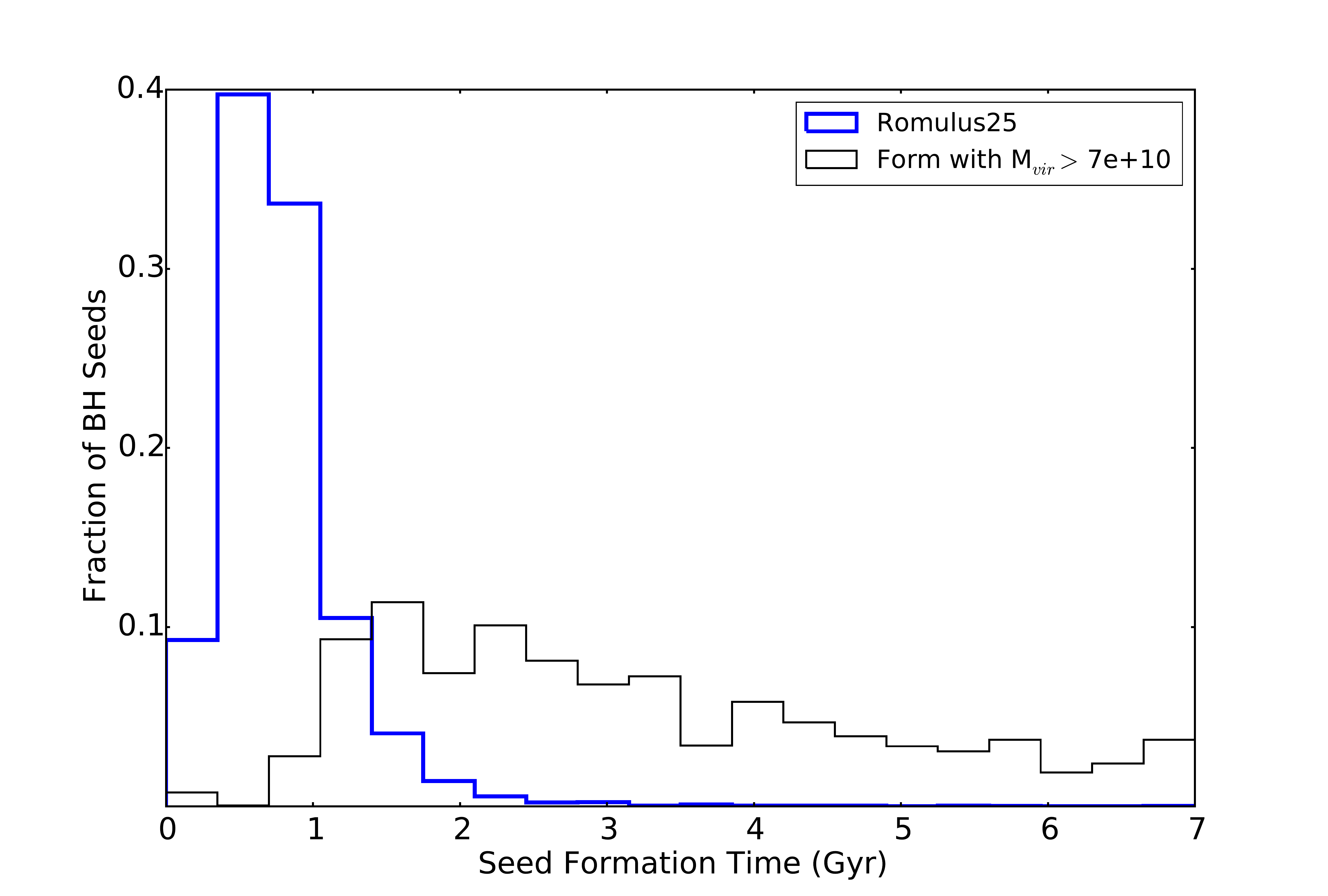}
\caption{{\sc Seed Formation Times.} The distribution of black hole seed formation times using our approach applied to a 25 Mpc run (Romulus25; blue line) compared to the seed formation if we applied a threshold halo mass criterion similar to other common approaches to seed formation in large simulation of this type \citep{dimatteoBH03,IllustrisBH15,eagle15}. Using our scheme, black hole seeds form much earlier, the vast majority forming within the first Gyr of the simulation, similar to the expected formation epoch for SMBHs \citep{volonteri12}.  We compare to the halo threshold scheme, meant to approximate that used in \citet{IllustrisBH15}, where halos are seeded once a halo reaches a critical mass of $7\times10^{10}\mathrm{M}_{\odot}$. Using this, black holes are seeded at much later times, even in the most massive halos, which would cause the earliest periods of SMBH growth to be missed.}
\label{bhform} 
\end{figure}

The metallicity threshold of $3\times 10^{-4}$ was chosen to select
gas that had seen very little chemical evolution. We found that
choosing more strict (lower) metallicity criteria or colder gas,
biases SMBH formation {\it away} from the densest regions of the early
Universe, an undesired outcome due to the finite resolution of our
runs, that we specifically decided to avoid.  SF will
often form stars nearly simultaneously with SMBH particles. 
As stars form and massive stars give off stellar winds and 
SNe explode, metal rich gas permeates throughout the halo and beyond,
effectively shutting down any potential seed formation within the
parent halo as well as nearby halos. Metal diffusion in SPH codes is explicitly regulated by a
diffusion equation; here we follow the implementation in
\citet{shen10} with coefficients for both metal and thermal diffusion
both set to 0.03, which give realistic values for galaxy metallicity
gradients in high resolution dwarfs (Brooks et al., in preparation.).

Once formed, the SMBH seed mass is set to $10^6$ M$_{\odot}$. To
attain this mass, the newly formed SMBH accretes as much mass as it
needs from surrounding gas particles (total mass is then explicitly
conserved), representing rapid, unresolved growth. The initial mass, while somewhat higher than most
theoretical estimates \citep{JohnsonBH12,volonteri12}, is motivated by the fact that
much of the early growth onto SMBH seeds, or the exotic objects that may proceed them,
can exceed $0.1$ M$_{\odot}$ yr$^{-1}$ and be governed by the environment and physical processes well
below the resolution limit of our simulations \citep[e.g.][]{hosokawa13,schleicher13}.
 In reality, SMBH seeds would likely attain a spectrum of masses early on,
  but since such processes are unresolved, we cannot 
differentiate between where a larger SMBH seed should grow. 
This mass is also sufficiently large compared to our DM and gas particle masses that
the dynamics of all SMBHs will be well resolved \citep{tremmel15}.
 We verified that even with this initial mass, SMBH seeds that exist in
unfavorable environments (i.e. dwarf galaxies) naturally have limited
growth, with  about 50\% of SMBH seeds having less than 10\% mass growth over a
Hubble time.  We also verify that SMBHs that grow to more than $10^7$ M$_{\odot}$ 
have grown enough through accretion to be insensitive to initial conditions.

In much of our future analysis, including that presented in \S 6, we take 
growth occurring in SMBHs with mass less than 110\% of their initial mass as still
undergoing their initial growth phases, a process which has not yet been observed and
 the physics of which is highly uncertain. Therefore, we will exclude these systems from analysis
 where appropriate.
 

 Figure~\ref{bhform} plots the distribution of formation times of all SMBH
 seeds formed using the above approach within the {\sc Romulus25} volume. As a comparison, we plot the distribution of seed formation times we would
 have using a halo mass threshold of $7\times10^{10}$
 M$_{\odot}$. This is meant to approximate a more common seeding
 mechanism utilized in other large cosmological simulations. Specifically, 
 our threshold approximates that from the Illustris Simulation \citep{genel14,IllustrisBH15}.
 Our approach
 forms SMBHs much earlier, closer to what would be expected in SMBH seed
 formation scenarios \citep{Volonteri2010AARV, volonteri12, habouzit17}. 
 We note that similar halo threshold techniques that have lower threshold masses will
 form seeds earlier, though they will still have a more substantial tail toward low redshift formation times. In {\sc Romulus25}
 SMBH seeds still form out to low redshift in some rare cases within small, unenriched halos. 
 These SMBHs constitute a very small fraction ($<<$1\%) of the overall SMBH population in the simulation.
 
 
 \subsection{Black Hole Mergers}
 
SMBHs are allowed to merge based on the same criteria as \citet{bellovary11}. Once SMBHs
become closer than two softening lengths in relative distance, they merge if they have low
enough relative velocities such that they would be considered gravitationally bound to one
another, i.e. $\frac{1}{2}\Delta \textbf{v} < \Delta \textbf{a} \cdot \Delta \textbf{r}$, where $\Delta \textbf{v}$, $\Delta \textbf{a}$, and $\Delta \textbf{r}$
are the relative velocity, acceleration, and distance vectors between two SMBH particles.

\subsection{Black Hole Dynamics }

Dynamical friction (DF), the force exerted by the gravitational wake
caused by a massive object moving in an extended
medium \citep{DF43,BinneyTremaine} causes the orbits of SMBHs to decay
towards the center of massive galaxies \citep{G94,stelios05}. However,
this effect is difficult to resolve in cosmological simulations due to
numerical noise and limited gravitational force resolution.
Our implementation includes a sub-grid approach for
modeling unresolved dynamical friction that has been shown to produce
realistically sinking SMBHs \citep{tremmel15} . This allows us to follow
the dynamics of SMBHs without assuming they should always be stable at
the centers of galaxies. As described in detail in \citet{tremmel15}
our approach assumes that within $\epsilon_{g}$ from
the black hole the velocity distribution is isotropic, giving
Chandrasekhar's dynamical friction formula \citep{DF43} for
a BH of mass M and surrounding particle mass $\mathrm{m}_a$ with
velocity distribution $f(\textbf{v})$.

\begin{equation}
\mathbf{a}_{DF} = -4\pi\mathrm{G}^2\mathrm{M}\mathrm{m}_a\mathrm{ln}\Lambda\frac{\mathbf{v}_{BH}}{\mathrm{v}_{BH}^3}\int_{0}^{\mathrm{v_{BH}}} d\mathrm{v}_a\mathrm{v}_a^2 f(\textbf{v}_a).
\end{equation}
\noindent The velocities of the BH and surrounding particles ($\mathrm{v}_{BH}$ and $\mathrm{v}_a$ respectively) 
are both taken relative to the local center of mass velocity within the smoothing kernel and  $\mathrm{ln}\Lambda$ is the Coulomb logarithm.
This equation can be further simplified by substituting the integral for $\rho(<\mathrm{v}_{BH})$, 
which is the density of particles moving slower than the black hole.

\begin{equation}
\mathbf{a}_{DF} = -4\pi\mathrm{G}^2\mathrm{M}\rho(<\mathrm{v}_{BH})\mathrm{ln}\Lambda\frac{\mathbf{v}_{BH}}{\mathrm{v}_{BH}^3}.
\end{equation}

Taking $\mathrm{ln}\Lambda \sim \mathrm{ln}(\frac{\mathrm{b}_{max}}{\mathrm{b}_{min}})$, we set b$_{max}$ = $\epsilon_{g}$ to avoid double counting frictional forces that are already occurring on larger scales, which are well resolved due to the high mass and spatial resolution of our simulations. We take the minimum impact parameter, b$_{min}$ to be the $90^{\circ}$ deflection radius, with a lower limit set to the Schwarzschild Radius, $\mathrm{R}_{Sch}$. The calculation is done using 64 collision-less particles (i.e. dark matter and star particles) closest to the black hole, with velocities taken relative to the COM velocity of all 64 particles.

A common technique in cosmological simulations is to
reposition or push the SMBH along its local potential gradient
\citep[e.g.][]{dimatteoBH05,sijacki07,IllustrisBH15}. However, 
these techniques (broadly referred to as `advection' from here
on) fail to property reflect what is often a significant
characteristic timescale for sinking SMBHs \citep[see][and references therein]{tremmel15}. During galaxy mergers,
`advection' techniques will result in a nearly immediate SMBH merger. It
also prevents SMBHs from becoming perturbed away from galactic center,
 which can affect SMBH growth during galaxy interactions and mergers

With our approach instead, we are able to resolve the dynamics of
SMBHs during and after galaxy mergers down to sub-kpc scales. The merger
rates of SMBHs will be realistically decoupled from galaxy mergers. This
will result in realistic SMBH growth and new predictions for
gravitational wave observations. Our approach will also naturally
produce dual and offset AGN down to sub-kpc distances, allowing us to
study and understand these transient events in a broader evolutionary
context (see \S9).

\begin{figure*}
\centering
\includegraphics[trim=20mm 37mm -33mm 45mm, clip, width=220mm]{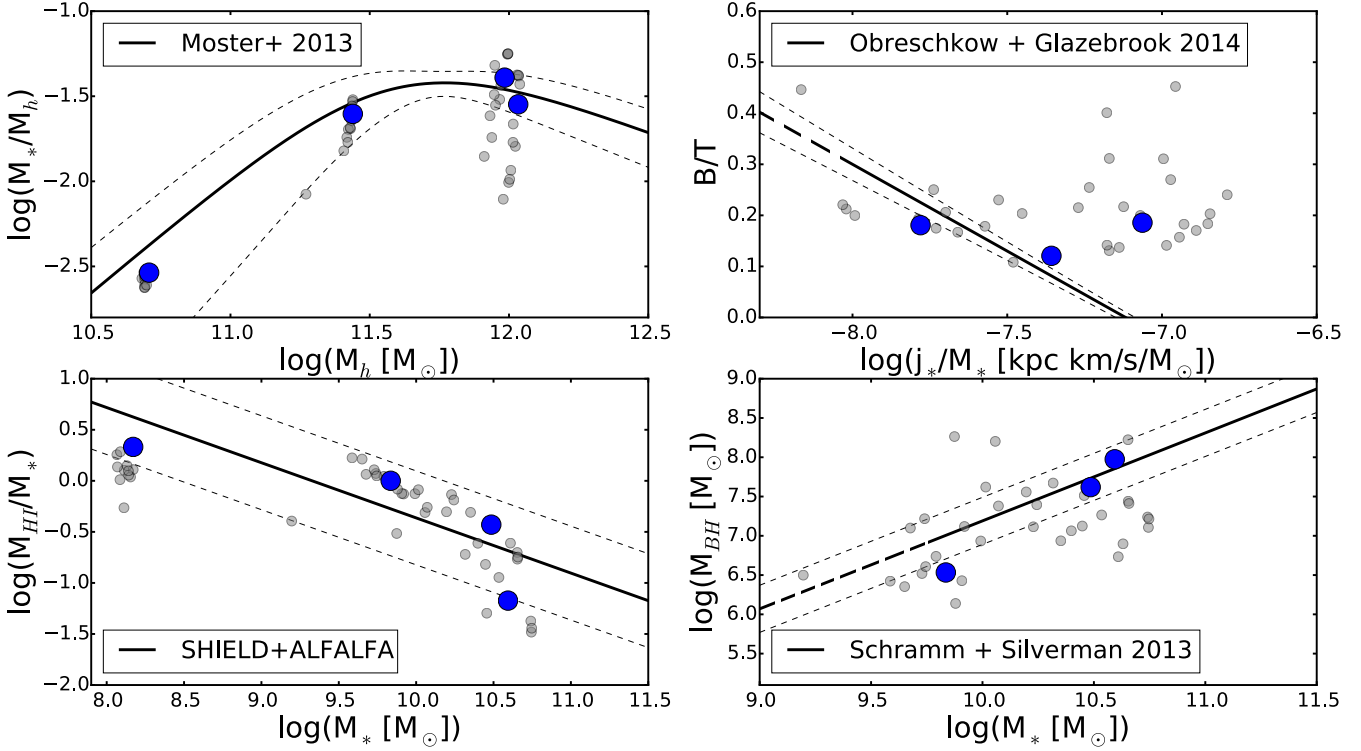}
\caption{{\sc SMBH Parameter Optimization.} Results from the search for optimal free parameters related to SMBH
 accretion and feedback. 12 realizations of accretion boost factor ($\beta$) and feedback efficiency ($\epsilon_f$)
 for SMBHs were run, each with four zoomed in runs of galaxies. All of the models are shown in light grey points and the
  best fitting model (the one that best matches overall to the four relations shown) is in blue. Each model is compared to
   different empirical relations governing star formation efficiency \citep[upper left,][]{moster13}, angular momentum
    \citep[upper right,][]{obreschkow14}, HI content \citep[lower left, derived from SHIELD and ALFALFA data, see ][]{cannon11,AlfAlfaHaynes11} and black 
     hole growth \citep[lower left,][]{schramm2013}. The thin dashed lines represent $1-\sigma$ errors. 
     The thick dashed lines represent where each relation has been extrapolated beyond observations.
     The blue points have the parameters, $\beta = 2, \epsilon_f = 0.02$,
      which are what we implement in the {\sc Romulus} models as well as the other simulations listed in Table 1. Note that for the angular momentum and SMBH mass
      tests, the dwarf galaxy was excluded. The former is due to the fact that angular momentum decomposition is difficult for a galaxy of this size. The latter is because
      observed SMBH masses are uncertain for dwarf galaxies and in our simulations, including in these parameter search runs, not every dwarf galaxy forms a SMBH.}
\label{params} 
\end{figure*}

\subsection{Accretion and feedback}

Black holes are allowed to grow by accreting mass from nearby gas
particles.  Energy from accretion is then isotropically imparted to the 32
nearest gas particles, distributing the energy among them according to the smoothing kernel. To ensure that the feedback energy is
realistically dissipated, gas particles that receive
energy from a SMBH are not allowed to cool for a time equal to the
timestep of the SMBH (typically $10^3$ to $10^4$ yrs), which is meant to represent the continuous
transfer of energy during each SMBH timestep. This is a similar technique that is used in the Blastwave
supernova feedback prescription, though here we utilize a different cooling shutoff time meant to 
approximate the continuous accretion and subsequent feedback that should occur during a timestep. The amount of energy
coupled to surrounding gas particles is given by

\begin{equation}
E = \epsilon_r \epsilon_f \dot{M} c^2dt,
\end{equation}

\noindent where the radiative efficiency, $\epsilon_r$, is assumed to be $10\%$ and the efficiency that energy couples to 
gas, $\epsilon_f$ is set to $2\%$ (see below for discussion on free parameter calibration). The accretion rate is assumed to be constant throughout one black hole timestep, $dt$.

The underlying assumption of these approaches is that the state of the gas at the smallest resolved scales drives the evolution of the unresolved physics on timescales relevant to the simulation.

The accretion rate, $\dot{M}$, is estimated via a modified Bondi-Hoyle
prescription applied to the smoothed properties of the 32 nearest gas
particles. The initial derivation of our approach is exactly the same as Bondi accretion.
If we define some accretion radius, $R$, relative to the SMBH beyond which gas
is bound to the black hole, and assume that
mass continuity is roughly upheld on long time scales, the accretion rate onto the SMBH
should be similar to the rate of mass flowing through a spherical surface of that radius:

\begin{equation}
\dot{M} \sim \pi R^2 \rho v.
\end{equation}

\noindent Here $v$ is the characteristic velocity of gas through the surface and
$\rho$ is the density of the ambient gas. In Bondi-Hoyle accretion, 
the calculation of the accretion radius, $R$, balances the SMBH's gravitational 
potential and both the internal and bulk kinetic energies of the gas.  In order to
avoid underestimating the accretion rate due to resolution effects
when calculating the density and temperature of nearby gas, we apply a
density dependent boost factor to this accretion rate,
following the prescription of \citep{BoothBH2009} where the
  standard Bondi rate is multiplied by a density dependent factor,
  $\left (\frac{n_{gas}}{n_{*}} \right )^{\beta} $, where $\beta$
  is a free parameter and $n_{*}$ is the star formation density
  threshold.

However, even with a well motivated (but often poorly constrained)
density boost, Bondi-Hoyle accretion is unable to account for angular
momentum support, which often dominates the dynamics of cold gas at
resolved scales, as in the disks of star forming galaxies
\citep{hopkins2010,hopkinsqso11}. Past efforts have focused on sub-grid models for
angular momentum transport on sub-galactic scales \citep{anglesalcazar17}
or within the SMBH's accretion torus \citep{rosasGuevara2015}.

To take advantage of the improved
spatial resolution of modern simulations, we implement an accretion
algorithm that accounts for the angular momentum of gas {\it at resolved scales}.
Our approach avoids any additional assumptions of sub-grid physics or free parameters
beyond those required by the conventional Bondi-Hoyle prescription. Namely that
the accretion rate, averaged over timescales relevant to the simulation, is a direct consequence
of mass flux across the accretion radius, defined as the radial distance at which the gravitational potential
of the SMBH balances the internal and bulk energetics of the gas as measured at the smallest resolved
scales of the simulation.



In the reference frame of rotating gas, angular momentum provides an effectively lower gravitational 
potential such that $U_{eff}(\textbf{r}) \sim -\frac{GM}{r} + \frac{j(r)^2}{2r^2}$, where $j(r)$ is the angular momentum per unit mass of the gas at distance $r$ from the SMBH. We can replace $
j(r)^2/r^2$ with $v_{\theta}^2$, the rotational velocity of the surrounding gas. It is important to note that $v_{\theta}$ is 
distinct from the bulk velocity, which we will refer to as $v_{bulk}$, in the Bondi-Hoyle formula, which accounts for a flow of 
gas, not a coherent rotational motion.

If the dominant motion of the gas is rotational rather than a bulk flow, we can 
use the effective potential above and solve for $R$, ignoring order unity terms, such that the effective potential balances 
with the thermal energy of the gas, i.e $U_{eff} \sim c_s^2$. By definition the tangential motion must not contribute to the 
mass flux through our area. Returning to the simple equation for $\dot{M}$ above we get the following relation.

\begin{equation}
\dot{M} \sim \frac{\pi(GM)^2 \rho c_s}{(v_{\theta}^2+c_s^2)^2}.
\end{equation}

Note that we do not assume $v_{\theta}$ is constant on unresolved scales, only
that its value should inform the radius, $R$, at which the gravity of the SMBH dominates the gas dynamics.
This is similar to the original Bondi-Hoyle formalism, where the energetics of gas far from the black hole
 are used to approximate the accretion radius. In this case, $v_{\theta}$ encapsulates the amount of angular momentum
 support the gas has on the smallest resolved scales, translating to a smaller accretion radius and therefore lower accretion rate.
 

To avoid uncertainties in particle dynamics below the force softening scale, we calculate the specific angular
momentum, $j$, relative to a target black hole for gas particles that
are between $3$ and $4$ softening lengths away (with our spline kernel
softening, Newtonian forces are followed exactly at $2$ $\epsilon_g$). We
then calculate the tangential velocity that gas one softening
length,$\epsilon_g$, away from the SMBH would have if the angular momentum
on the larger scales was conserved, $v_{\theta}(\epsilon_g) \sim
j/\epsilon_g$.  The smallest relative velocity of the $32$ gas particles closest to the SMBH, which
we take as a proxy to $v_{\mathrm{bulk}}$, is compared to $v_{\theta}$.
 If $v_{\theta} > v_{bulk}$ we use equation (6) to
calculate $\dot{M}$. Otherwise, we use the normal Bondi rate. Both
calculations include the density-dependent boost factor, resulting in:

\begin{equation}
\dot{M} = \alpha \times \begin{cases}
\frac{\pi(GM)^2 \rho}{(v_{\mathrm{bulk}}^2+c_s^2)^{3/2}} & \text{ if } v_{\mathrm{bulk}}>v_{\theta} \\ \\
\frac{\pi(GM)^2 \rho c_s}{(v_{\theta}^2+c_s^2)^{2}} & \text{ if }  v_{\mathrm{bulk}}<v_{\theta}
\end{cases};
 \alpha = \begin{cases}
\left ( \frac{n}{n_{th,*}} \right )^\beta & \text{if } n \geq n_{th,*}\\ \\
1 &  \text{if } n < n_{th, \star}
\end{cases}.
\end{equation}

Unlike \citet{rosasGuevara2015}, we do not implement a viscosity parameter in our accretion rate calculation. This was an explicit choice made to avoid the inclusion of an additional free parameter and is justified by the fact that we are not attempting to approximate the behavior of an accretion torus, as in \citet{rosasGuevara2015}, where viscous timescales can be more critical. Still, there is uncertainty in the normalization of equation (7) when $v_{\theta}>>c_s$, which will be explored in future work. It should also be noted that equation (7) is not continuous at  $v_{\mathrm{bulk}}=v_{\theta}$. We find this effect is sub-dominant compared to variations in density and velocity inherent to discreet calculations. This is shown in practice in figure 14, where our approach produces a less bursty accretion history in MW-mass halos compared to normal Bondi accretion.

For the density dependent boost factor, we compare the local density to the star formation density threshold, $n_*$, meant to represent the limit beyond which the simulation fails to resolve the multiphase ISM. The exponent $\beta$ is a free parameter which we take to have a value of $2$ (see next section). Equation (7) 
is then compared to the Eddington rate, $\dot{M}_{edd}(M)$, given the SMBH's mass at time $t$ such that $\dot{M}_{BH,final}(t) = \text{min( }\dot{M}(t)\text{, }\dot{M}_{edd}(M_{BH}(t)\text{ ) )}$. 

\begin{figure}
\includegraphics[trim=13mm 0mm -50mm 28mm, clip, width=100mm]{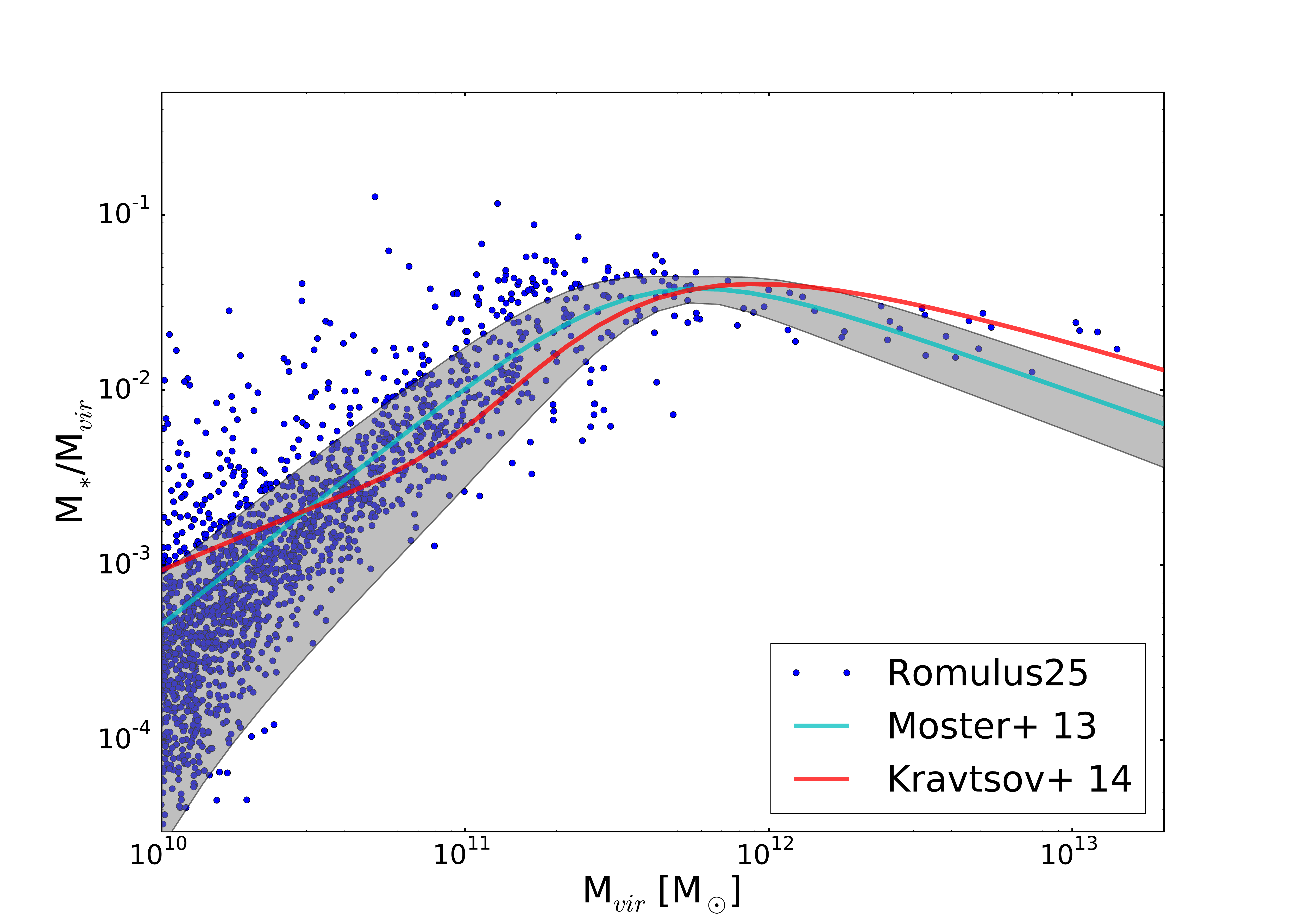}
\caption{{\sc Stellar Mass Halo Mass (SMHM) Relation.} Data from {\sc Romulus25} at z = 0.25 is shown in blue, plotted against two abundance matching relations from \citet{moster13} and \citet{kravtsov14}. Any halo at least partially within the virial radius of a larger halo is not counted in this analysis in order to exclude satellites and interacting systems. The grey region shows the error in the \citet{moster13} relation, calculated from the errors reported for the best fit parameters. The stellar and virial masses for each halo are corrected to make them more directly comparable to observations following  \citet{munshi13} (see \S2.3). 
Our results match well with those from abundance matching. Of particular interest are the high mass galaxies (M$_{vir}>10^{12}$M$_{\odot}$), which indicate that SMBH feedback is correctly regulating their growth.}
\label{smhm25}
\end{figure}

\subsection{Calibration of SMBH Free Parameters}

Our model of SMBH accretion and feedback has two free parameters
controlling the accretion rate ($\beta$) and the efficiency at which
radiated energy is transferred to surrounding gas ($\epsilon_f$). In a
similar approach as for the star formation parameters (see \S3 and Appendix A)
we run 48 zoom-in cosmological simulations, in identical sets of four
galaxies ranging from dwarf to Milky Way masses over several choices
of these two parameters.  Each set of simulations was run using 
the same set of star formation parameters, optimized in a separate parameter
search without SMBH physics (see section 3). This ensures that we start with a
model that performs as well as possible before the inclusion of SMBH physics.
 Figure~\ref{params} shows the results
of this search graphically. We tested values for $\beta$ between 1.5 and 3 and values for $\epsilon_f$ between 0.005 and 0.1. Our parameter
space exploration was guided by the Kriging algorithm (see Appendix A). Each parameter set was graded in the 
same way as described in \S3, each galaxy being compared to each scaling relation.
Changing the parameters just for SMBH physics has enough
of an affect to clearly isolate a `best' set of parameters, i.e. the one in which the summed deviation
of each galaxy from each scaling relation was the least.
We find that the model that performs the best overall has $\beta = 2.0$ and $\epsilon_f = 0.02$ and we
adopted those values for all the production runs. While no explicit assumption has been made in the model
 for the mass scale at which SMBH feedback becomes important, the dwarf galaxy stellar mass and HI content exhibit
 minimal dependence on SMBH model nor a dependence on the inclusion of SMBHs at all, as several iterations of
 the dwarf galaxy simulation never form a central SMBH.

\begin{figure}
\includegraphics[trim=5mm 1mm -40mm 15mm, clip, width=105mm]{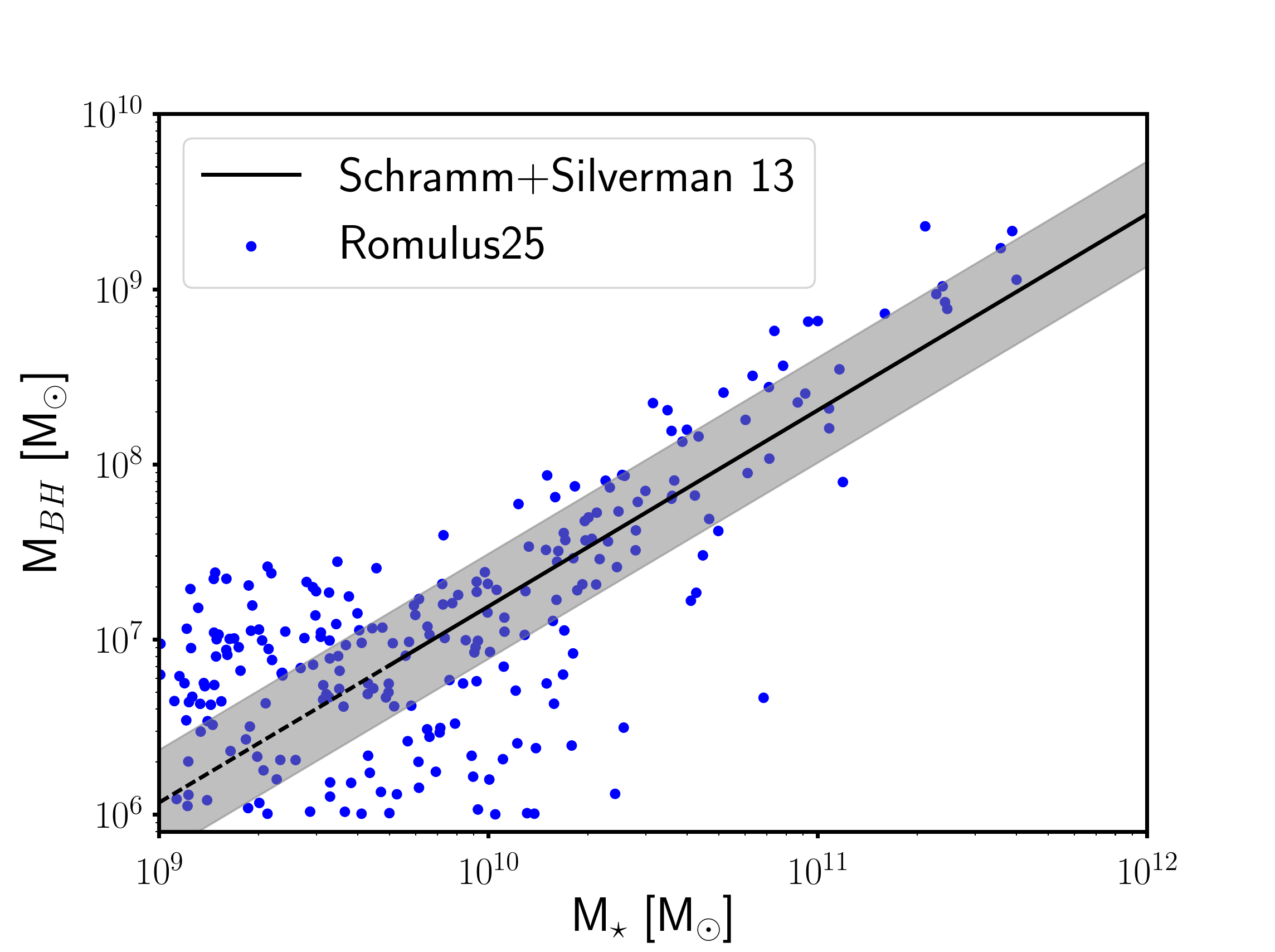}
\caption{{\sc The SMBH Mass Stellar Mass Relation}. Each point plots the mass of the largest black hole in each galaxy against each galaxy's stellar mass, corrected by a factor of 0.6 from the total stellar mass in each halo (see \S2.3). Also shown is the empirical relation from \citet{schramm2013}, where the grey region represents the $1-\sigma$ scatter and the dashed part of the line is where the relation has been extrapolated past observations. The overall match to the data is good, particularly at higher masses. High mass galaxies tend to exhibit less scatter and lie near the relation, though slightly biased toward higher mass SMBHs. Less massive systems show a broader scatter in black hole mass. The relation from \citet{schramm2013} was derived from higher mass galaxies and there is evidence that smaller, star forming galaxies lie on different relations \citep{ReinesVolonteri15,savorgnan16}}
\label{bhmstar}
\end{figure}

\section{First Results from {\sc Romulus25}: The Build-up of Stars and Black Holes}

In this section we present initial results from our flagship {\sc Romulus25} uniform volume simulation, run to $z = 0$. It should be noted
that such a small volume will miss some of the effects of large-scale structure and will not include 
the population of satellite galaxies in large halos. We see this effect most strongly in regards to downsizing of both
star formation and SMBH accretion (see below). In future work, we will include the cluster simulations
in our analysis as well. Within the scope of this paper, we find the {\sc Romulus25} simulation to be
sufficient as a proof of concept that our method produces realistic galaxies and SMBHs at $z = 0$.


\begin{figure}
\includegraphics[trim=10mm 0mm -50mm 20mm, clip, width=105mm]{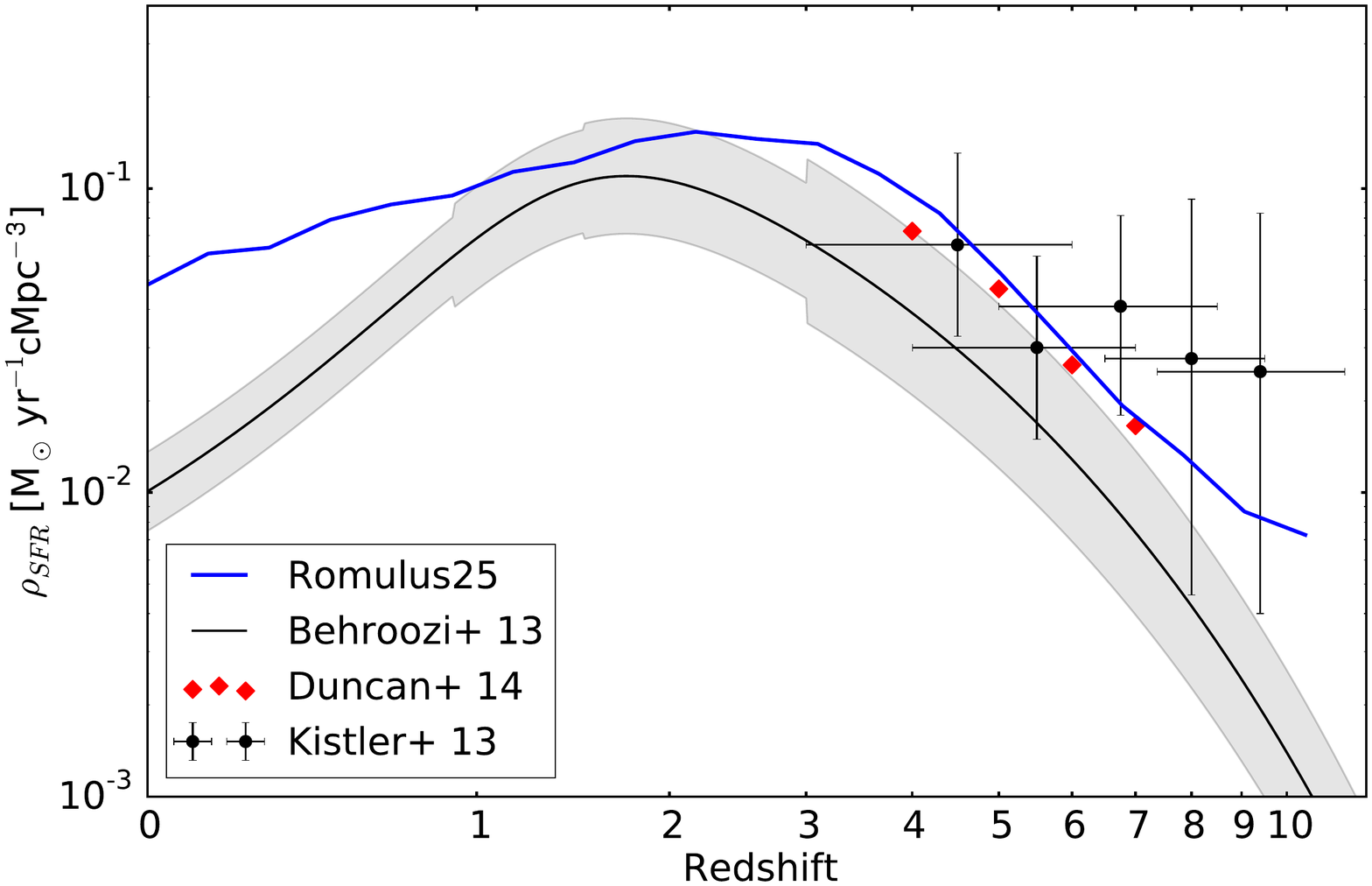}
\caption{{\sc Cosmic Star Formation History.} The solid blue line shows the total cosmic star formation history in {\sc Romulus25} plotted against a fit to observation data from \citep{behroozi13b} as well as recent high redshift observations \citep{kistler13,duncan14}. The grey region represents the spread in observational data for different redshift bins, as reported by \citet{behroozi13b}. {\sc Romulus25} accurately reproduces the evolution of the cosmic star formation rate density at high redshift, reaching a maximum at z = 2 and declining toward lower redshift. The overproduction of stars at low redshift, which is in stark contrast with observations, is due to only a handful of high SFR systems, a result of our relatively small volume. A 25 Mpc volume lacks larger systems that would better sample the effect of cosmic downsizing at late times. At $z > 5$ a significant portion (50\%-90\%) of star formation in {\sc Romulus25} occurs galaxies with stellar masses less than $10^8$ M$_{\odot}$, a regime where the observed luminosity function is not well constrained \citep{anderson17}.}
\label{csfh}
\end{figure}

Figure~\ref{smhm25} shows the stellar mass halo mass (SMHM) relationship in {\sc Romulus25} at z = 0 after removing all satellite galaxies from the sample. Our results are consistent with results from \citet{moster13}, which our model has been calibrated to reproduce, as well as \citet{kravtsov14} for halos spanning more than three decades in mass. It should be noted that while these results are in part due to our parameter calibration, the results for high mass halos (M$_{vir} > 10^{12}$ M$_{\odot}$) have not been calibrated and can be considered predictions of our model. At M$_{vir} > 10^{12.5}$, the {\sc Romulus25} halos match better to the \citet{kravtsov14} results. As discussed in \S2.3, we utilize the corrections from \citet{munshi13} for the stellar and virial masses to attain a more `apples-to-apples' comparison. The correction, particularly when applied to larger group-size halos, accounts for the mass that exists in extended stellar halos and satellites.

Figure~\ref{bhmstar} plots the mass of SMBHs in {\sc Romulus25} against the stellar masses of their host galaxies, again applying the correction from \citet{munshi13}. Satellite galaxies have also been removed from this sample. This is another empirical relation that we had used to constrain our sub-grid model, so the fact that the simulation data matches the relation from \citet{schramm2013} is a success of our parameter search technique. High mass galaxies show less scatter than low mass galaxies, but are slightly biased to higher SMBH mass compared to the empirical relation. At low mass, we see a lot of scatter, both above and below the relation. While it is beyond the scope of this paper to examine in detail the nature of this scatter, it follows from recent observations that low mass, star forming galaxies have significantly more scatter in SMBH mass than higher mass galaxies, indicative  that not all galaxies should lie on the same relation \citep{ReinesVolonteri15, savorgnan16}. The significant scatter above the relation could be explained by tidal stripping 
\citep{volonteri08b, volonteri16, barber16}, but we have removed satellite galaxies, making this connection less obvious. Likely it is due to stochastic SMBH growth in smaller galaxies. We will explore this further in future work.

The parameter search was meant to ensure that stars and SMBHs form and grow in the correct places. 
This is achieved in {\sc Romulus} out to mass scales beyond those that the parameter search probed.  Of particular interest are the high mass halos (M$_{vir}>10^{12}$ M$_{\odot}$) that were not explicitly constrained with our parameter search and represent the regime in which feedback from SMBHs dominates stellar feedback in regulating star formation \citep{croton2006,keller16}. The fact that these halos produce galaxies with stellar masses very similar to abundance matching results as 
well as SMBH masses that are consistent with empirical scaling relations is a very promising result.
How and when the growth of stars and SMBHs occur in {\sc Romulus25} is also
a testable prediction of the model.

\begin{figure}
\includegraphics[trim=20mm 5mm -50mm 25mm, clip, width=105mm]{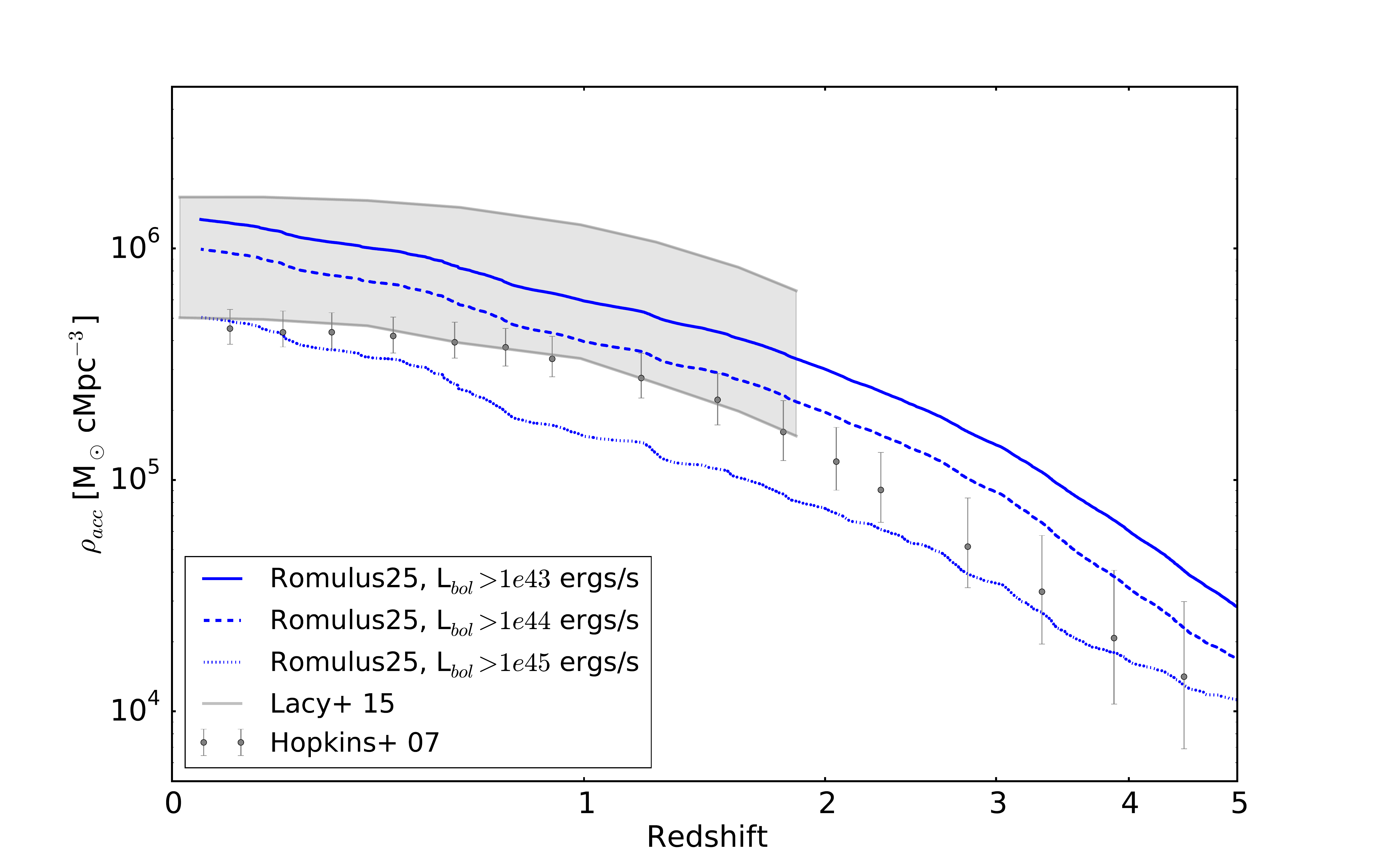}
\caption{{\sc SMBH Accretion History.} The cumulative mass density accumulated in luminous SMBH accretion events in {\sc Romulus25} across cosmic time. SMBH growth is faster at high redshift and slows down at later times. Higher luminosity systems (L$_{bol} > 10^{44}$ ergs/s) account for 50-80\% of the accreted mass density at all times. The late time evolution at z < 1 is driven by a small number ($\sim1-5$) of systems with L$_{bol}>10^{45}$ ergs/s. These results are consistent with the integration of AGN luminosity functions out to high redshift \citep[][shown as the grey region for a range of different values of radiative efficiency]{lacy15}. Also shown are the results from \citet{hopkinsBH07}, which are the result of different assumptions regarding absorption and bolometric corrections.} 
\label{rhoacc}
\end{figure}


\begin{figure*}
\includegraphics[trim=40mm 0mm -30mm 20mm, clip, width=210mm]{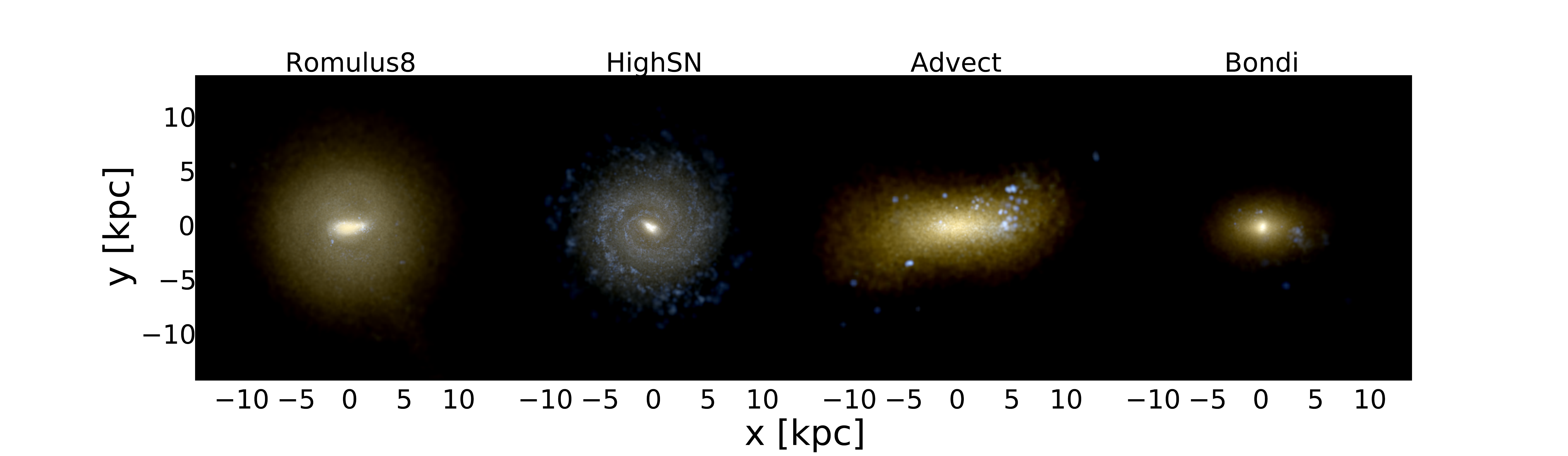}
\caption{{\sc Mock Images of Stars} in the largest galaxy from the Romulus8, HighSN, Bondi, and Advect simulations at z = 0.5. The virial mass of the host halo is $\sim2\times10^{12}$ M$_{\odot}$. On average, galaxies of this size should be quenched by this time \citep{papovich15}. Colors are based on the contribution of different bands within each pixel using U (blue), V (green), J (red) assuming a Kroupa IMF, so young stars look blue and older stars look yellow. These images are indicative of the importance of physically motivated SMBH physics implementations on the evolution of large galaxies. It is clear that the inclusion of only SN feedback (HighSN) is not enough to quench the galaxy. SMBH feedback is able to quench in all cases, but the morphology and star formation history (see figure~\ref{sfh1}) are noticeably affected by the details of the implementation.}
\label{rom8image}
\end{figure*}


Figure~\ref{csfh} shows the cosmic star formation history in {\sc Romulus25}, which matches nicely with observations at high (z > 2) redshift, reaching a maximum just before $z = 2$ and then dropping off accordingly toward $z = 0$. At high redshift (z > 5), we find the bulk of star formation is occurring in small galaxies (M$_{\star}<10^8$ M$_{\odot}$), likely missed by high redshift observations \citep{anderson17}. This explains why {\sc Romulsu25} lies above the derived star formation history from \citet{behroozi13b} but is more similar to estimates using more recent data that are more sensitive to lower mass galaxies. At low redshift (z < 2) {\sc Romulus25} lies far above the observed star formation rates. This overproduction of stars at low redshift is due to only a handful of high SFR systems, a result of our relatively small volume which does not properly sample the higher density environments needed to recover the behavior of cosmic downsizing at late times.

Figure~\ref{rhoacc} plots the cumulative mass density accumulated in luminous SMBH accretion events across time. We only include data from SMBHs with mass greater than 110\% of their initial seed mass (see \S 5.1). We verify that excluding these systems does not substantially change our results. The cuts in  luminosity are meant to not only show the contribution of different varieties of active SMBHs, but also ensure that we only sample the portion of the luminosity function that can be accurately constrained by observations. We verify that for each luminosity cut, the contribution from low Eddington ratio SMBHs, where accretion is thought to become radiatively inefficient ($\lambda_{fedd} < 0.01$) is negligible. At early times, the black hole population grows more rapidly, slowing as it gets to lower redshifts. At all times the overall growth is dominated ($\sim 50-80$\%) by the more luminous SMBHs (L$_{bol}$ $> 10^{44}$ ergs/s). Below z = 1, a significant fraction of this growth is taking place in a small number (1-5) of very luminous SMBHs (L$_{bol}$ $> 10^{45}$ ergs/s). This is similar to the effect we see with star formation, where our small volume is unable to appropriately sample AGN downsizing.

The overall growth of the black hole population in {\sc Romulus25} is consistent with observations. The grey region is from \citet{lacy15} and is obtained from integrating the observed AGN luminosity function between z = 0 and z = 5 assuming a radiative efficiency, $\epsilon_r$ between 0.06 (upper limit) and 0.18 (lower limit) and the data points with error bars are from \citet{hopkinsBH07}. The data from \citet{lacy15} were obtained from Spitzer observations in the mid-infrared. This makes them less sensitive to absorption, which can significantly impact optical and X-ray observations across all redshifts \citep{treister10a,lansbury15,buchner15, lacy15}. However, the data from \citet{lacy15} are poorly constrained at redshifts higher than $\sim2$, which is why we limit this region to $z < 2$. The higher luminosity data from {\sc Romulus25} (L$_{bol} > 10^{44}$ ergs/s) fits well with both observational data sets shown. The divergence away from the \citet{hopkinsBH07} data at $z < 1$ is due to a small number of bright SMBHs, a consequence of our relatively small volume. Bolometric luminosities less than $10^{44}$ ergs/s represent a regime in which the observed luminosity functions are poorly constrained, particularly at high redshift, and sensitive to assumptions regarding the redshift and luminosity dependencies of absorption and bolometric correction \citep{merloni16}.

These initial results show that {\sc Romulus25} 1) produces galaxies with stellar and black hole masses that are consistent with observations at low redshift (Figures~\ref{smhm25} and~\ref{bhmstar}) and 2) produces high redshift star formation and SMBH accretion histories that are consistent with observations, where differences arising at low redshift (z < 2) are due to our small volume not being able to properly capture downsizing for high mass galaxies. These results show the strength of both our SMBH sub-grid model and our method for free parameter calibration. We leave the analysis of gas content and kinematics in {\sc Romulus} for future work.


\section{Black Hole Feedback Compared to Stellar Feedback}

In this section, we wish to explore the differences in SMBH and supernovae (SN) feedback mechanisms.
It is often possible to tune parameters in order to reproduce observations of galaxies of a certain mass.
During our parameter search (see \S3 and Appendix A) we found that the models for star formation and SN feedback without SMBHs
that produced the most realistic galaxies in MW-mass halos did not work well in reproducing realistic smaller galaxies. 
However, in this section we go beyond this to show that SMBH feedback is not only
a crucial ingredient for reproducing scaling relations across all mass scales, it also has important consequences
for reproducing the evolution of galaxies. In this case, we focus on MW-mass halos (M$_{vir} \sim 10^{12}$ M$_{\odot}$).

We compare two $8$ Mpc uniform volume simulations, Romulus8 and HighSN (see Table 1), in order to gain insight into how the addition of extra feedback in the form of black holes compares to simply increasing the efficiency of SN feedback. The feedback efficiency in HighSN
was chosen based off of the value we found to best reproduce scaling relations for galaxies in $10^{12}$ M$_{\odot}$ halos. 
The simulations are run to z = 0.5 to avoid some of the biases such a small volume will introduce into the evolution at later times.

\begin{figure}
\includegraphics[trim=8mm 0mm -50mm 25mm, clip, width=105mm]{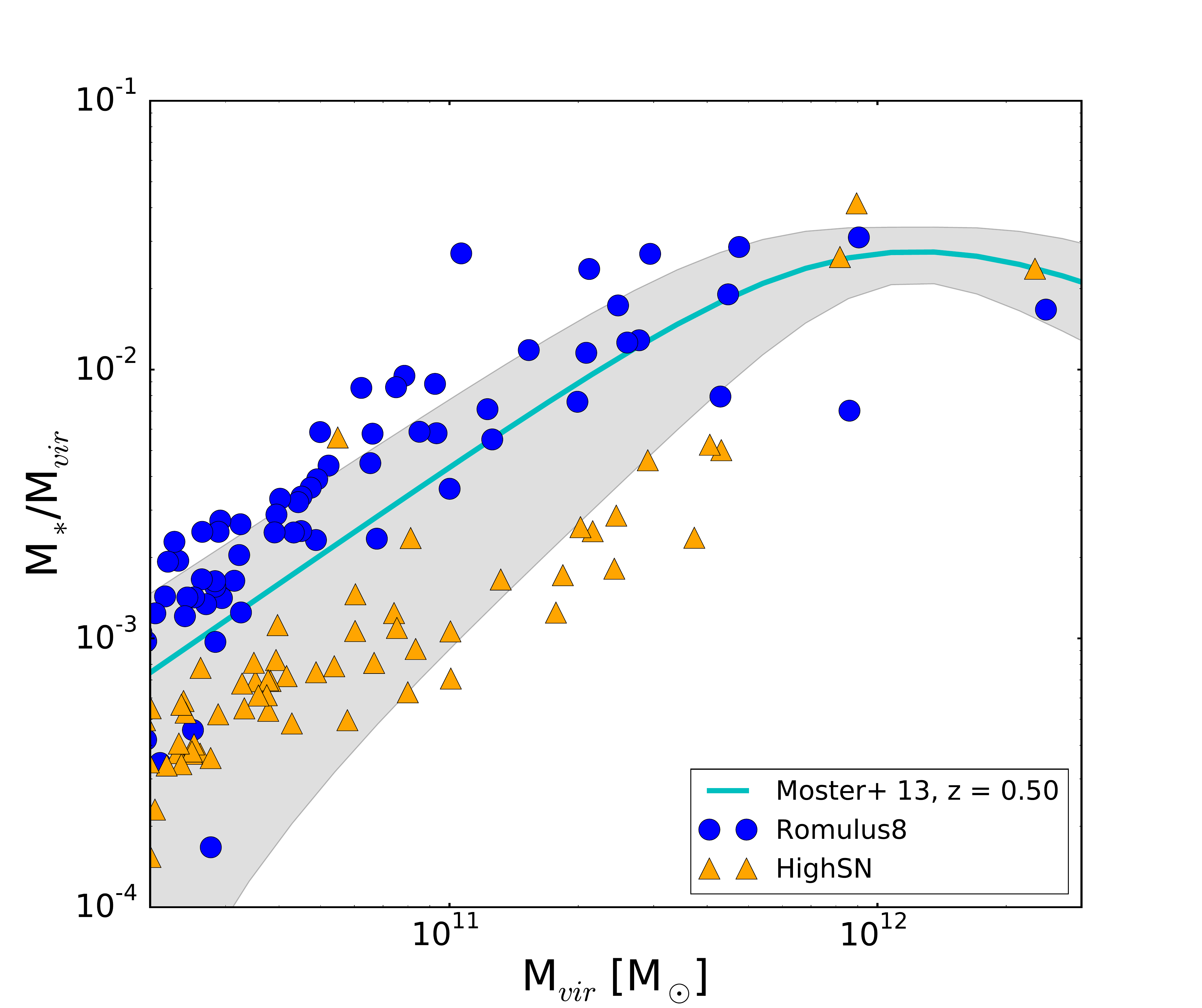}
\caption{ {\sc How Feedback from SMBHs and SN Affect SF Efficiency.} The SMHM relation for the Romulus8 (blue) and HighSN (orange) simulations. Increasing the efficiency of stellar feedback to produce stellar masses that match observations for higher mass galaxies (HighSN) causes and underproduction of stars in low mass systems. The high mass galaxies match the observed relations well in the HighSN simulation, but this success is misleading, as the galaxies maintain significant star formation through the end of the simulation (see figures~\ref{rom8image} and~\ref{sfh1}). The inclusion of black hole feedback combined with a lower stellar feedback efficiency (see table 1) produces realistic stellar masses in halos ranging from dwarfs to MW-mass.}
\label{smhm1} 
\end{figure}

Figure~\ref{smhm1} shows the stellar mass halo mass relationship for the two simulations, plotted against the $z = 0.5$ best fit relationship from \citet{moster13}, applying the correction to stellar and halo masses from \citet{munshi13}. The Romulus8 model fits the data well. The highSN model drastically under-produces stars in intermediate mass halos. This is, of course, due to the fact that SN feedback is much more efficient in lower mass halos that exhibit a shallower potential well \citep{G10,brook11}. Such high efficiencies are necessary, however, to reproduce observed stellar masses in higher mass halos without SMBH feedback. Because SMBH growth naturally depends on the host galaxy mass, SMBH feedback is able to preferentially limit the growth of higher mass galaxies, while not quenching the star formation in low mass halos.

\begin{figure}
\includegraphics[trim=30mm 0mm -50mm 25mm, clip, width=105mm]{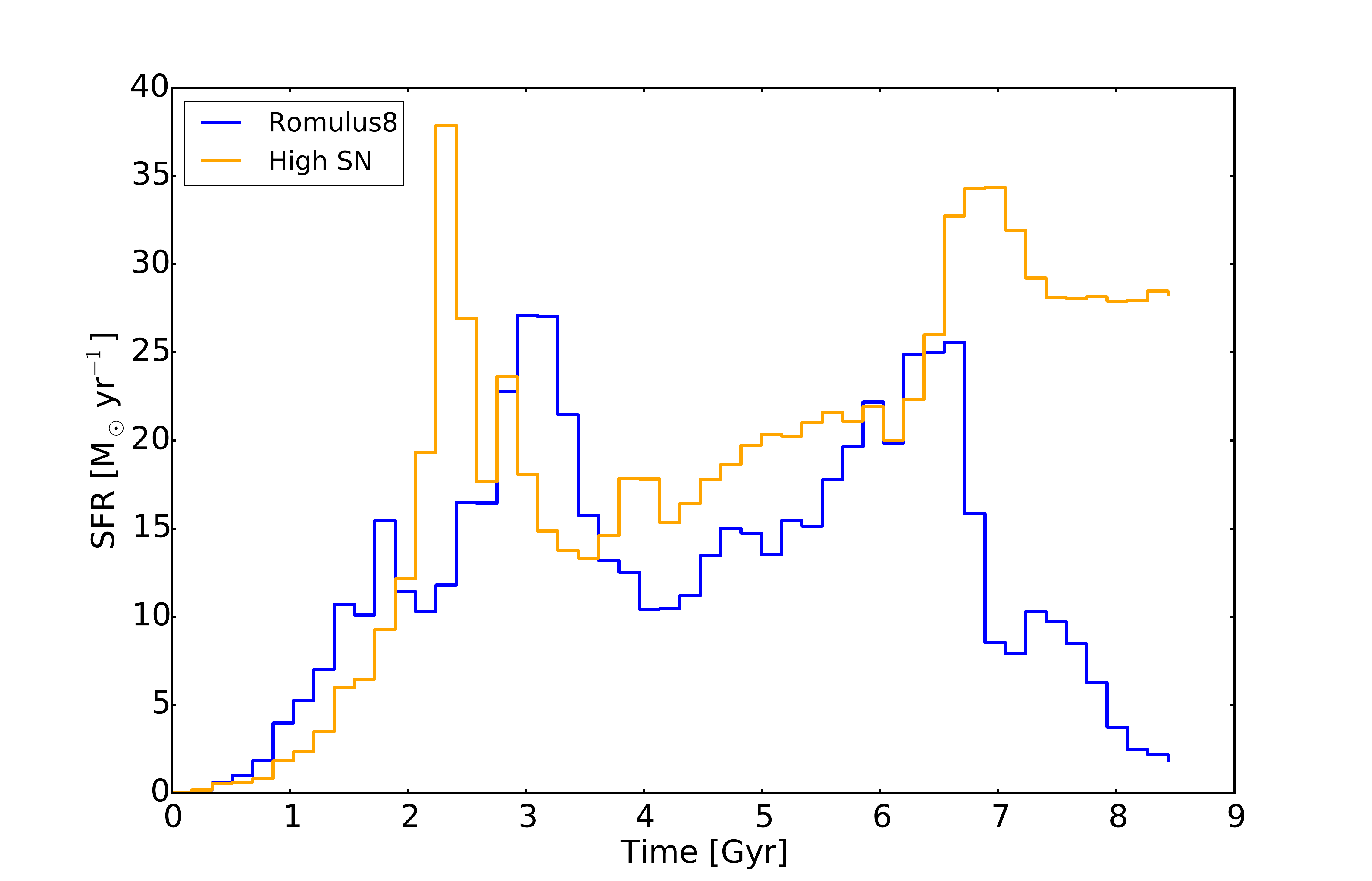}
\caption{{\sc SMBHs and Galaxy Quenching.} The star formation rate as a function of time for the most massive halo in the 8 Mpc volume, run with both the Romulus8 and HighSN models. The halo mass is consistent with being a Milky Way progenitor. The star formation histories are similar up until about 2 Gyr prior to the end of the simulation. While the enhanced SN feedback is able to make stellar masses consistent with observations (see Figure~\ref{smhm1}) the feedback from stars alone is unable to turn off star formation at late times, which is expected for systems of this mass \citep{papovich15}. With lower SN feedback but the inclusion of black hole accretion and feedback (Romulus8), the galaxy is able to attain both a realistic stellar mass and have star formation quench before z = 0.5.}
\label{sfh1} 
\end{figure}

Figure~\ref{sfh1} shows the star formation history of the most massive halo (M$_{vir}(z=0) \sim 2 \times 10^{12} \mathrm{M}_{\odot}$) in the volume for each simulation. 
While the final stellar masses are within realistic bounds in both simulations, the galaxy in Romulus8 has very low star formation by the end of the simulation while the same galaxy in highSN fails to quench. The majority of galaxies (70-80\%) in this mass range should be quenched by $z = 0.5$ \citep{papovich15}.

\begin{figure}
\includegraphics[trim=5mm 5mm -50mm 25mm, clip, width=110mm]{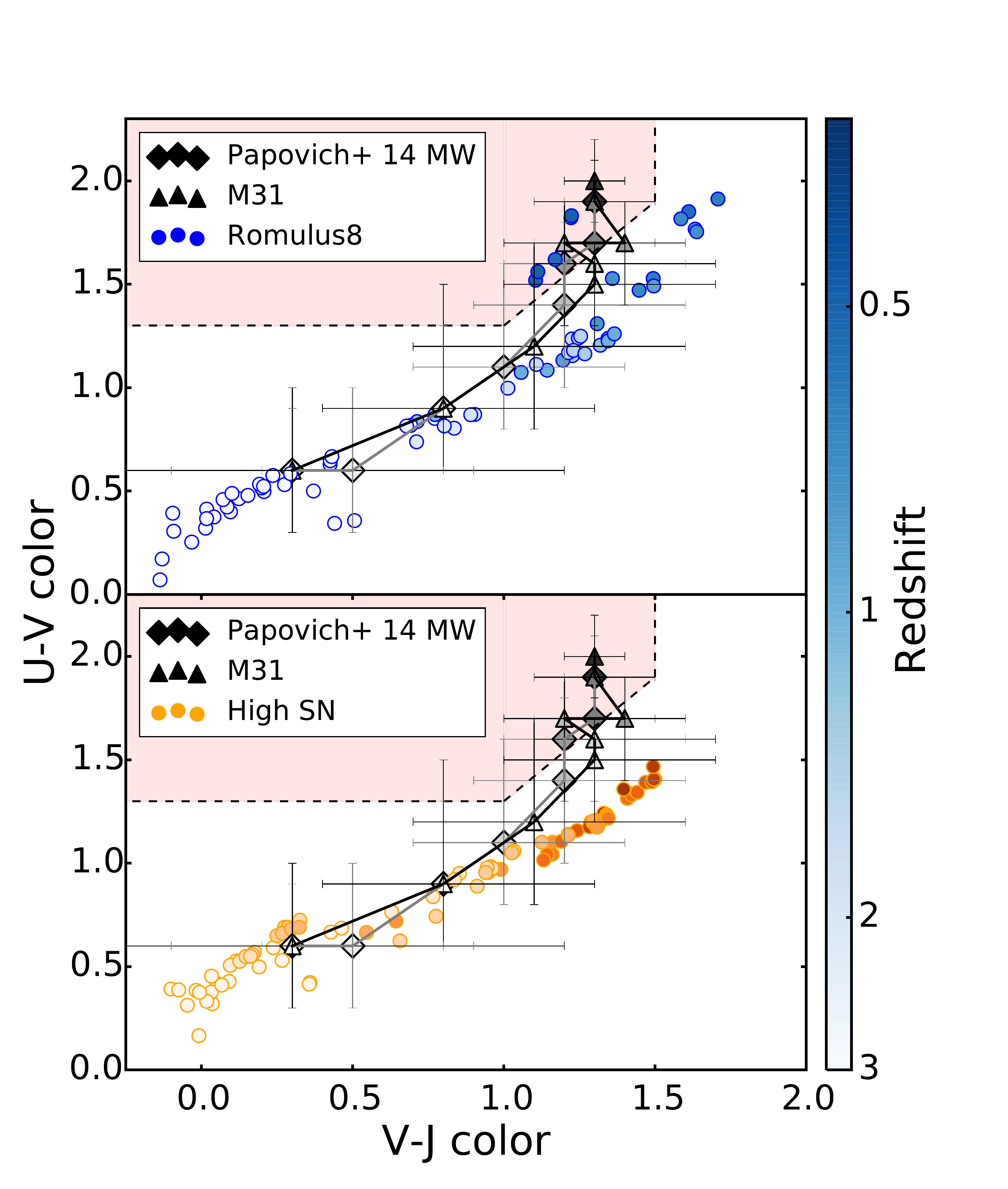}
\caption{{\sc A Color-Color History of MW halos} in the Romulus8 and HighSN simulations with a simple prescription for the average dust attenuation (see Appendix B). Darker points represent lower redshifts. The observed data points (black) are from CANDELS and ZFOURGE, using abundance matching techniques to define Milky Way and M31 progenitors across cosmic time \citep{papovich15}. In Romulus8 (blue), the two Milky Way progenitors follow closely the average observed evolution, becoming quenched by z = 0.5. In the HighSN simulation (orange), the galaxy remains in the realm where color is dominated by dust attenuation and ultimately fails to quench by z = 0.5. Without black hole feedback, Milky Way mass halos remain very gaseous and dusty, with star formation continuing at high levels.}
\label{ccplot} 
\end{figure}

Figure~\ref{ccplot}  shows the color evolution of the two most massive galaxies in the simulations run with SMBH physics (Romulus8) compared to that run only with enhanced SN feedback (HighSN). The galaxies show different color evolution, with Romulus8 following much more closely the results from the CANDELS and ZFOURGE data \citep{papovich15}.  Colors from stellar emission are calculated using tables generated from population synthesis models using http://stev.oapd.inaf.it/cgi-bin/cmd \citep{marigo08,girardi10}. Dust is accounted for using a simple approach based on metallicity and cold gas content of a galaxy (see Appendix B). In the highSN simulation, the colors of the galaxies remain dominated by dust at late times, never falling into the `quenched' regime. The color evolution also fails to follow the evolutionary path seen in the multi-epoch observations.

SMBH feedback, because it is more concentrated than SN feedback, is able to drive more powerful winds, which can disrupt inflowing
 material and lead to galaxy quenching \citep{volonteri16IAU,pontzen16}. Here we have shown that this effect is important
 for reproducing the observed evolution of MW-mass progenitor galaxies. One of the failures of simulations without SMBH feedback is
 the inability to quench galaxies in MW-mass halos, something that our SMBH model is able to produce. 
 Quenching galaxies in halos of $\sim 10^{12}$ M$_{\odot}$ 
 has generally been challenging for modern cosmological simulations \citep[e.g.][]{bluck16}.

\section{Results from Different Black Hole Physics Implementations}

In this section we compare our implementation for SMBH dynamics and accretion (model Romulus8) against more common implementations found in large cosmological simulations (models Bondi and Advect). It is instructive to note that our parameter optimization was done using our SMBH implementation. While it may be possible to find a combination of parameters that create galaxies that fall on various empirical relations using these other models, the point of this section is to explore the effects that the additional physics our implementation includes have on galaxy evolution. 

\begin{figure}
\includegraphics[trim=10mm 0mm -50mm 10mm, clip, width=105mm]{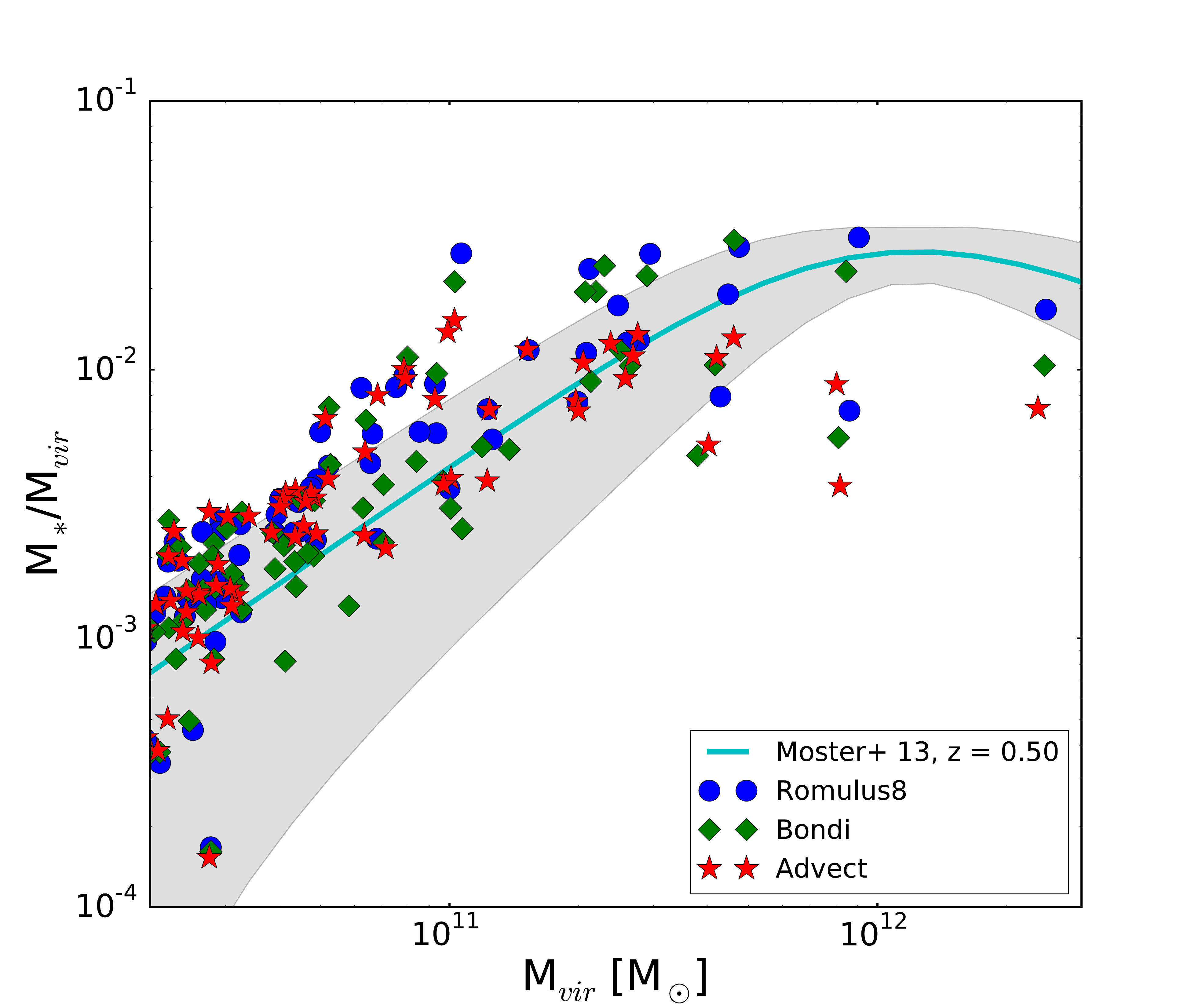}
\caption{{\sc The Effect of SMBH Implementation on SF Efficiency.} Same as Figure~\ref{smhm1} but for the Romulus8, Advect, and Bondi simulations. The stellar mass halo mass relation changes little between the simulations for low mass halos, but noticeable differences can be seen for halos with virial masses above $\sim2\times10^{11} \mathrm{M}_{\odot}$. SMBHs do exist in smaller halos in this simulation (see section 3.1) but, regardless of the SMBH physics implemented, small galaxies will not experience much black hole growth or feedback. For higher mass galaxies, artificial advection and Bondi accretion not limited by gas dynamics work to increase the effect of SMBHs on star formation compared to our implementation utilized in the Romulus8 simulation.}
\label{smhm2}
\end{figure}

We are again using a smaller 8 Mpc uniform volume realizations of our main simulation suite. Figure~\ref{smhm2} shows the SMHM relationship of the three simulations. The high mass end of the relationship is the only part noticeably  affected by the different models, indicating that a lack of SMBH growth in low mass galaxies is a natural consequence of the environment and not greatly affected by choice of sub-grid SMBH physics. Both aspects of our implementation (described in section 3) work to soften the effect of SMBHs on their host galaxy, as both Advect and Bondi have lower stellar masses at a given halo mass. 

\begin{figure}
\includegraphics[trim=0mm 5mm -50mm 10mm, clip, width=105mm, height=90mm]{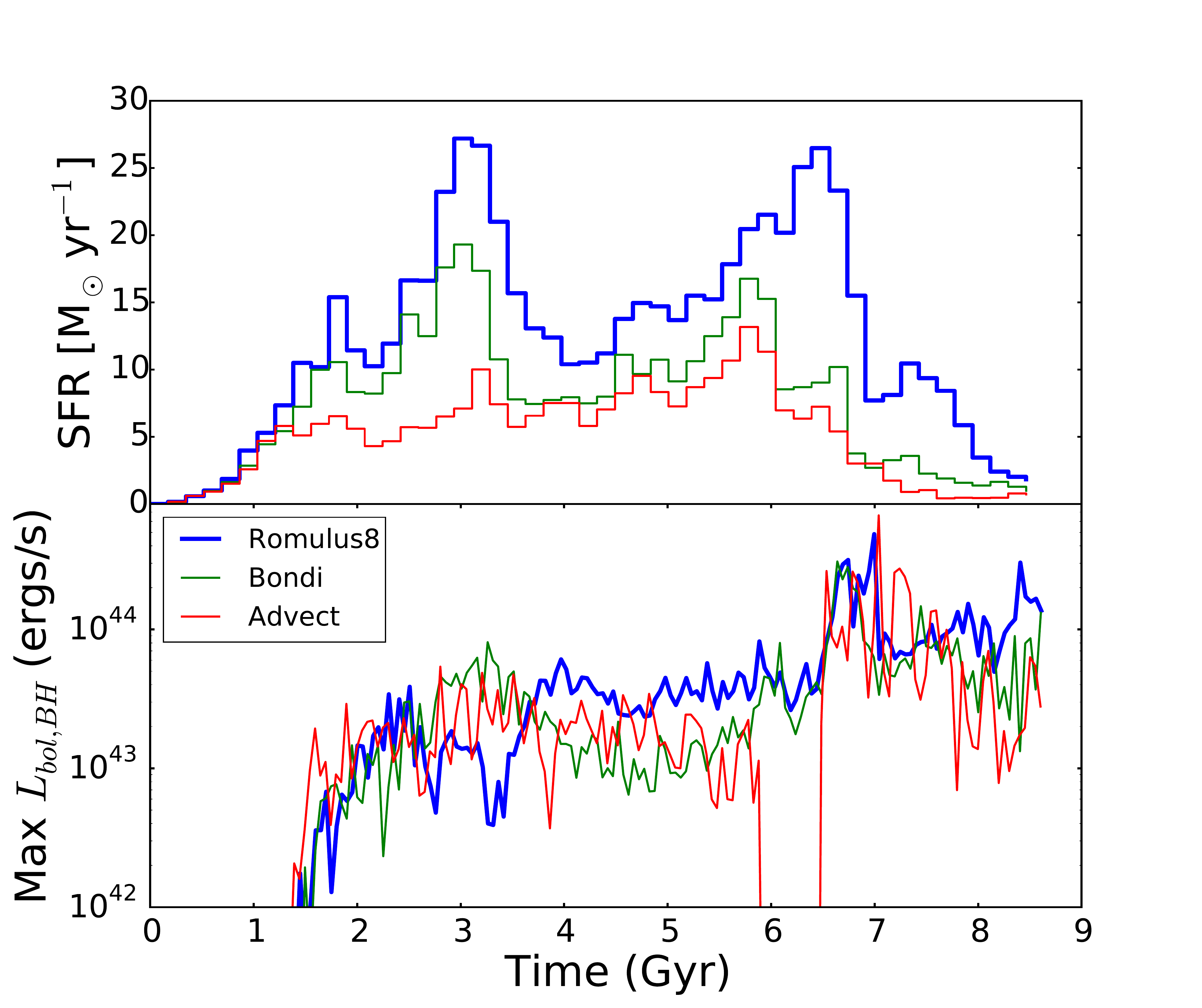}
\caption{{\sc The Effect of SMBH Implementation on the SF History of Massive Galaxies} (Top) The star formation history of the most massive halo in the 8 Mpc simulations, taken from the total stellar population of the galaxy at z = 0.5. A clear difference can be seen between Romulus8 (blue), Advect (red) and Bondi (green). (Bottom) For the same galaxy, the luminosity of the most luminous black hole across time within the galaxy's main progenitor branch. At later times Romulus8 has more active black holes. The values of luminosity are averaged over 50 Myr intervals. The strong dip in the red curve is due to the active black hole instantaneously transferring between two halos during a major merger.}
\label{sfh2}
\end{figure}

Synthetic images of the stars in the central galaxy of the most massive halo in the volume are shown in Figure~\ref{rom8image}, where a clear distinction between the three models can be seen. In Figure~\ref{sfh2}, we plot the star formation history of the most massive halo in the volume and the luminosity of the brightest black hole in that halo throughout time, averaged over 50 Myr intervals. While the star formation histories are quite different between models, the accretion history of black holes in the halo are not strikingly different and at later times the Romulus8 model is the most active of the three.

\begin{figure}
\includegraphics[trim=20mm 00mm -30mm 25mm, clip, width=100mm]{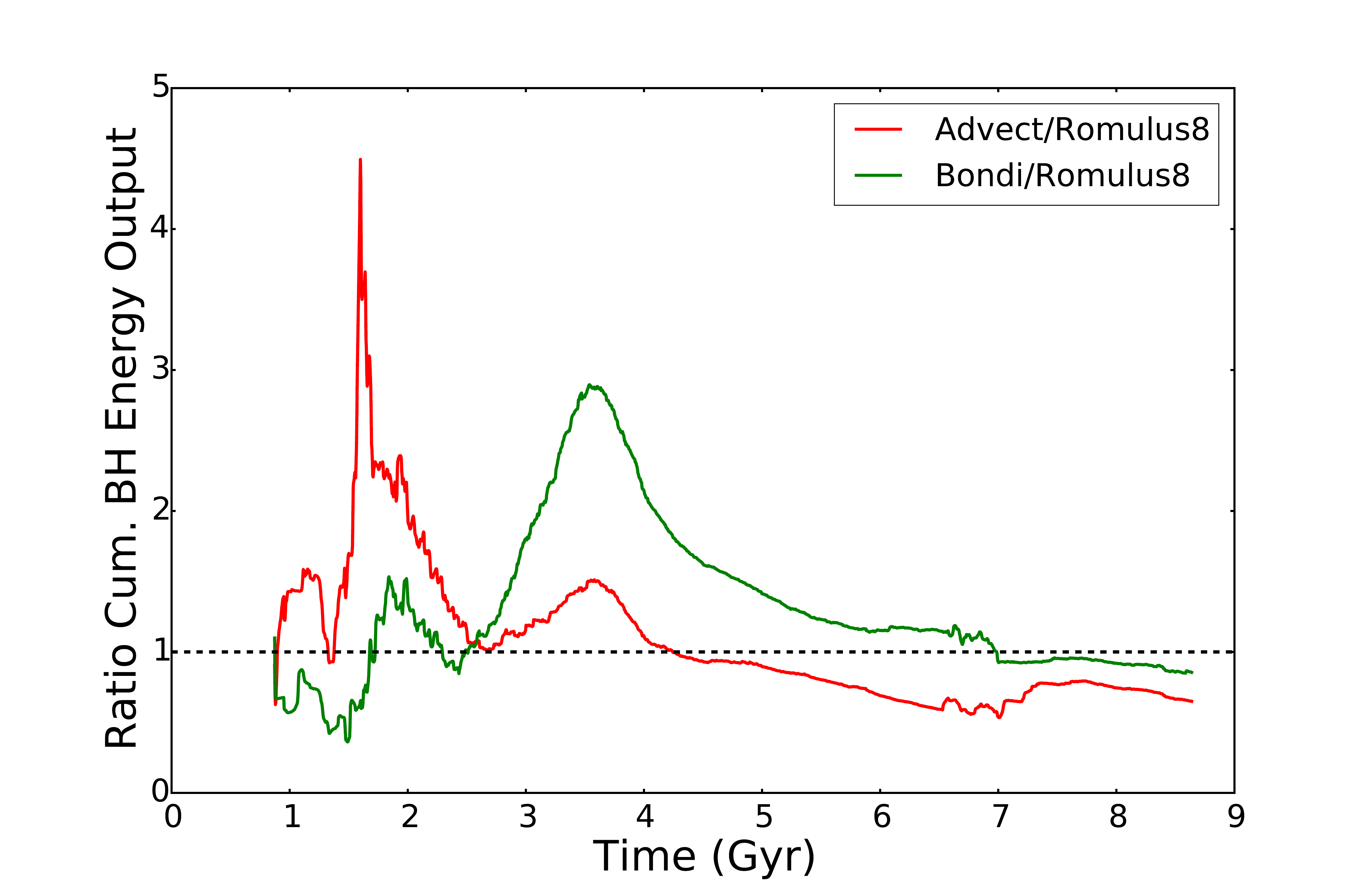}
\caption{{\sc The Cumulative Energy Output from SMBHs} within the most massive halo in the 8 Mpc simulations. The Advect (red) and Bondi (green) models compared with Romulus8 across cosmic time. During the first 4 Gyr of the simulation, the Romulus8 halo experiences less feedback from SMBHs.}
\label{cumbhfb} 
\end{figure}

The important difference in how black holes regulate the star formation of their host galaxies occurs at high redshift. Figure~\ref{cumbhfb} plots the cumulative energy output of black holes within the central galaxy, tracking the halo backward in time along its main progenitor branch. The energies are reported relative to that in the Romulus8 model. We find that both Bondi and Advect experience more activity during the first several billion years of the simulation. The implications from this are 1) early black hole activity can have important consequences for later galaxy evolution and 2) black hole dynamics and angular momentum limited accretion play an important role in determining accretion in the early Universe. It makes sense that the former is true, as the environment in which the black holes are active is different, namely the host halo is smaller, which would allow feedback from black holes to play a more drastic role in shaping the host galaxy. At early times, the black holes will exist in smaller galaxies that are undergoing more interactions, thus the black holes are more likely to become perturbed away from the galaxy center if they are allowed. In addition, at earlier times, when star formation is climbing toward its peak, one would expect to see more cold, disk dominated galaxies. 

\begin{figure}
\includegraphics[trim=25mm 0mm -30mm 25mm, clip, width=100mm]{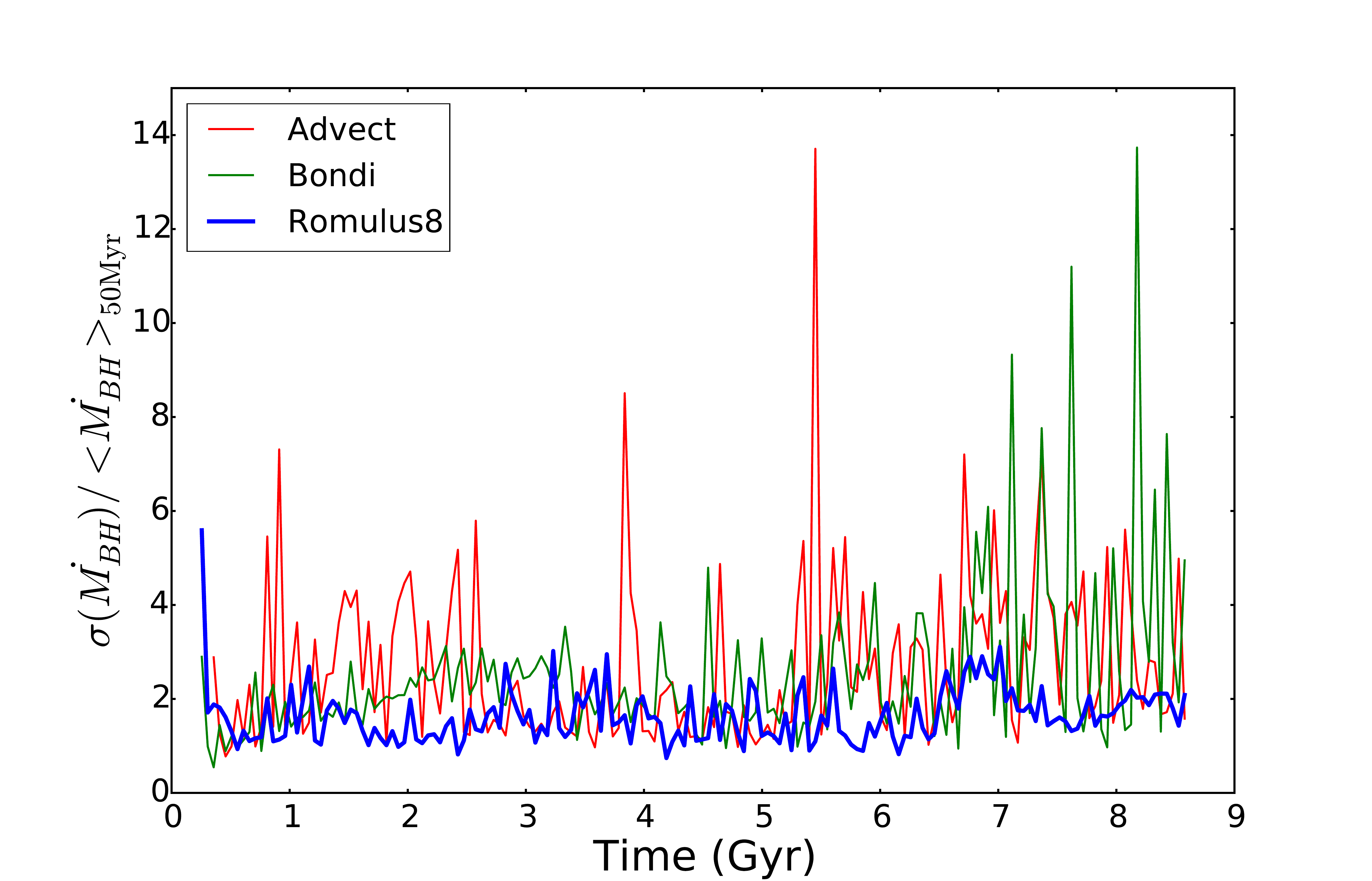}
\caption{{\sc The `Burstiness' of SMBH Accretion} for the most massive black hole in the most massive halo in the three 8 Mpc simulations: Romulus8 (blue), Advect(red), and Bondi (green), defined to be the ratio of the standard deviation to the mean accretion rate over 50 Myr timescales. In both Advect and Bondi we see that the black hole experiences a much more bursty accretion history.}
\label{bhsigma} 
\end{figure}

The black hole model not only affects where and when accretion takes place, but also how the accretion rate varies on smaller timescales. Figure~\ref{bhsigma} plots the standard deviation of the accretion rate for the most massive black hole in the most massive halo of the simulation, taken over intervals of 50 Myr and normalized by the average accretion rate throughout that time. For the entire simulation, both Advect and Bondi experience a significantly more bursty accretion history. So, while the smoothed accretion rate may look relatively similar between the models (see figure~\ref{sfh2}) there is a more bursty process occurring on smaller timescales. For Advect, the cause is numerical, as repositioning each timestep can cause the black hole to feel numerical noise, as the location of the potential minimum fluctuates \citep{wurster13a,tremmel15}. In the Bondi simulation, the reason for such bursty accretion is that, without regulation, the accretion rate will rise quickly with gas density, which in turn will create a stronger feedback event that will drive gas back temporarily. The black hole then waits for the gas to relax again and the process continues. Including the gas dynamics in the accretion calculation softens this process because dense gas tends to also be in a disk, which will feel rotational support.

A future paper is planned to look in more detail of the relative effects of angular momentum limited accretion and stellar feedback on the evolution of Milky Way and sub-Milky-Way mass galaxies. Within the scope of this paper, the important result is that both black hole dynamics and angular momentum regulated accretion have an appreciable effect on galaxy evolution for galaxies in higher mass halos (M$_{vir} > 10^{11.5}$).

\section{Application: Understanding Dual AGN in a Larger Context}

Dual AGN, systems with multiple active black holes, are beginning to be observed in the local Universe \citep{comerford11, comerford13, comerford15} and represent an important regime for studies of SMBH-galaxy co-evolution, as they are a transient state possibly connected  to a recent or on-going galaxy merger. Being able to reproduce such systems in simulations is necessary in order to gain a theoretical understanding of their place in the broader context of SMBH-galaxy co-evolution. Some important work has already been done to that end \citep{vanwassenhove12, Hirschmann14,steinborn15DualAGN} and the methods presented in this paper represent the logical next step.

Our approach to black hole physics is particularly well suited for realistically modeling dual AGN because we are able to accurately follow the dynamics of black holes within their host galaxies as they get perturbed away from center or fall into a new host following a galaxy merger event. We are able to track the black hole orbits to an accuracy of the simulation's resolution limit (250 pc) and without making assumptions regarding the larger scale structure of the galaxy or halo in which the black hole resides.  We are therefore not only able to create dual AGN down to a separation of $< 1$ kpc, we can follow the evolution of the system accurately throughout the parent system's evolution.

\begin{figure*}
\includegraphics[trim=10mm 0mm -30mm 30mm, clip, width=210mm]{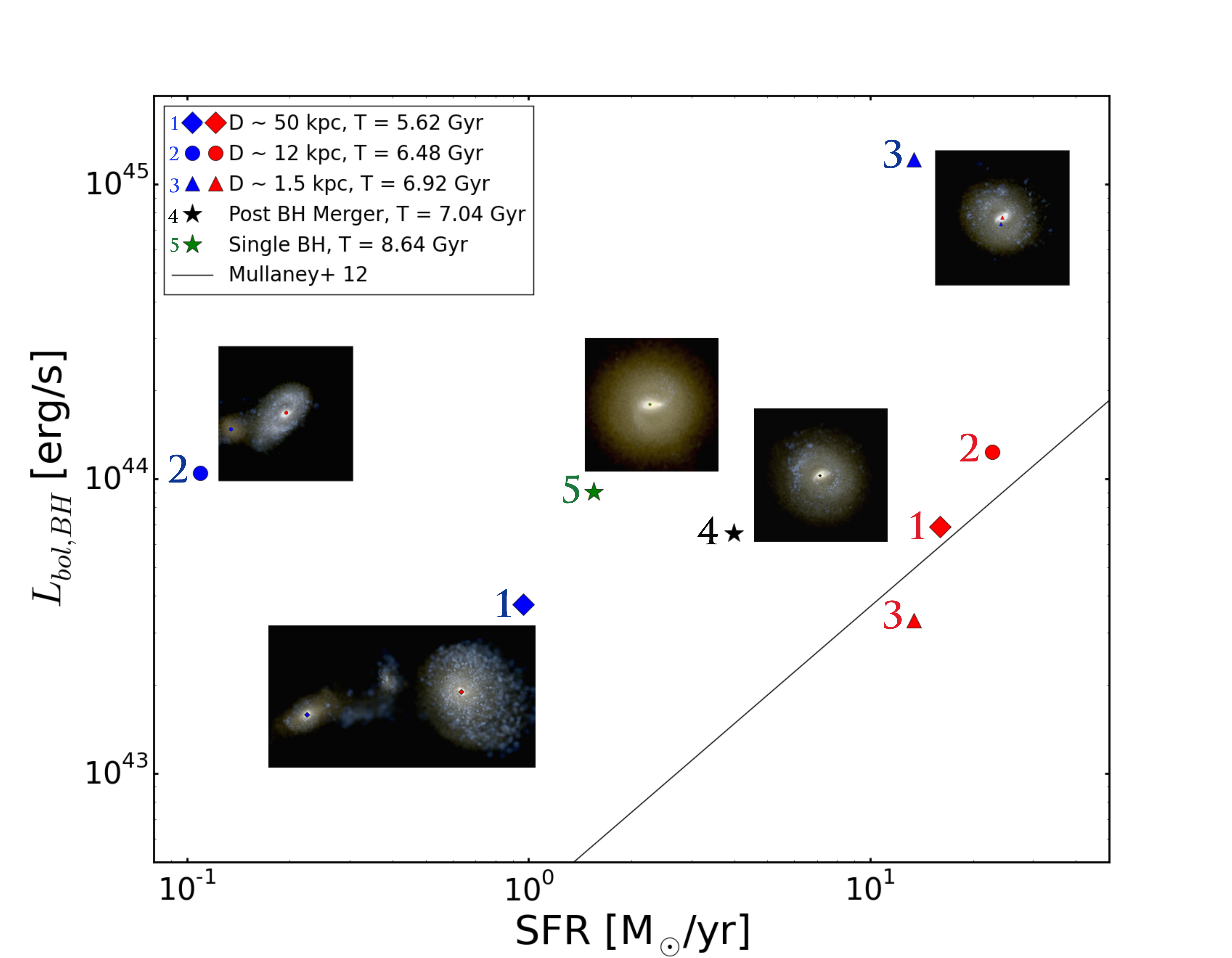}
\caption{{\sc The Evolution of Dual AGN} The evolution a merging galaxy pair and resulting remnant galaxy in terms of star formation rate and black hole luminosity. Thumbnails showing the stars of the galaxies are shown along with each data point set. The different colored points in each thumbnail represent the positions of the active black hole(s) at each time. The data points and thumbnails shown were chosen to encapsulate several important phases of evolution: 1) the beginning of the interaction, when the smaller galaxy has just entered the virial radius of the larger galaxy and is being stripped and environmentally quenched 2) the end of the galaxy merger phase, where there are two distinct galaxies, but the smaller one has been completely stripped. 3) the remnant resulting from the galaxy merger, still with two separate, bright black holes 4) just after the two black holes merge. 5) the merger remnant after it has been given time to relax, showing the galaxy quenching under the influence of a single, still very active black hole.}
\label{dualagn} 
\end{figure*}

An example of dual AGN created using our approach, taken from the Romulus8 simulation during the last major merger of the most massive halo in the volume, is shown in Figure~\ref{dualagn} (the same halo used for analysis in the previous two sections; see Figures~\ref{rom8image},~\ref{sfh1}, and others in \S7 and \S8).  We show 5 snapshots in time of a single galaxy merger that results in three instances of a dual AGN with separations of $\sim50$ kpc, $\sim 12$ kpc, and $\sim 1.5$ kpc. These are progenitor events leading up to a black hole merger and the quenching of the host galaxy, which by the end of the simulation has halo and stellar masses similar to the Milky Way. By looking at each snapshot, we gain insight into how the simulation is evolving. The entire process takes less than $1.5$ Gyr from the initial dual AGN event until black hole merger. Two of these dual AGN  events (snapshots 2 and 3 on the plot) look analogous to systems found by \citet{comerford15}. When searching for these events, we defined `active' to mean a bolometric luminosity of more than $10^{43}$ ergs/s.

To give the events more context, we plot the black hole luminosity as a function of star formation rate for the merging galaxies in each snapshot  (Figure~\ref{dualagn}). The smaller galaxy is in the process of being stripped by the larger galaxy. The original baryonic masses of the galaxies before the merger was $M_{1}/M_{2}\sim1.2$ and the ratio of black hole masses was $M_{BH,1}/M_{BH,2}\sim0.5$, where the less massive galaxy, denoted by 2, is the one that is being stripped and the one that hosts the more massive black hole.

The stripped galaxy is clearly in the process of being quenched by a combination of its environment and the active black hole within it. As the galaxies get closer, the black holes become more active. The star formation rate of the larger galaxy remains roughly constant while the stripped galaxy is further quenched. Throughout the interaction, the black hole activity and star formation rate of the more massive galaxy matches well with the relation derived from observations of z = 1-2 galaxies \citep{mullaney2012}. The stripped galaxy always lies above the relation. After the two galaxies merge, the black hole originally in the stripped galaxy becomes even more active, with a luminosity much higher than expected given the star formation rate in its new host. After the black holes merge, the central black hole remains very active and the galaxy moves further to the left on the plot as it quenches. This merger event, over the course of $\sim1.5$ Gyr and resulting in different instances of dual AGN, is the progenitor to a newly quenched galaxy. The heightened black hole activity corresponds with the quick decay of the star formation rate over the next billion years.

In the example given here from the Romulus8 simulation, we show that the Dual AGN event is a direct result of a major merger taking place between the galaxies. The black hole in the smaller galaxy becomes active as its galaxy quenches, with activity increasing as it moves closer to the more massive galaxy. In this case, having multiple black holes was indicative of a future black hole merger and would result in the quenching of what originally was a gas rich, star forming galaxy.

This is only one example, but it shows the level of detail with which we can approach the problem of Dual AGN. It is also indicative that these systems are not necessarily very rare across cosmic time, as we were able to generate a relatively long lived event in a volume of only 578 Mpc$^3$. In a future paper we will search both the 25 Mpc volume ({\sc Romulus25}) and the cluster ({\sc RomulusC}) for Dual AGN events across cosmic time, giving us a much larger sample to look at and understand better the physical processes necessary to generate Dual AGN. 

\section{Summary}

In this Paper, we present a novel approach for modeling SMBH formation, dynamics, and feedback that represents a marked improvement over currently common approaches utilized in most cosmological simulation to date. Our approach, combined with a new method of parameter optimization, has been applied to a new set of cosmological simulations called {\sc Romulus}.

We presented the initial results from our flagship simulation, {\sc Romulus25}, showing that our model reproduces the observed SMHM and M$_{BH}$-M$_{\star}$ relations for z = 0 galaxies. We also show that both the star formation and SMBH accretion histories are consistent with observations at high redshift,
though both suffer from our small volume's inability to capture cosmic downsizing. Using a set of smaller simulations, we also show how SMBH physics is a necessary component for quenching star formation in massive galaxies and reproducing the observed evolution of MW-mass galaxies. We also show that our implementation gives appreciably different results for galaxies in massive ($10^{12}$ M$_{\odot}$) halos compared with more common approaches. This highlights not only the importance of including SMBH physics in cosmological simulations, but also that the details of the implementation are imprinted on the evolution of massive galaxies.

Finally, we present an illustrative example of how our implementation will result not only in realistic SMBH mergers, but also allow us to study the dual AGN that may often precede such events with unprecedented detail. This will be explored more thoroughly in future work, but represents an important proof of concept that our model will provide new data to put transient events such as dual AGN and SMBH mergers into a broader context.

 The {\sc Romulus} simulation suite, with resolution on par with the highest resolution cosmological simulations run to date, will provide a crucial dataset with which to study the evolution of galaxies with halo mass $10^{9}-10^{13}$ M$_{\odot}$. The inclusion of {\sc RomulusC} and {\sc Romulus50} will provide further insight into galaxy evolution in rarer, high density regions not sampled by {\sc Romulus25} alone. The high resolution of these simulations is necessary not only to study the structure of galaxies, but also to properly follow the dynamics of SMBHs  \citep{tremmel15}. The SMBH implementation we presented in this Paper will allow SMBHs to form in the early Universe and exist in both large galaxies and dwarfs, while ensuring that they respond realistically to their changing environment. This is the first set of simulations of this size and resolution to simultaneously provide physically motivated sub-grid models for SMBH formation (\S 5.1)  and dynamics \citep[\S 5.2 and][]{tremmel15} while also accounting for resolution effects \citep{BoothBH2009} and dynamically supported gas (\S 5.3) when calculating SMBH accretion. {\sc Romulus} represents a natural next step for cosmological simulations to provide more detailed insight into the evolving structure of galaxies, the co-evolution of galaxies and SMBHs, and transient events such as Dual AGN and SMBH mergers.
 
 \section*{Acknowledgments}

We would like to thank Richard Bower,  the referee,  for their helpful and thorough comments. FG, TQ, and Lauren Anderson were partially supported by NSF award AST-1311956 and HST award AR-13264. FG, TQ and MT were partially supported by NSF award AST-1514868.  AP was supported by the Royal Society. This research is part of the Blue Waters sustained-petascale computing project, which is supported by the National Science Foundation (awards OCI-0725070 and ACI-1238993) and the state of Illinois. Blue Waters is a joint effort of the University of Illinois at Urbana-Champaign and its National Center for Supercomputing Applications. This work is also part of a PRAC allocation support by the National Science Foundation (award number OCI-1144357). MV acknowledges funding from the European Research Council under the European Community's Seventh Framework Programme (FP7/2007-2013 Grant Agreement no. 614199, project `BLACK'). Much of the analysis done in this work was done using the Pynbody package \citep{pynbody}

\bibliography{bibref_mjt.bib}
\bibliographystyle{mn2e}

\bsp

\medskip

\appendix
\section{Quantitative Parameter Search for SF and SMBHs physics}

\begin{figure*}
\includegraphics[trim=35mm 0mm -70mm 28mm, clip, width=220mm]{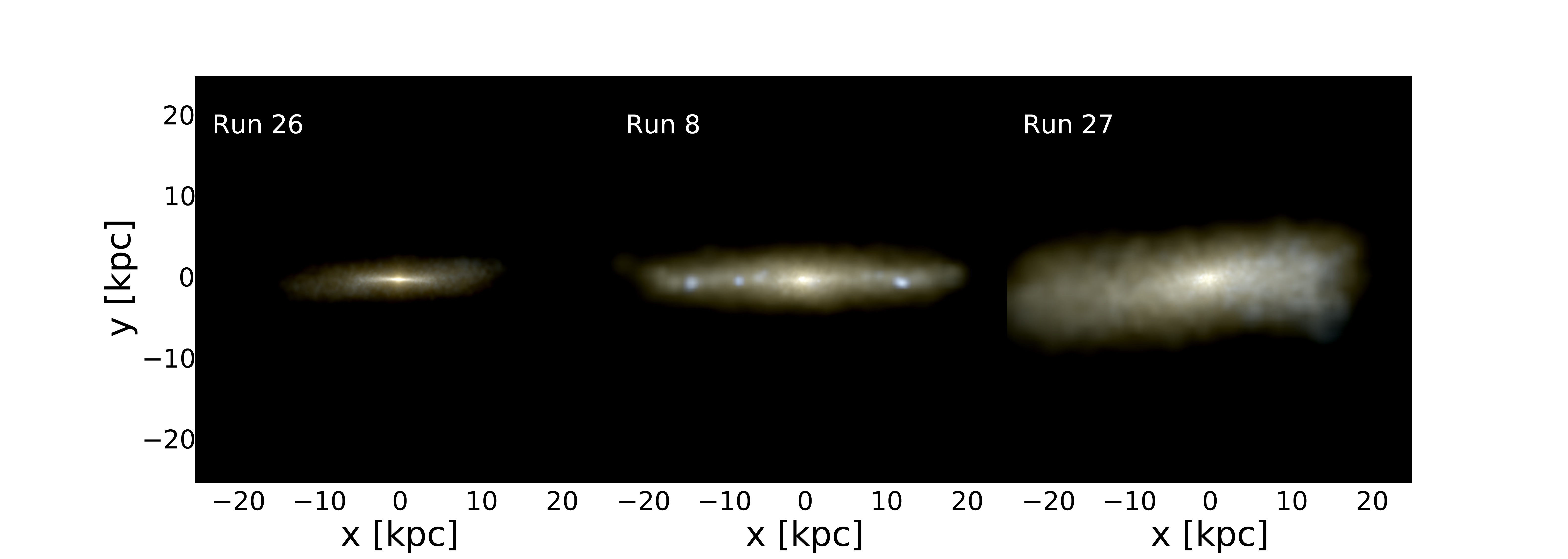}
\caption{{\sc Kriging Parameter Search in Practice.} Here we show three realizations of our zoomed-in run of a $10^{11.5}$ M$_{\odot}$ halo at z = 0. From left to right we show the best parameter set, a poor set, and the worst set based on our grading criteria. See Table 1 for the parameters for each of these simulations. This illustrates that the parameters we chose and the way we varied them throughout our search has a clear effect on galaxy properties. In this case, run 26 has a clear thin disk, run 8 has a more diffuse disk and run 27 fails to form a thin disk at all. Our approach is able to thoroughly and efficiently search through the allowed parameter space and arrive at a set of parameters that results in realistic galaxies.}
\label{app1} 
\end{figure*}

Large simulations require proportionally vast computational resources
and face two main problems: limited force and mass resolution and the
extensive need for sub- grid physics, as the modeling of physical
processes happening below the resolved scales. Examples for such
sub-grid physics parameters are the density at which SF should form,
the fraction of energy form SNe and SMBHs that couples to the
surrounding gas the speed at which metals diffuse in the Intergalactic
medium (IGM). Note that the same points hold even in simulations that
claim no free parameters,
for numerical parameters such as the precision of the step integration, the
value of the force softening or the adopted IMF.

A common problem in simulations has been how to design an efficient
strategy to {\it quantitatively} optimize, in a statistically
controlled way these physical, but poorly constrained parameters,
hence optimizing the results for the chosen physical model \citep{G07}.
A similar problem is faced by the so-called semi analytical models
\citep{monaco02,somerville08}. However parameter searches for SAM are
computationally cheaper and can be performed using different
statistical approaches such as emulation \citep{Bower2010}, Monte
Carlo Markov Chain (MCMC) \citep{benson14} or Particle Swarm
Optimization \citep{ruiz15}.

As described in \S3, in this work we have implemented a novel
optimization technique to optimally choose sub-grid parameters
associated with the implementations of 1) SF and SNe feedback and then
2) SMBHs accretion and feedback. To optimize the SF and SNe feedback
parameters we proceeded in the way described in \S2.2. Here we
describe in a  more detail some of the choices we made and the so--called
Kriging techniques (see below) that we
used to map out the suitability of the parameter space explored.  The
kriging algorithm penalizes parameter values that lead to simulations
that deviate from the properties of real galaxies and then searches
for parameter values that instead minimize this deviation. Runs are
repeated with the same galaxies set, but with the updated parameters
until the desired `convergence' to the SF values listed in \S4.

To summarize, our approach introduces a number of desirable qualities
compared when only a limited number of experiments, as typical of
numerical simulations, can be carried out. It presents several
advantages over shutting off or including individual physics modules
\citep{genel14} or to running a small cosmological volume multiple
times \citep{eagle15,schaye10}.  Namely the non linear effect of
changing more than one parameter at the time can now be followed
\citep{eagle15} and the search for best parameters can cover a mass
range similar to that of the final, large scale simulation (which tend
to have more massive halos than small test volumes).

\begin{itemize}

\item  1) Minimal resources are wasted in 'bad' regions of parameter space.

\item 2) There is no need to wait for convergence, every simulation is
  useful immediately (unlike Markov chain Monte Carlo and many
  optimization techniques) and

\item 3) Kriging is robust to changes in model choice and penalization/weighting
methods as suitability values can easily be recalculated.

\end{itemize}

Figure~\ref{app1} illustrates the results of this process, showing
images of the stars
of a disk galaxy at z=0 using the best, poor, and worst star formation parameters.

\subsection{Grading Parameter Realizations}

Each parameter set realization is graded against a set of z = 0 empirical scaling relations that govern
star formation efficiency \citep{moster13}, the gas depletion time \citep{cannon11,AlfAlfaHaynes11}, 
galaxy size and angular momentum \citep{obreschkow14}, and SMBH growth  \citep{schramm2013}.
The stellar mass fraction for our simulated galaxies is obtained following
\citep{munshi13}, a procedure that includes the effects of a fixed
aperture and the under-weighting of older, redder stellar populations.
The HI fractions are measured directly from the simulations, which track the HI content of each gas particle. 
To calculate the bulge to disk ratio, galaxies are decomposed into their different dynamical components based on the energy and angular momentum of each particle. Then the total angular momentum is calculated from every star particle not considered to be dynamically a
part of the halo. Black hole masses are taken directly from the most massive black hole in each halo.

The SMHM relation constrains the SF
efficiency over the whole Hubble time. SF efficiency also affects many
other structural relations such as the M$_{\star}$-V$_{peak}$, and the
stellar mass - metallicity relation.  The
J$_{star}$/M$_{\star}$ relation and the HI/stellar mass relation were
included as good proxies of the effect of feedback processes on low redshift SF and
the angular momentum distribution and size of a galaxy. Finally the
M$_{BH}$-M$_{\star}$ is an important constraint on SMBHs
processes, in particular the coeval growth of stars and SMBHs within galaxies.
These grading choices are by no means unique, but allow us to be confident in the success
of a given parameter set in creating galaxies that match what is observed in the local Universe, while
still leaving room to make predictions for the evolution of various galaxy properties over cosmic time.

\subsection{Finding the Optimal Parameters}

\begin{table}
\centering
\begin{tabular}{@{}cccccc}
\hline
Run        & n$_{\star}$ & c$_{\star}$ & $\epsilon_{\mathrm{SN}}$   \\
\hline
1   & 1.000 & 0.2000 &  2.000\\
 2  & 0.100 &  0.1000 & 1.000\\
 3  & 0.100 &  0.1000 & 4.000\\
 4  & 0.100 &  0.4000 & 1.000\\
 5  & 0.100 &  0.4000 & 4.000\\
 6  & 4.000 &  0.1000 & 1.000\\
 7  & 4.000 &  0.1000 & 4.000\\
 8  & 4.000 &  0.4000 & 1.000\\
 9  & 4.000 &  0.4000 & 4.000\\
10  & 0.1   &  0.1   &  1.5  \\
11 & 0.1   &  0.1   &  2.0  \\
12  & 0.1  &   0.2   &  1.0  \\
13  & 0.1  &   0.2  &   1.5  \\
14  & 0.1  &   0.2  &   2.0  \\
15  & 1.0  &   0.1  &   1.0  \\
16  & 1.0  &   0.1  &   1.5  \\
17  & 1.0  &   0.1  &   2.0  \\
18  & 1.0  &   0.2  &   1.0  \\
19  & 1.0  &   0.2  &   1.5  \\
20  & 0.05 &   0.05  &  0.5  \\
21  & 0.05 &   0.05  &  1.5  \\
22  & 0.05 &   0.15  &  0.5  \\
23  & 0.05 &   0.15 &   1.5  \\
24  & 0.2  &   0.05  &  0.5  \\
25  & 0.2  &   0.05  &  1.5  \\
26* &  0.2 &    0.15 &   0.5  \\ 
27  & 0.2 &    0.15 &   1.5  \\
\hline
\end{tabular}
\caption{{\sc Example set of parameter space realizations}. The free parameters tested are the SN efficiency, $\epsilon_{\mathrm{SN}}$, the threshold density for star formation, $n_{\star}$, and the star formation efficiency, $c_{\star}$. Different sets of parameters chosen based on the Kriging technique until a `best' set of parameters is converged upon (run 26 here). Note that these runs were done with lower DM mass resolution compared to {\sc Romulus} (see text).}
\end{table}

In order to avoid a 5-dimensional parameter space calculation, we first 
performed the full analysis, using the Kriging technique, on galaxies with no SMBH physics.
This allowed us to converge upon the set of SF parameters that created the most realistic
galaxies possible without the inclusion of SMBHs. A series of 27 parameter realizations (see Table 1)
was run for sets of 3 halos with z = 0 virial masses of $10^{10.5}$, $10^{11.5}$, and $10^{12}$ M$_{\odot}$.
Each set was graded by summing up the logarithmic distance
of each galaxy from each scaling relation, though the angular momentum of the dwarf galaxy was excluded
due to the fact that the dynamical decomposition technique becomes unreliable at low masses. Each galaxy is weighted evenly in the
final grade for each parameter realization. The best model converged upon by
this approach is marked with a star in Table 1.

Once the SF parameters were chosen, another set of 12 simulations were run with SMBH physics to find the best parameters
for accretion and feedback strength (see \S5.4). The same general approach was used, though a more hands-on approach was used to
dictate how we traversed the available parameter space (see below). Because SMBH physics is thought to preferentially affect more massive galaxies, we include a fourth halo, with virial mass $10^{12}$ M$_{\odot}$, in each set of simulations. When grading each parameter set, the average deviation of these two halos is used instead of their individual deviations. Again, each galaxy is weighted evenly though the dwarf galaxy is again excluded from the SMBH relation due to the fact that the fraction of dwarfs hosting a central SMBH is not well known (see \citet{Volonteri2010AARV} for theoretical arguments and \citep{reinesComastri16} for an observational review). Furthermore, as noted in section 5.5, the inclusion of a SMBH does not have a significant impact on the scaling relationships.

\subsection{The Kriging Approach to Parameter Search}

\begin{figure*}
\includegraphics[trim=0mm 0mm -50mm 0mm, clip, width=220mm]{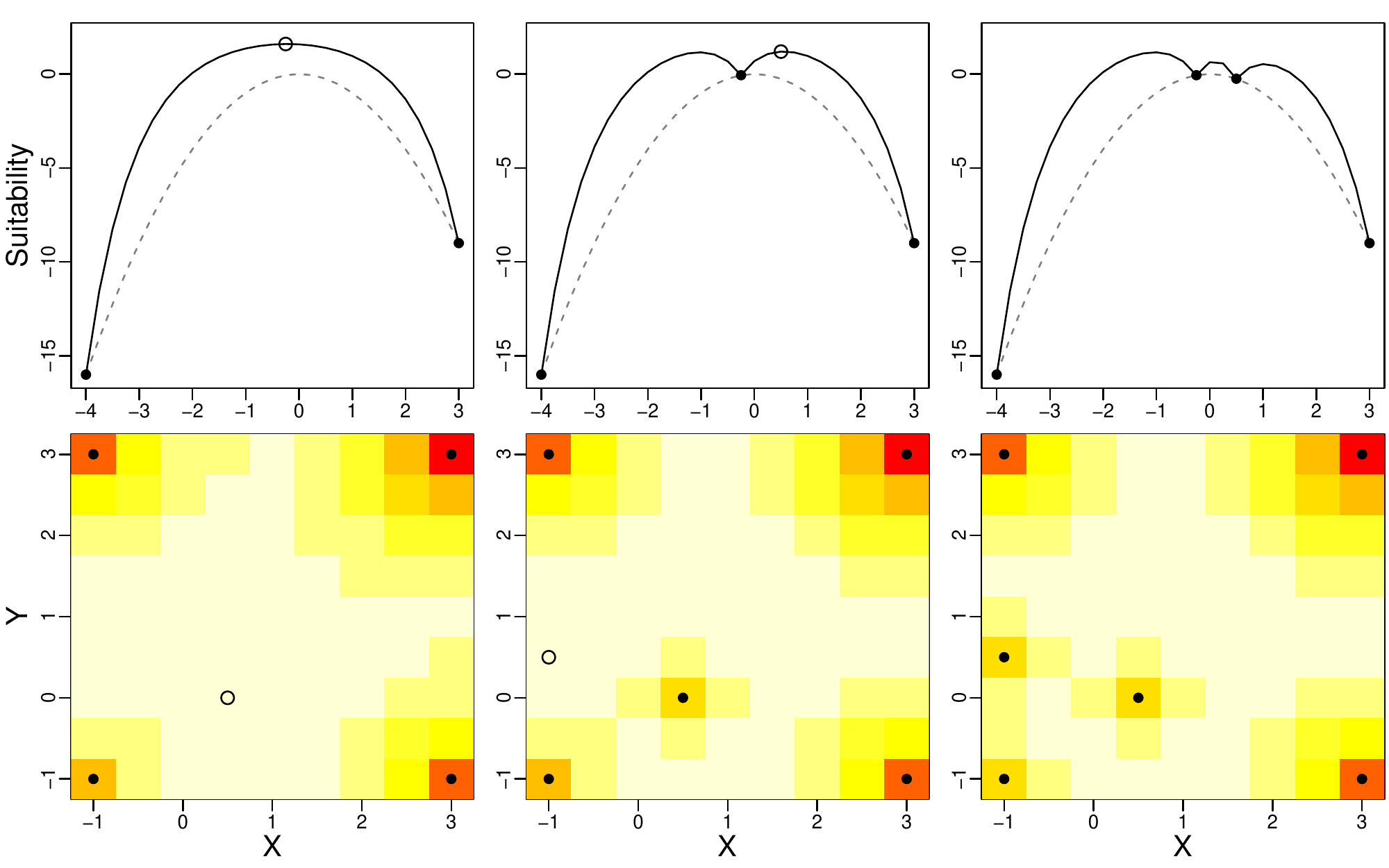}
\caption{{\sc Kriging parameter optimization technique example.} Two iterations of the Kriging search algorithm on a 1-dimensional example (first row) and a 2-dimensional example (second row). In the 1-dimensional scenario, we are attempting to optimize the suitability function $-x^2$, shown as the dashed gray line. The algorithm starts by interpolating a pseudo-confidence manifold from two known points (filled points), and finding the greatest value (unfilled point). The algorithm then calculates the true suitability value for that point, and repeats the process. The 2-dimensional scenario is similar---we attempt to optimize the suitability function $-\mathbf{x}^2$. Here, the pseudo-confidence manifold is shown as a heatmap, with red/darker representing lower suitability and white/brighter representing higher suitability. The point selection and evaluation process (filled and unfilled points) is identical to the 1-dimensional scenario.}
\label{kriging} 
\end{figure*}

The Kriging Approach allows us to efficiently traverse parameter space and know when we have
 converged on the `best' set of parameters without the use of a large number of simulations as would be
 required of other techniques such as MCMC.

MCMC requires 1) a joint prior distribution on the parameter
space, from which initial points can be drawn, 2) a likelihood function
describing the distribution of the observables given a particular
parameter set, and 3) a proposal distribution that generates the next
parameter values to examine, given the current values. MCMC then uses
these functions to iterate over the parameter space, deciding whether
or not to jump to the next point depending on how likely the next
point is to explain the data relative to the current values.  After
a very large number of iterations (sometimes millions), the accepted points
become a sample from the posterior distribution of parameter values
given the observables, and useful inferences can thereby be derived,
including point estimates of the best parameter values, and 95\%
credible regions for where the best parameter values may lie.

Because cosmological simulations consume a large amount of computing
resources, simulating so many iterations is not possible. Our approach trades the
unattainable statistical properties of MCMC for the ability
to make direct use of human expertise and intuition, the flexibility
to adapt to changing measures of fitness, and keeping the certainty of
knowing that every iteration makes a distinguishable contribution to
our knowledge of the parameter space. While we lose access to the
posterior distribution (i.e. a full sampling and ranking of parameter
space), that is not really necessary. Instead, we gain an efficient
means of finding the region of the parameter space that produces the
most realistic galaxies, which is our goal.

We achieve all this by adapting Gaussian process Kriging
techniques into a more intelligent and efficient grid search
algorithm (see Figure~\ref{kriging}). We start by constructing a suitability function---a
function that takes in a simulation and compares it to observed
relationships and returns a score
describing how realistic the simulation is (see above).

Using the following formula \citep{mackay1998introduction},
  we then interpolate the suitability function between all of our
  simulated points and put pseudo-confidence bounds around
  where the suitability function will actually fall. We see,
\begin{align*}
\mathbf{f_*} | X_*, X, \mathbf{f} & \sim N( K(X_*, X) K(X, X)^{-1} \mathbf{f},\\
 & K(X_*, X_*) - K(X_*, X)K(X,X)^{-1}K(X, X_*) ),
\end{align*}
where $X$ is the matrix of already-simulated parameter values,
$\mathbf{f}$ is the corresponding vector of known suitability
values, $X_*$ is a matrix of new parameter values that we wish
to examine, $\mathbf{f_*}$ is the corresponding as yet unknown
suitabilities, and $K(\cdot, \cdot)$ is a covariance matrix 
derived from a pre-specified covariance function $k(x_1, x_2)$. 

Since we aren't seeking statistical properties, only utilitarian properties,
 we don't estimate the covariance scale so much as choose one that spreads 
 the first few suggested points away from the initial points, in our case a 99\% `pseudo confidence surface'
  with covariance scale of 1, meant to ensure parameter space is widely sampled. 
  Since it is a suggestion algorithm rather than a statistical method, 
  the covariance scale can even be adjusted freely before or after points have been selected and tested.

If we wish to take a hands-off approach, we would then examine the
upper pseudo-confidence manifold, and instruct the algorithm to find
the point with the highest potential suitability (at a fixed
confidence level), and then numerically simulate that point. This
  is the approach we used for the initial search where we optimized
  the SF parameters (see above). However, if human intuition can be sufficient,
we may also examine the pseudo-confidence manifold manually and select
the next point ourselves without concern over losing statistical rigor.
This is the approach we used for tuning the SMBHs parameters.
One complication is that in regions of the parameter space where the
Kriging process is extrapolating rather than interpolating, the
confidence regions become extraordinarily wide, leading a naive
algorithm to always select an extrapolated point. This has at least
two solutions. One is to restrict any automation to the convex hull of
already simulated points and use manual intervention to select points
outside the convex hull if it becomes clear that such a point would
make a good candidate. The second is to only calculate the Kriging
bounds for a predefined, \textit{a priori} reasonable region of the
parameter space. The algorithm will quickly explore the outer boundary
and then turn inward. From experience we learned that a good
  approach is to start the parameter exploration from a coarse grid of
  parameters values, including a range over which simulations will
  provide 'bad' results (e.g testing SN efficiency ranging from 0 to
  4, values that will surely over and under produce stars).  An option
for future work would be to include higher-z constraints from the
progenitors of massive present day halos, this would allow to
constraint the high end of the present day galaxy stellar mass function
using a limited amount of computational resources.

A sample result of this process is in Table 1. By starting with a coarse grid of values for each of our 3 parameters,
we utilize Kriging to traverse parameter space. After each iteration, Kriging sees both the current `best' point
and the algorithm will then run a simulation in a region not yet well enough constrained. With time, each parameter space
realization gets closer to the `best' values until Kriging tells us it has sufficiently converged. Regions of parameter space that
behave the worst are then sampled much less often while regions nearby the `best' parameter set are sampled in more detail. The
results presented in Table 1 are from simulations that do not oversample DM particles and therefore have lower mass resolution for DM
than the {\sc Romulus} simulations. We find that this increased resolution results in higher star formation in dwarf galaxies. Thus, the {\sc Romulus}
simulations use the values from run 26, but with a higher SN efficiency of 0.75, a combination we find results in final properties very similar 
to run 26 in Table 1.

\section{Dust Extinction Approximation}

When comparing the colors of simulated galaxies to observations, it is important to account for dust attenuation. Because we only care about the average attenuation across all lines of sight integrated over all stars in a given galaxy, we utilize a simple `spherical cow' approach similar to \citet{shimizuDust2011}.

For a given dust distribution, the amount of attenuation can be calculated at any wavelength using the Calzetti Law \citep{calzetti00}, but first it must be properly normalized. For this, it is convenient to use far UV light, since the extinction cross section is roughly equal to the dust grain size. We choose 1600 Angstroms as our normalizing far UV wavelength. We then make the assumption that the dust is uniformly distributed in a sheet around the stars, which allows us to relate the dust extinction by a simple function of the dust optical depth \citep{calzetti01}.

\begin{equation}
A_{\lambda} \sim \frac{\tau_{\lambda}}{0.921}
\end{equation}

This is obviously not true in reality, but is not a bad assumption if we think of this calculation as an average over all lines of sight. Assuming spherical symmetry also makes the optical depth a simple function of average dust properties.

\begin{equation}
\tau{\lambda} = \int \sigma_{d}(\lambda)n(r)r \sim \frac {\sigma_{d}(\lambda)\Sigma_{d}}{m_p}
\end{equation}

In the above equation $\sigma_d$ is the dust cross section, $\Sigma_d$ is the column density, and $m_p$ is the mass per dust grain. When dealing with far UV light, the cross section is just the cross sectional area of the average dust particle. Because we are not accounting for structure within the gas, we can use instead estimate the average column density using the total mass, $\mathrm{M}_d$ of dust within the galaxy and the half mass radius, $\mathrm{R}_{1/2,d}$ of the dust.

\begin{equation}
\Sigma_d \sim \frac{(1/2)\mathrm{M}_d}{\pi \mathrm{R}^2_{1/2,d}}
\end{equation}

The total mass in dust for a halo is given by the following relation from \citet{draine07} summed over the HI mass, $m_{HI,i}$ of every gas particle in a halo. We follow \citet{shimizuDust2011} and normalize instead to the solar metallicity, rather than galactic O/H values as in the original paper.

\begin{equation}
\mathrm{M}_d \sim \sum^{N_{gas}}_{i=1} 0.01 \frac{Z}{Z_{\odot}}m_{HI,i}
\end{equation}

These equations, put together with the physical properties of dust grains and applied to 1600 Angstroms, gives us A$_{1600}$, which we can use to set the normalization of the Calzetti Law. We take the dust particle size to be $0.1 \mu\mathrm{m}$ and density to be $2.5 g/cc$ \citep{todini01,nozawa03}. We cap A$_{1600}$ at a value of 2, given that more advanced dust models show that attenuation deviates significantly from its linear relationship with optical depth as column densities increase due to the fact that dustier systems will tend to be clumpier \citep{calzetti01}. This normalization, combined with our adopted value of $R_v = 4.0$, gives us the ability to estimate the dust attenuation at any wavelength. When dealing with bands of wavelengths, we calculate attenuation using the central wavelength of the band.

\label{lastpage}
\end{document}

%% file: rgb.tex
  \definecolor{snow}{rgb}{1.000000,0.980392,0.980392}
  \definecolor{ghost white}{rgb}{0.972549,0.972549,1.000000}
  \definecolor{GhostWhite}{rgb}{0.972549,0.972549,1.000000}
  \definecolor{white smoke}{rgb}{0.960784,0.960784,0.960784}
  \definecolor{WhiteSmoke}{rgb}{0.960784,0.960784,0.960784}
  \definecolor{gainsboro}{rgb}{0.862745,0.862745,0.862745}
  \definecolor{floral white}{rgb}{1.000000,0.980392,0.941176}
  \definecolor{FloralWhite}{rgb}{1.000000,0.980392,0.941176}
  \definecolor{old lace}{rgb}{0.992157,0.960784,0.901961}
  \definecolor{OldLace}{rgb}{0.992157,0.960784,0.901961}
  \definecolor{linen}{rgb}{0.980392,0.941176,0.901961}
  \definecolor{antique white}{rgb}{0.980392,0.921569,0.843137}
  \definecolor{AntiqueWhite}{rgb}{0.980392,0.921569,0.843137}
  \definecolor{papaya whip}{rgb}{1.000000,0.937255,0.835294}
  \definecolor{PapayaWhip}{rgb}{1.000000,0.937255,0.835294}
  \definecolor{blanched almond}{rgb}{1.000000,0.921569,0.803922}
  \definecolor{BlanchedAlmond}{rgb}{1.000000,0.921569,0.803922}
  \definecolor{bisque}{rgb}{1.000000,0.894118,0.768627}
  \definecolor{peach puff}{rgb}{1.000000,0.854902,0.725490}
  \definecolor{PeachPuff}{rgb}{1.000000,0.854902,0.725490}
  \definecolor{navajo white}{rgb}{1.000000,0.870588,0.678431}
  \definecolor{NavajoWhite}{rgb}{1.000000,0.870588,0.678431}
  \definecolor{moccasin}{rgb}{1.000000,0.894118,0.709804}
  \definecolor{cornsilk}{rgb}{1.000000,0.972549,0.862745}
  \definecolor{ivory}{rgb}{1.000000,1.000000,0.941176}
  \definecolor{lemon chiffon}{rgb}{1.000000,0.980392,0.803922}
  \definecolor{LemonChiffon}{rgb}{1.000000,0.980392,0.803922}
  \definecolor{seashell}{rgb}{1.000000,0.960784,0.933333}
  \definecolor{honeydew}{rgb}{0.941176,1.000000,0.941176}
  \definecolor{mint cream}{rgb}{0.960784,1.000000,0.980392}
  \definecolor{MintCream}{rgb}{0.960784,1.000000,0.980392}
  \definecolor{azure}{rgb}{0.941176,1.000000,1.000000}
  \definecolor{alice blue}{rgb}{0.941176,0.972549,1.000000}
  \definecolor{AliceBlue}{rgb}{0.941176,0.972549,1.000000}
  \definecolor{lavender}{rgb}{0.901961,0.901961,0.980392}
  \definecolor{lavender blush}{rgb}{1.000000,0.941176,0.960784}
  \definecolor{LavenderBlush}{rgb}{1.000000,0.941176,0.960784}
  \definecolor{misty rose}{rgb}{1.000000,0.894118,0.882353}
  \definecolor{MistyRose}{rgb}{1.000000,0.894118,0.882353}
  \definecolor{white}{rgb}{1.000000,1.000000,1.000000}
  \definecolor{black}{rgb}{0.000000,0.000000,0.000000}
  \definecolor{dark slate gray}{rgb}{0.184314,0.309804,0.309804}
  \definecolor{DarkSlateGray}{rgb}{0.184314,0.309804,0.309804}
  \definecolor{dark slate grey}{rgb}{0.184314,0.309804,0.309804}
  \definecolor{DarkSlateGrey}{rgb}{0.184314,0.309804,0.309804}
  \definecolor{dim gray}{rgb}{0.411765,0.411765,0.411765}
  \definecolor{DimGray}{rgb}{0.411765,0.411765,0.411765}
  \definecolor{dim grey}{rgb}{0.411765,0.411765,0.411765}
  \definecolor{DimGrey}{rgb}{0.411765,0.411765,0.411765}
  \definecolor{slate gray}{rgb}{0.439216,0.501961,0.564706}
  \definecolor{SlateGray}{rgb}{0.439216,0.501961,0.564706}
  \definecolor{slate grey}{rgb}{0.439216,0.501961,0.564706}
  \definecolor{SlateGrey}{rgb}{0.439216,0.501961,0.564706}
  \definecolor{light slate gray}{rgb}{0.466667,0.533333,0.600000}
  \definecolor{LightSlateGray}{rgb}{0.466667,0.533333,0.600000}
  \definecolor{light slate grey}{rgb}{0.466667,0.533333,0.600000}
  \definecolor{LightSlateGrey}{rgb}{0.466667,0.533333,0.600000}
  \definecolor{gray}{rgb}{0.745098,0.745098,0.745098}
  \definecolor{grey}{rgb}{0.745098,0.745098,0.745098}
  \definecolor{light grey}{rgb}{0.827451,0.827451,0.827451}
  \definecolor{LightGrey}{rgb}{0.827451,0.827451,0.827451}
  \definecolor{light gray}{rgb}{0.827451,0.827451,0.827451}
  \definecolor{LightGray}{rgb}{0.827451,0.827451,0.827451}
  \definecolor{midnight blue}{rgb}{0.098039,0.098039,0.439216}
  \definecolor{MidnightBlue}{rgb}{0.098039,0.098039,0.439216}
  \definecolor{navy}{rgb}{0.000000,0.000000,0.501961}
  \definecolor{navy blue}{rgb}{0.000000,0.000000,0.501961}
  \definecolor{NavyBlue}{rgb}{0.000000,0.000000,0.501961}
  \definecolor{cornflower blue}{rgb}{0.392157,0.584314,0.929412}
  \definecolor{CornflowerBlue}{rgb}{0.392157,0.584314,0.929412}
  \definecolor{dark slate blue}{rgb}{0.282353,0.239216,0.545098}
  \definecolor{DarkSlateBlue}{rgb}{0.282353,0.239216,0.545098}
  \definecolor{slate blue}{rgb}{0.415686,0.352941,0.803922}
  \definecolor{SlateBlue}{rgb}{0.415686,0.352941,0.803922}
  \definecolor{medium slate blue}{rgb}{0.482353,0.407843,0.933333}
  \definecolor{MediumSlateBlue}{rgb}{0.482353,0.407843,0.933333}
  \definecolor{light slate blue}{rgb}{0.517647,0.439216,1.000000}
  \definecolor{LightSlateBlue}{rgb}{0.517647,0.439216,1.000000}
  \definecolor{medium blue}{rgb}{0.000000,0.000000,0.803922}
  \definecolor{MediumBlue}{rgb}{0.000000,0.000000,0.803922}
  \definecolor{royal blue}{rgb}{0.254902,0.411765,0.882353}
  \definecolor{RoyalBlue}{rgb}{0.254902,0.411765,0.882353}
  \definecolor{blue}{rgb}{0.000000,0.000000,1.000000}
  \definecolor{dodger blue}{rgb}{0.117647,0.564706,1.000000}
  \definecolor{DodgerBlue}{rgb}{0.117647,0.564706,1.000000}
  \definecolor{deep sky blue}{rgb}{0.000000,0.749020,1.000000}
  \definecolor{DeepSkyBlue}{rgb}{0.000000,0.749020,1.000000}
  \definecolor{sky blue}{rgb}{0.529412,0.807843,0.921569}
  \definecolor{SkyBlue}{rgb}{0.529412,0.807843,0.921569}
  \definecolor{light sky blue}{rgb}{0.529412,0.807843,0.980392}
  \definecolor{LightSkyBlue}{rgb}{0.529412,0.807843,0.980392}
  \definecolor{steel blue}{rgb}{0.274510,0.509804,0.705882}
  \definecolor{SteelBlue}{rgb}{0.274510,0.509804,0.705882}
  \definecolor{light steel blue}{rgb}{0.690196,0.768627,0.870588}
  \definecolor{LightSteelBlue}{rgb}{0.690196,0.768627,0.870588}
  \definecolor{light blue}{rgb}{0.678431,0.847059,0.901961}
  \definecolor{LightBlue}{rgb}{0.678431,0.847059,0.901961}
  \definecolor{powder blue}{rgb}{0.690196,0.878431,0.901961}
  \definecolor{PowderBlue}{rgb}{0.690196,0.878431,0.901961}
  \definecolor{pale turquoise}{rgb}{0.686275,0.933333,0.933333}
  \definecolor{PaleTurquoise}{rgb}{0.686275,0.933333,0.933333}
  \definecolor{dark turquoise}{rgb}{0.000000,0.807843,0.819608}
  \definecolor{DarkTurquoise}{rgb}{0.000000,0.807843,0.819608}
  \definecolor{medium turquoise}{rgb}{0.282353,0.819608,0.800000}
  \definecolor{MediumTurquoise}{rgb}{0.282353,0.819608,0.800000}
  \definecolor{turquoise}{rgb}{0.250980,0.878431,0.815686}
  \definecolor{cyan}{rgb}{0.000000,1.000000,1.000000}
  \definecolor{light cyan}{rgb}{0.878431,1.000000,1.000000}
  \definecolor{LightCyan}{rgb}{0.878431,1.000000,1.000000}
  \definecolor{cadet blue}{rgb}{0.372549,0.619608,0.627451}
  \definecolor{CadetBlue}{rgb}{0.372549,0.619608,0.627451}
  \definecolor{medium aquamarine}{rgb}{0.400000,0.803922,0.666667}
  \definecolor{MediumAquamarine}{rgb}{0.400000,0.803922,0.666667}
  \definecolor{aquamarine}{rgb}{0.498039,1.000000,0.831373}
  \definecolor{dark green}{rgb}{0.000000,0.392157,0.000000}
  \definecolor{DarkGreen}{rgb}{0.000000,0.392157,0.000000}
  \definecolor{dark olive green}{rgb}{0.333333,0.419608,0.184314}
  \definecolor{DarkOliveGreen}{rgb}{0.333333,0.419608,0.184314}
  \definecolor{dark sea green}{rgb}{0.560784,0.737255,0.560784}
  \definecolor{DarkSeaGreen}{rgb}{0.560784,0.737255,0.560784}
  \definecolor{sea green}{rgb}{0.180392,0.545098,0.341176}
  \definecolor{SeaGreen}{rgb}{0.180392,0.545098,0.341176}
  \definecolor{medium sea green}{rgb}{0.235294,0.701961,0.443137}
  \definecolor{MediumSeaGreen}{rgb}{0.235294,0.701961,0.443137}
  \definecolor{light sea green}{rgb}{0.125490,0.698039,0.666667}
  \definecolor{LightSeaGreen}{rgb}{0.125490,0.698039,0.666667}
  \definecolor{pale green}{rgb}{0.596078,0.984314,0.596078}
  \definecolor{PaleGreen}{rgb}{0.596078,0.984314,0.596078}
  \definecolor{spring green}{rgb}{0.000000,1.000000,0.498039}
  \definecolor{SpringGreen}{rgb}{0.000000,1.000000,0.498039}
  \definecolor{lawn green}{rgb}{0.486275,0.988235,0.000000}
  \definecolor{LawnGreen}{rgb}{0.486275,0.988235,0.000000}
  \definecolor{green}{rgb}{0.000000,1.000000,0.000000}
  \definecolor{chartreuse}{rgb}{0.498039,1.000000,0.000000}
  \definecolor{medium spring green}{rgb}{0.000000,0.980392,0.603922}
  \definecolor{MediumSpringGreen}{rgb}{0.000000,0.980392,0.603922}
  \definecolor{green yellow}{rgb}{0.678431,1.000000,0.184314}
  \definecolor{GreenYellow}{rgb}{0.678431,1.000000,0.184314}
  \definecolor{lime green}{rgb}{0.196078,0.803922,0.196078}
  \definecolor{LimeGreen}{rgb}{0.196078,0.803922,0.196078}
  \definecolor{yellow green}{rgb}{0.603922,0.803922,0.196078}
  \definecolor{YellowGreen}{rgb}{0.603922,0.803922,0.196078}
  \definecolor{forest green}{rgb}{0.133333,0.545098,0.133333}
  \definecolor{ForestGreen}{rgb}{0.133333,0.545098,0.133333}
  \definecolor{olive drab}{rgb}{0.419608,0.556863,0.137255}
  \definecolor{OliveDrab}{rgb}{0.419608,0.556863,0.137255}
  \definecolor{dark khaki}{rgb}{0.741176,0.717647,0.419608}
  \definecolor{DarkKhaki}{rgb}{0.741176,0.717647,0.419608}
  \definecolor{khaki}{rgb}{0.941176,0.901961,0.549020}
  \definecolor{pale goldenrod}{rgb}{0.933333,0.909804,0.666667}
  \definecolor{PaleGoldenrod}{rgb}{0.933333,0.909804,0.666667}
  \definecolor{light goldenrod yellow}{rgb}{0.980392,0.980392,0.823529}
  \definecolor{LightGoldenrodYellow}{rgb}{0.980392,0.980392,0.823529}
  \definecolor{light yellow}{rgb}{1.000000,1.000000,0.878431}
  \definecolor{LightYellow}{rgb}{1.000000,1.000000,0.878431}
  \definecolor{yellow}{rgb}{1.000000,1.000000,0.000000}
  \definecolor{gold}{rgb}{1.000000,0.843137,0.000000}
  \definecolor{light goldenrod}{rgb}{0.933333,0.866667,0.509804}
  \definecolor{LightGoldenrod}{rgb}{0.933333,0.866667,0.509804}
  \definecolor{goldenrod}{rgb}{0.854902,0.647059,0.125490}
  \definecolor{dark goldenrod}{rgb}{0.721569,0.525490,0.043137}
  \definecolor{DarkGoldenrod}{rgb}{0.721569,0.525490,0.043137}
  \definecolor{rosy brown}{rgb}{0.737255,0.560784,0.560784}
  \definecolor{RosyBrown}{rgb}{0.737255,0.560784,0.560784}
  \definecolor{indian red}{rgb}{0.803922,0.360784,0.360784}
  \definecolor{IndianRed}{rgb}{0.803922,0.360784,0.360784}
  \definecolor{saddle brown}{rgb}{0.545098,0.270588,0.074510}
  \definecolor{SaddleBrown}{rgb}{0.545098,0.270588,0.074510}
  \definecolor{sienna}{rgb}{0.627451,0.321569,0.176471}
  \definecolor{peru}{rgb}{0.803922,0.521569,0.247059}
  \definecolor{burlywood}{rgb}{0.870588,0.721569,0.529412}
  \definecolor{beige}{rgb}{0.960784,0.960784,0.862745}
  \definecolor{wheat}{rgb}{0.960784,0.870588,0.701961}
  \definecolor{sandy brown}{rgb}{0.956863,0.643137,0.376471}
  \definecolor{SandyBrown}{rgb}{0.956863,0.643137,0.376471}
  \definecolor{tan}{rgb}{0.823529,0.705882,0.549020}
  \definecolor{chocolate}{rgb}{0.823529,0.411765,0.117647}
  \definecolor{firebrick}{rgb}{0.698039,0.133333,0.133333}
  \definecolor{brown}{rgb}{0.647059,0.164706,0.164706}
  \definecolor{dark salmon}{rgb}{0.913725,0.588235,0.478431}
  \definecolor{DarkSalmon}{rgb}{0.913725,0.588235,0.478431}
  \definecolor{salmon}{rgb}{0.980392,0.501961,0.447059}
  \definecolor{light salmon}{rgb}{1.000000,0.627451,0.478431}
  \definecolor{LightSalmon}{rgb}{1.000000,0.627451,0.478431}
  \definecolor{orange}{rgb}{1.000000,0.647059,0.000000}
  \definecolor{dark orange}{rgb}{1.000000,0.549020,0.000000}
  \definecolor{DarkOrange}{rgb}{1.000000,0.549020,0.000000}
  \definecolor{coral}{rgb}{1.000000,0.498039,0.313726}
  \definecolor{light coral}{rgb}{0.941176,0.501961,0.501961}
  \definecolor{LightCoral}{rgb}{0.941176,0.501961,0.501961}
  \definecolor{tomato}{rgb}{1.000000,0.388235,0.278431}
  \definecolor{orange red}{rgb}{1.000000,0.270588,0.000000}
  \definecolor{OrangeRed}{rgb}{1.000000,0.270588,0.000000}
  \definecolor{red}{rgb}{1.000000,0.000000,0.000000}
  \definecolor{hot pink}{rgb}{1.000000,0.411765,0.705882}
  \definecolor{HotPink}{rgb}{1.000000,0.411765,0.705882}
  \definecolor{deep pink}{rgb}{1.000000,0.078431,0.576471}
  \definecolor{DeepPink}{rgb}{1.000000,0.078431,0.576471}
  \definecolor{pink}{rgb}{1.000000,0.752941,0.796078}
  \definecolor{light pink}{rgb}{1.000000,0.713726,0.756863}
  \definecolor{LightPink}{rgb}{1.000000,0.713726,0.756863}
  \definecolor{pale violet red}{rgb}{0.858824,0.439216,0.576471}
  \definecolor{PaleVioletRed}{rgb}{0.858824,0.439216,0.576471}
  \definecolor{maroon}{rgb}{0.690196,0.188235,0.376471}
  \definecolor{medium violet red}{rgb}{0.780392,0.082353,0.521569}
  \definecolor{MediumVioletRed}{rgb}{0.780392,0.082353,0.521569}
  \definecolor{violet red}{rgb}{0.815686,0.125490,0.564706}
  \definecolor{VioletRed}{rgb}{0.815686,0.125490,0.564706}
  \definecolor{magenta}{rgb}{1.000000,0.000000,1.000000}
  \definecolor{violet}{rgb}{0.933333,0.509804,0.933333}
  \definecolor{plum}{rgb}{0.866667,0.627451,0.866667}
  \definecolor{orchid}{rgb}{0.854902,0.439216,0.839216}
  \definecolor{medium orchid}{rgb}{0.729412,0.333333,0.827451}
  \definecolor{MediumOrchid}{rgb}{0.729412,0.333333,0.827451}
  \definecolor{dark orchid}{rgb}{0.600000,0.196078,0.800000}
  \definecolor{DarkOrchid}{rgb}{0.600000,0.196078,0.800000}
  \definecolor{dark violet}{rgb}{0.580392,0.000000,0.827451}
  \definecolor{DarkViolet}{rgb}{0.580392,0.000000,0.827451}
  \definecolor{blue violet}{rgb}{0.541176,0.168627,0.886275}
  \definecolor{BlueViolet}{rgb}{0.541176,0.168627,0.886275}
  \definecolor{purple}{rgb}{0.627451,0.125490,0.941176}
  \definecolor{medium purple}{rgb}{0.576471,0.439216,0.858824}
  \definecolor{MediumPurple}{rgb}{0.576471,0.439216,0.858824}
  \definecolor{thistle}{rgb}{0.847059,0.749020,0.847059}
  \definecolor{snow1}{rgb}{1.000000,0.980392,0.980392}
  \definecolor{snow2}{rgb}{0.933333,0.913725,0.913725}
  \definecolor{snow3}{rgb}{0.803922,0.788235,0.788235}
  \definecolor{snow4}{rgb}{0.545098,0.537255,0.537255}
  \definecolor{seashell1}{rgb}{1.000000,0.960784,0.933333}
  \definecolor{seashell2}{rgb}{0.933333,0.898039,0.870588}
  \definecolor{seashell3}{rgb}{0.803922,0.772549,0.749020}
  \definecolor{seashell4}{rgb}{0.545098,0.525490,0.509804}
  \definecolor{AntiqueWhite1}{rgb}{1.000000,0.937255,0.858824}
  \definecolor{AntiqueWhite2}{rgb}{0.933333,0.874510,0.800000}
  \definecolor{AntiqueWhite3}{rgb}{0.803922,0.752941,0.690196}
  \definecolor{AntiqueWhite4}{rgb}{0.545098,0.513726,0.470588}
  \definecolor{bisque1}{rgb}{1.000000,0.894118,0.768627}
  \definecolor{bisque2}{rgb}{0.933333,0.835294,0.717647}
  \definecolor{bisque3}{rgb}{0.803922,0.717647,0.619608}
  \definecolor{bisque4}{rgb}{0.545098,0.490196,0.419608}
  \definecolor{PeachPuff1}{rgb}{1.000000,0.854902,0.725490}
  \definecolor{PeachPuff2}{rgb}{0.933333,0.796078,0.678431}
  \definecolor{PeachPuff3}{rgb}{0.803922,0.686275,0.584314}
  \definecolor{PeachPuff4}{rgb}{0.545098,0.466667,0.396078}
  \definecolor{NavajoWhite1}{rgb}{1.000000,0.870588,0.678431}
  \definecolor{NavajoWhite2}{rgb}{0.933333,0.811765,0.631373}
  \definecolor{NavajoWhite3}{rgb}{0.803922,0.701961,0.545098}
  \definecolor{NavajoWhite4}{rgb}{0.545098,0.474510,0.368627}
  \definecolor{LemonChiffon1}{rgb}{1.000000,0.980392,0.803922}
  \definecolor{LemonChiffon2}{rgb}{0.933333,0.913725,0.749020}
  \definecolor{LemonChiffon3}{rgb}{0.803922,0.788235,0.647059}
  \definecolor{LemonChiffon4}{rgb}{0.545098,0.537255,0.439216}
  \definecolor{cornsilk1}{rgb}{1.000000,0.972549,0.862745}
  \definecolor{cornsilk2}{rgb}{0.933333,0.909804,0.803922}
  \definecolor{cornsilk3}{rgb}{0.803922,0.784314,0.694118}
  \definecolor{cornsilk4}{rgb}{0.545098,0.533333,0.470588}
  \definecolor{ivory1}{rgb}{1.000000,1.000000,0.941176}
  \definecolor{ivory2}{rgb}{0.933333,0.933333,0.878431}
  \definecolor{ivory3}{rgb}{0.803922,0.803922,0.756863}
  \definecolor{ivory4}{rgb}{0.545098,0.545098,0.513726}
  \definecolor{honeydew1}{rgb}{0.941176,1.000000,0.941176}
  \definecolor{honeydew2}{rgb}{0.878431,0.933333,0.878431}
  \definecolor{honeydew3}{rgb}{0.756863,0.803922,0.756863}
  \definecolor{honeydew4}{rgb}{0.513726,0.545098,0.513726}
  \definecolor{LavenderBlush1}{rgb}{1.000000,0.941176,0.960784}
  \definecolor{LavenderBlush2}{rgb}{0.933333,0.878431,0.898039}
  \definecolor{LavenderBlush3}{rgb}{0.803922,0.756863,0.772549}
  \definecolor{LavenderBlush4}{rgb}{0.545098,0.513726,0.525490}
  \definecolor{MistyRose1}{rgb}{1.000000,0.894118,0.882353}
  \definecolor{MistyRose2}{rgb}{0.933333,0.835294,0.823529}
  \definecolor{MistyRose3}{rgb}{0.803922,0.717647,0.709804}
  \definecolor{MistyRose4}{rgb}{0.545098,0.490196,0.482353}
  \definecolor{azure1}{rgb}{0.941176,1.000000,1.000000}
  \definecolor{azure2}{rgb}{0.878431,0.933333,0.933333}
  \definecolor{azure3}{rgb}{0.756863,0.803922,0.803922}
  \definecolor{azure4}{rgb}{0.513726,0.545098,0.545098}
  \definecolor{SlateBlue1}{rgb}{0.513726,0.435294,1.000000}
  \definecolor{SlateBlue2}{rgb}{0.478431,0.403922,0.933333}
  \definecolor{SlateBlue3}{rgb}{0.411765,0.349020,0.803922}
  \definecolor{SlateBlue4}{rgb}{0.278431,0.235294,0.545098}
  \definecolor{RoyalBlue1}{rgb}{0.282353,0.462745,1.000000}
  \definecolor{RoyalBlue2}{rgb}{0.262745,0.431373,0.933333}
  \definecolor{RoyalBlue3}{rgb}{0.227451,0.372549,0.803922}
  \definecolor{RoyalBlue4}{rgb}{0.152941,0.250980,0.545098}
  \definecolor{blue1}{rgb}{0.000000,0.000000,1.000000}
  \definecolor{blue2}{rgb}{0.000000,0.000000,0.933333}
  \definecolor{blue3}{rgb}{0.000000,0.000000,0.803922}
  \definecolor{blue4}{rgb}{0.000000,0.000000,0.545098}
  \definecolor{DodgerBlue1}{rgb}{0.117647,0.564706,1.000000}
  \definecolor{DodgerBlue2}{rgb}{0.109804,0.525490,0.933333}
  \definecolor{DodgerBlue3}{rgb}{0.094118,0.454902,0.803922}
  \definecolor{DodgerBlue4}{rgb}{0.062745,0.305882,0.545098}
  \definecolor{SteelBlue1}{rgb}{0.388235,0.721569,1.000000}
  \definecolor{SteelBlue2}{rgb}{0.360784,0.674510,0.933333}
  \definecolor{SteelBlue3}{rgb}{0.309804,0.580392,0.803922}
  \definecolor{SteelBlue4}{rgb}{0.211765,0.392157,0.545098}
  \definecolor{DeepSkyBlue1}{rgb}{0.000000,0.749020,1.000000}
  \definecolor{DeepSkyBlue2}{rgb}{0.000000,0.698039,0.933333}
  \definecolor{DeepSkyBlue3}{rgb}{0.000000,0.603922,0.803922}
  \definecolor{DeepSkyBlue4}{rgb}{0.000000,0.407843,0.545098}
  \definecolor{SkyBlue1}{rgb}{0.529412,0.807843,1.000000}
  \definecolor{SkyBlue2}{rgb}{0.494118,0.752941,0.933333}
  \definecolor{SkyBlue3}{rgb}{0.423529,0.650980,0.803922}
  \definecolor{SkyBlue4}{rgb}{0.290196,0.439216,0.545098}
  \definecolor{LightSkyBlue1}{rgb}{0.690196,0.886275,1.000000}
  \definecolor{LightSkyBlue2}{rgb}{0.643137,0.827451,0.933333}
  \definecolor{LightSkyBlue3}{rgb}{0.552941,0.713726,0.803922}
  \definecolor{LightSkyBlue4}{rgb}{0.376471,0.482353,0.545098}
  \definecolor{SlateGray1}{rgb}{0.776471,0.886275,1.000000}
  \definecolor{SlateGray2}{rgb}{0.725490,0.827451,0.933333}
  \definecolor{SlateGray3}{rgb}{0.623529,0.713726,0.803922}
  \definecolor{SlateGray4}{rgb}{0.423529,0.482353,0.545098}
  \definecolor{LightSteelBlue1}{rgb}{0.792157,0.882353,1.000000}
  \definecolor{LightSteelBlue2}{rgb}{0.737255,0.823529,0.933333}
  \definecolor{LightSteelBlue3}{rgb}{0.635294,0.709804,0.803922}
  \definecolor{LightSteelBlue4}{rgb}{0.431373,0.482353,0.545098}
  \definecolor{LightBlue1}{rgb}{0.749020,0.937255,1.000000}
  \definecolor{LightBlue2}{rgb}{0.698039,0.874510,0.933333}
  \definecolor{LightBlue3}{rgb}{0.603922,0.752941,0.803922}
  \definecolor{LightBlue4}{rgb}{0.407843,0.513726,0.545098}
  \definecolor{LightCyan1}{rgb}{0.878431,1.000000,1.000000}
  \definecolor{LightCyan2}{rgb}{0.819608,0.933333,0.933333}
  \definecolor{LightCyan3}{rgb}{0.705882,0.803922,0.803922}
  \definecolor{LightCyan4}{rgb}{0.478431,0.545098,0.545098}
  \definecolor{PaleTurquoise1}{rgb}{0.733333,1.000000,1.000000}
  \definecolor{PaleTurquoise2}{rgb}{0.682353,0.933333,0.933333}
  \definecolor{PaleTurquoise3}{rgb}{0.588235,0.803922,0.803922}
  \definecolor{PaleTurquoise4}{rgb}{0.400000,0.545098,0.545098}
  \definecolor{CadetBlue1}{rgb}{0.596078,0.960784,1.000000}
  \definecolor{CadetBlue2}{rgb}{0.556863,0.898039,0.933333}
  \definecolor{CadetBlue3}{rgb}{0.478431,0.772549,0.803922}
  \definecolor{CadetBlue4}{rgb}{0.325490,0.525490,0.545098}
  \definecolor{turquoise1}{rgb}{0.000000,0.960784,1.000000}
  \definecolor{turquoise2}{rgb}{0.000000,0.898039,0.933333}
  \definecolor{turquoise3}{rgb}{0.000000,0.772549,0.803922}
  \definecolor{turquoise4}{rgb}{0.000000,0.525490,0.545098}
  \definecolor{cyan1}{rgb}{0.000000,1.000000,1.000000}
  \definecolor{cyan2}{rgb}{0.000000,0.933333,0.933333}
  \definecolor{cyan3}{rgb}{0.000000,0.803922,0.803922}
  \definecolor{cyan4}{rgb}{0.000000,0.545098,0.545098}
  \definecolor{DarkSlateGray1}{rgb}{0.592157,1.000000,1.000000}
  \definecolor{DarkSlateGray2}{rgb}{0.552941,0.933333,0.933333}
  \definecolor{DarkSlateGray3}{rgb}{0.474510,0.803922,0.803922}
  \definecolor{DarkSlateGray4}{rgb}{0.321569,0.545098,0.545098}
  \definecolor{aquamarine1}{rgb}{0.498039,1.000000,0.831373}
  \definecolor{aquamarine2}{rgb}{0.462745,0.933333,0.776471}
  \definecolor{aquamarine3}{rgb}{0.400000,0.803922,0.666667}
  \definecolor{aquamarine4}{rgb}{0.270588,0.545098,0.454902}
  \definecolor{DarkSeaGreen1}{rgb}{0.756863,1.000000,0.756863}
  \definecolor{DarkSeaGreen2}{rgb}{0.705882,0.933333,0.705882}
  \definecolor{DarkSeaGreen3}{rgb}{0.607843,0.803922,0.607843}
  \definecolor{DarkSeaGreen4}{rgb}{0.411765,0.545098,0.411765}
  \definecolor{SeaGreen1}{rgb}{0.329412,1.000000,0.623529}
  \definecolor{SeaGreen2}{rgb}{0.305882,0.933333,0.580392}
  \definecolor{SeaGreen3}{rgb}{0.262745,0.803922,0.501961}
  \definecolor{SeaGreen4}{rgb}{0.180392,0.545098,0.341176}
  \definecolor{PaleGreen1}{rgb}{0.603922,1.000000,0.603922}
  \definecolor{PaleGreen2}{rgb}{0.564706,0.933333,0.564706}
  \definecolor{PaleGreen3}{rgb}{0.486275,0.803922,0.486275}
  \definecolor{PaleGreen4}{rgb}{0.329412,0.545098,0.329412}
  \definecolor{SpringGreen1}{rgb}{0.000000,1.000000,0.498039}
  \definecolor{SpringGreen2}{rgb}{0.000000,0.933333,0.462745}
  \definecolor{SpringGreen3}{rgb}{0.000000,0.803922,0.400000}
  \definecolor{SpringGreen4}{rgb}{0.000000,0.545098,0.270588}
  \definecolor{green1}{rgb}{0.000000,1.000000,0.000000}
  \definecolor{green2}{rgb}{0.000000,0.933333,0.000000}
  \definecolor{green3}{rgb}{0.000000,0.803922,0.000000}
  \definecolor{green4}{rgb}{0.000000,0.545098,0.000000}
  \definecolor{chartreuse1}{rgb}{0.498039,1.000000,0.000000}
  \definecolor{chartreuse2}{rgb}{0.462745,0.933333,0.000000}
  \definecolor{chartreuse3}{rgb}{0.400000,0.803922,0.000000}
  \definecolor{chartreuse4}{rgb}{0.270588,0.545098,0.000000}
  \definecolor{OliveDrab1}{rgb}{0.752941,1.000000,0.243137}
  \definecolor{OliveDrab2}{rgb}{0.701961,0.933333,0.227451}
  \definecolor{OliveDrab3}{rgb}{0.603922,0.803922,0.196078}
  \definecolor{OliveDrab4}{rgb}{0.411765,0.545098,0.133333}
  \definecolor{DarkOliveGreen1}{rgb}{0.792157,1.000000,0.439216}
  \definecolor{DarkOliveGreen2}{rgb}{0.737255,0.933333,0.407843}
  \definecolor{DarkOliveGreen3}{rgb}{0.635294,0.803922,0.352941}
  \definecolor{DarkOliveGreen4}{rgb}{0.431373,0.545098,0.239216}
  \definecolor{khaki1}{rgb}{1.000000,0.964706,0.560784}
  \definecolor{khaki2}{rgb}{0.933333,0.901961,0.521569}
  \definecolor{khaki3}{rgb}{0.803922,0.776471,0.450980}
  \definecolor{khaki4}{rgb}{0.545098,0.525490,0.305882}
  \definecolor{LightGoldenrod1}{rgb}{1.000000,0.925490,0.545098}
  \definecolor{LightGoldenrod2}{rgb}{0.933333,0.862745,0.509804}
  \definecolor{LightGoldenrod3}{rgb}{0.803922,0.745098,0.439216}
  \definecolor{LightGoldenrod4}{rgb}{0.545098,0.505882,0.298039}
  \definecolor{LightYellow1}{rgb}{1.000000,1.000000,0.878431}
  \definecolor{LightYellow2}{rgb}{0.933333,0.933333,0.819608}
  \definecolor{LightYellow3}{rgb}{0.803922,0.803922,0.705882}
  \definecolor{LightYellow4}{rgb}{0.545098,0.545098,0.478431}
  \definecolor{yellow1}{rgb}{1.000000,1.000000,0.000000}
  \definecolor{yellow2}{rgb}{0.933333,0.933333,0.000000}
  \definecolor{yellow3}{rgb}{0.803922,0.803922,0.000000}
  \definecolor{yellow4}{rgb}{0.545098,0.545098,0.000000}
  \definecolor{gold1}{rgb}{1.000000,0.843137,0.000000}
  \definecolor{gold2}{rgb}{0.933333,0.788235,0.000000}
  \definecolor{gold3}{rgb}{0.803922,0.678431,0.000000}
  \definecolor{gold4}{rgb}{0.545098,0.458824,0.000000}
  \definecolor{goldenrod1}{rgb}{1.000000,0.756863,0.145098}
  \definecolor{goldenrod2}{rgb}{0.933333,0.705882,0.133333}
  \definecolor{goldenrod3}{rgb}{0.803922,0.607843,0.113725}
  \definecolor{goldenrod4}{rgb}{0.545098,0.411765,0.078431}
  \definecolor{DarkGoldenrod1}{rgb}{1.000000,0.725490,0.058824}
  \definecolor{DarkGoldenrod2}{rgb}{0.933333,0.678431,0.054902}
  \definecolor{DarkGoldenrod3}{rgb}{0.803922,0.584314,0.047059}
  \definecolor{DarkGoldenrod4}{rgb}{0.545098,0.396078,0.031373}
  \definecolor{RosyBrown1}{rgb}{1.000000,0.756863,0.756863}
  \definecolor{RosyBrown2}{rgb}{0.933333,0.705882,0.705882}
  \definecolor{RosyBrown3}{rgb}{0.803922,0.607843,0.607843}
  \definecolor{RosyBrown4}{rgb}{0.545098,0.411765,0.411765}
  \definecolor{IndianRed1}{rgb}{1.000000,0.415686,0.415686}
  \definecolor{IndianRed2}{rgb}{0.933333,0.388235,0.388235}
  \definecolor{IndianRed3}{rgb}{0.803922,0.333333,0.333333}
  \definecolor{IndianRed4}{rgb}{0.545098,0.227451,0.227451}
  \definecolor{sienna1}{rgb}{1.000000,0.509804,0.278431}
  \definecolor{sienna2}{rgb}{0.933333,0.474510,0.258824}
  \definecolor{sienna3}{rgb}{0.803922,0.407843,0.223529}
  \definecolor{sienna4}{rgb}{0.545098,0.278431,0.149020}
  \definecolor{burlywood1}{rgb}{1.000000,0.827451,0.607843}
  \definecolor{burlywood2}{rgb}{0.933333,0.772549,0.568627}
  \definecolor{burlywood3}{rgb}{0.803922,0.666667,0.490196}
  \definecolor{burlywood4}{rgb}{0.545098,0.450980,0.333333}
  \definecolor{wheat1}{rgb}{1.000000,0.905882,0.729412}
  \definecolor{wheat2}{rgb}{0.933333,0.847059,0.682353}
  \definecolor{wheat3}{rgb}{0.803922,0.729412,0.588235}
  \definecolor{wheat4}{rgb}{0.545098,0.494118,0.400000}
  \definecolor{tan1}{rgb}{1.000000,0.647059,0.309804}
  \definecolor{tan2}{rgb}{0.933333,0.603922,0.286275}
  \definecolor{tan3}{rgb}{0.803922,0.521569,0.247059}
  \definecolor{tan4}{rgb}{0.545098,0.352941,0.168627}
  \definecolor{chocolate1}{rgb}{1.000000,0.498039,0.141176}
  \definecolor{chocolate2}{rgb}{0.933333,0.462745,0.129412}
  \definecolor{chocolate3}{rgb}{0.803922,0.400000,0.113725}
  \definecolor{chocolate4}{rgb}{0.545098,0.270588,0.074510}
  \definecolor{firebrick1}{rgb}{1.000000,0.188235,0.188235}
  \definecolor{firebrick2}{rgb}{0.933333,0.172549,0.172549}
  \definecolor{firebrick3}{rgb}{0.803922,0.149020,0.149020}
  \definecolor{firebrick4}{rgb}{0.545098,0.101961,0.101961}
  \definecolor{brown1}{rgb}{1.000000,0.250980,0.250980}
  \definecolor{brown2}{rgb}{0.933333,0.231373,0.231373}
  \definecolor{brown3}{rgb}{0.803922,0.200000,0.200000}
  \definecolor{brown4}{rgb}{0.545098,0.137255,0.137255}
  \definecolor{salmon1}{rgb}{1.000000,0.549020,0.411765}
  \definecolor{salmon2}{rgb}{0.933333,0.509804,0.384314}
  \definecolor{salmon3}{rgb}{0.803922,0.439216,0.329412}
  \definecolor{salmon4}{rgb}{0.545098,0.298039,0.223529}
  \definecolor{LightSalmon1}{rgb}{1.000000,0.627451,0.478431}
  \definecolor{LightSalmon2}{rgb}{0.933333,0.584314,0.447059}
  \definecolor{LightSalmon3}{rgb}{0.803922,0.505882,0.384314}
  \definecolor{LightSalmon4}{rgb}{0.545098,0.341176,0.258824}
  \definecolor{orange1}{rgb}{1.000000,0.647059,0.000000}
  \definecolor{orange2}{rgb}{0.933333,0.603922,0.000000}
  \definecolor{orange3}{rgb}{0.803922,0.521569,0.000000}
  \definecolor{orange4}{rgb}{0.545098,0.352941,0.000000}
  \definecolor{DarkOrange1}{rgb}{1.000000,0.498039,0.000000}
  \definecolor{DarkOrange2}{rgb}{0.933333,0.462745,0.000000}
  \definecolor{DarkOrange3}{rgb}{0.803922,0.400000,0.000000}
  \definecolor{DarkOrange4}{rgb}{0.545098,0.270588,0.000000}
  \definecolor{coral1}{rgb}{1.000000,0.447059,0.337255}
  \definecolor{coral2}{rgb}{0.933333,0.415686,0.313726}
  \definecolor{coral3}{rgb}{0.803922,0.356863,0.270588}
  \definecolor{coral4}{rgb}{0.545098,0.243137,0.184314}
  \definecolor{tomato1}{rgb}{1.000000,0.388235,0.278431}
  \definecolor{tomato2}{rgb}{0.933333,0.360784,0.258824}
  \definecolor{tomato3}{rgb}{0.803922,0.309804,0.223529}
  \definecolor{tomato4}{rgb}{0.545098,0.211765,0.149020}
  \definecolor{OrangeRed1}{rgb}{1.000000,0.270588,0.000000}
  \definecolor{OrangeRed2}{rgb}{0.933333,0.250980,0.000000}
  \definecolor{OrangeRed3}{rgb}{0.803922,0.215686,0.000000}
  \definecolor{OrangeRed4}{rgb}{0.545098,0.145098,0.000000}
  \definecolor{red1}{rgb}{1.000000,0.000000,0.000000}
  \definecolor{red2}{rgb}{0.933333,0.000000,0.000000}
  \definecolor{red3}{rgb}{0.803922,0.000000,0.000000}
  \definecolor{red4}{rgb}{0.545098,0.000000,0.000000}
  \definecolor{DeepPink1}{rgb}{1.000000,0.078431,0.576471}
  \definecolor{DeepPink2}{rgb}{0.933333,0.070588,0.537255}
  \definecolor{DeepPink3}{rgb}{0.803922,0.062745,0.462745}
  \definecolor{DeepPink4}{rgb}{0.545098,0.039216,0.313726}
  \definecolor{HotPink1}{rgb}{1.000000,0.431373,0.705882}
  \definecolor{HotPink2}{rgb}{0.933333,0.415686,0.654902}
  \definecolor{HotPink3}{rgb}{0.803922,0.376471,0.564706}
  \definecolor{HotPink4}{rgb}{0.545098,0.227451,0.384314}
  \definecolor{pink1}{rgb}{1.000000,0.709804,0.772549}
  \definecolor{pink2}{rgb}{0.933333,0.662745,0.721569}
  \definecolor{pink3}{rgb}{0.803922,0.568627,0.619608}
  \definecolor{pink4}{rgb}{0.545098,0.388235,0.423529}
  \definecolor{LightPink1}{rgb}{1.000000,0.682353,0.725490}
  \definecolor{LightPink2}{rgb}{0.933333,0.635294,0.678431}
  \definecolor{LightPink3}{rgb}{0.803922,0.549020,0.584314}
  \definecolor{LightPink4}{rgb}{0.545098,0.372549,0.396078}
  \definecolor{PaleVioletRed1}{rgb}{1.000000,0.509804,0.670588}
  \definecolor{PaleVioletRed2}{rgb}{0.933333,0.474510,0.623529}
  \definecolor{PaleVioletRed3}{rgb}{0.803922,0.407843,0.537255}
  \definecolor{PaleVioletRed4}{rgb}{0.545098,0.278431,0.364706}
  \definecolor{maroon1}{rgb}{1.000000,0.203922,0.701961}
  \definecolor{maroon2}{rgb}{0.933333,0.188235,0.654902}
  \definecolor{maroon3}{rgb}{0.803922,0.160784,0.564706}
  \definecolor{maroon4}{rgb}{0.545098,0.109804,0.384314}
  \definecolor{VioletRed1}{rgb}{1.000000,0.243137,0.588235}
  \definecolor{VioletRed2}{rgb}{0.933333,0.227451,0.549020}
  \definecolor{VioletRed3}{rgb}{0.803922,0.196078,0.470588}
  \definecolor{VioletRed4}{rgb}{0.545098,0.133333,0.321569}
  \definecolor{magenta1}{rgb}{1.000000,0.000000,1.000000}
  \definecolor{magenta2}{rgb}{0.933333,0.000000,0.933333}
  \definecolor{magenta3}{rgb}{0.803922,0.000000,0.803922}
  \definecolor{magenta4}{rgb}{0.545098,0.000000,0.545098}
  \definecolor{orchid1}{rgb}{1.000000,0.513726,0.980392}
  \definecolor{orchid2}{rgb}{0.933333,0.478431,0.913725}
  \definecolor{orchid3}{rgb}{0.803922,0.411765,0.788235}
  \definecolor{orchid4}{rgb}{0.545098,0.278431,0.537255}
  \definecolor{plum1}{rgb}{1.000000,0.733333,1.000000}
  \definecolor{plum2}{rgb}{0.933333,0.682353,0.933333}
  \definecolor{plum3}{rgb}{0.803922,0.588235,0.803922}
  \definecolor{plum4}{rgb}{0.545098,0.400000,0.545098}
  \definecolor{MediumOrchid1}{rgb}{0.878431,0.400000,1.000000}
  \definecolor{MediumOrchid2}{rgb}{0.819608,0.372549,0.933333}
  \definecolor{MediumOrchid3}{rgb}{0.705882,0.321569,0.803922}
  \definecolor{MediumOrchid4}{rgb}{0.478431,0.215686,0.545098}
  \definecolor{DarkOrchid1}{rgb}{0.749020,0.243137,1.000000}
  \definecolor{DarkOrchid2}{rgb}{0.698039,0.227451,0.933333}
  \definecolor{DarkOrchid3}{rgb}{0.603922,0.196078,0.803922}
  \definecolor{DarkOrchid4}{rgb}{0.407843,0.133333,0.545098}
  \definecolor{purple1}{rgb}{0.607843,0.188235,1.000000}
  \definecolor{purple2}{rgb}{0.568627,0.172549,0.933333}
  \definecolor{purple3}{rgb}{0.490196,0.149020,0.803922}
  \definecolor{purple4}{rgb}{0.333333,0.101961,0.545098}
  \definecolor{MediumPurple1}{rgb}{0.670588,0.509804,1.000000}
  \definecolor{MediumPurple2}{rgb}{0.623529,0.474510,0.933333}
  \definecolor{MediumPurple3}{rgb}{0.537255,0.407843,0.803922}
  \definecolor{MediumPurple4}{rgb}{0.364706,0.278431,0.545098}
  \definecolor{thistle1}{rgb}{1.000000,0.882353,1.000000}
  \definecolor{thistle2}{rgb}{0.933333,0.823529,0.933333}
  \definecolor{thistle3}{rgb}{0.803922,0.709804,0.803922}
  \definecolor{thistle4}{rgb}{0.545098,0.482353,0.545098}
  \definecolor{gray0}{rgb}{0.000000,0.000000,0.000000}
  \definecolor{grey0}{rgb}{0.000000,0.000000,0.000000}
  \definecolor{gray1}{rgb}{0.011765,0.011765,0.011765}
  \definecolor{grey1}{rgb}{0.011765,0.011765,0.011765}
  \definecolor{gray2}{rgb}{0.019608,0.019608,0.019608}
  \definecolor{grey2}{rgb}{0.019608,0.019608,0.019608}
  \definecolor{gray3}{rgb}{0.031373,0.031373,0.031373}
  \definecolor{grey3}{rgb}{0.031373,0.031373,0.031373}
  \definecolor{gray4}{rgb}{0.039216,0.039216,0.039216}
  \definecolor{grey4}{rgb}{0.039216,0.039216,0.039216}
  \definecolor{gray5}{rgb}{0.050980,0.050980,0.050980}
  \definecolor{grey5}{rgb}{0.050980,0.050980,0.050980}
  \definecolor{gray6}{rgb}{0.058824,0.058824,0.058824}
  \definecolor{grey6}{rgb}{0.058824,0.058824,0.058824}
  \definecolor{gray7}{rgb}{0.070588,0.070588,0.070588}
  \definecolor{grey7}{rgb}{0.070588,0.070588,0.070588}
  \definecolor{gray8}{rgb}{0.078431,0.078431,0.078431}
  \definecolor{grey8}{rgb}{0.078431,0.078431,0.078431}
  \definecolor{gray9}{rgb}{0.090196,0.090196,0.090196}
  \definecolor{grey9}{rgb}{0.090196,0.090196,0.090196}
  \definecolor{gray10}{rgb}{0.101961,0.101961,0.101961}
  \definecolor{grey10}{rgb}{0.101961,0.101961,0.101961}
  \definecolor{gray11}{rgb}{0.109804,0.109804,0.109804}
  \definecolor{grey11}{rgb}{0.109804,0.109804,0.109804}
  \definecolor{gray12}{rgb}{0.121569,0.121569,0.121569}
  \definecolor{grey12}{rgb}{0.121569,0.121569,0.121569}
  \definecolor{gray13}{rgb}{0.129412,0.129412,0.129412}
  \definecolor{grey13}{rgb}{0.129412,0.129412,0.129412}
  \definecolor{gray14}{rgb}{0.141176,0.141176,0.141176}
  \definecolor{grey14}{rgb}{0.141176,0.141176,0.141176}
  \definecolor{gray15}{rgb}{0.149020,0.149020,0.149020}
  \definecolor{grey15}{rgb}{0.149020,0.149020,0.149020}
  \definecolor{gray16}{rgb}{0.160784,0.160784,0.160784}
  \definecolor{grey16}{rgb}{0.160784,0.160784,0.160784}
  \definecolor{gray17}{rgb}{0.168627,0.168627,0.168627}
  \definecolor{grey17}{rgb}{0.168627,0.168627,0.168627}
  \definecolor{gray18}{rgb}{0.180392,0.180392,0.180392}
  \definecolor{grey18}{rgb}{0.180392,0.180392,0.180392}
  \definecolor{gray19}{rgb}{0.188235,0.188235,0.188235}
  \definecolor{grey19}{rgb}{0.188235,0.188235,0.188235}
  \definecolor{gray20}{rgb}{0.200000,0.200000,0.200000}
  \definecolor{grey20}{rgb}{0.200000,0.200000,0.200000}
  \definecolor{gray21}{rgb}{0.211765,0.211765,0.211765}
  \definecolor{grey21}{rgb}{0.211765,0.211765,0.211765}
  \definecolor{gray22}{rgb}{0.219608,0.219608,0.219608}
  \definecolor{grey22}{rgb}{0.219608,0.219608,0.219608}
  \definecolor{gray23}{rgb}{0.231373,0.231373,0.231373}
  \definecolor{grey23}{rgb}{0.231373,0.231373,0.231373}
  \definecolor{gray24}{rgb}{0.239216,0.239216,0.239216}
  \definecolor{grey24}{rgb}{0.239216,0.239216,0.239216}
  \definecolor{gray25}{rgb}{0.250980,0.250980,0.250980}
  \definecolor{grey25}{rgb}{0.250980,0.250980,0.250980}
  \definecolor{gray26}{rgb}{0.258824,0.258824,0.258824}
  \definecolor{grey26}{rgb}{0.258824,0.258824,0.258824}
  \definecolor{gray27}{rgb}{0.270588,0.270588,0.270588}
  \definecolor{grey27}{rgb}{0.270588,0.270588,0.270588}
  \definecolor{gray28}{rgb}{0.278431,0.278431,0.278431}
  \definecolor{grey28}{rgb}{0.278431,0.278431,0.278431}
  \definecolor{gray29}{rgb}{0.290196,0.290196,0.290196}
  \definecolor{grey29}{rgb}{0.290196,0.290196,0.290196}
  \definecolor{gray30}{rgb}{0.301961,0.301961,0.301961}
  \definecolor{grey30}{rgb}{0.301961,0.301961,0.301961}
  \definecolor{gray31}{rgb}{0.309804,0.309804,0.309804}
  \definecolor{grey31}{rgb}{0.309804,0.309804,0.309804}
  \definecolor{gray32}{rgb}{0.321569,0.321569,0.321569}
  \definecolor{grey32}{rgb}{0.321569,0.321569,0.321569}
  \definecolor{gray33}{rgb}{0.329412,0.329412,0.329412}
  \definecolor{grey33}{rgb}{0.329412,0.329412,0.329412}
  \definecolor{gray34}{rgb}{0.341176,0.341176,0.341176}
  \definecolor{grey34}{rgb}{0.341176,0.341176,0.341176}
  \definecolor{gray35}{rgb}{0.349020,0.349020,0.349020}
  \definecolor{grey35}{rgb}{0.349020,0.349020,0.349020}
  \definecolor{gray36}{rgb}{0.360784,0.360784,0.360784}
  \definecolor{grey36}{rgb}{0.360784,0.360784,0.360784}
  \definecolor{gray37}{rgb}{0.368627,0.368627,0.368627}
  \definecolor{grey37}{rgb}{0.368627,0.368627,0.368627}
  \definecolor{gray38}{rgb}{0.380392,0.380392,0.380392}
  \definecolor{grey38}{rgb}{0.380392,0.380392,0.380392}
  \definecolor{gray39}{rgb}{0.388235,0.388235,0.388235}
  \definecolor{grey39}{rgb}{0.388235,0.388235,0.388235}
  \definecolor{gray40}{rgb}{0.400000,0.400000,0.400000}
  \definecolor{grey40}{rgb}{0.400000,0.400000,0.400000}
  \definecolor{gray41}{rgb}{0.411765,0.411765,0.411765}
  \definecolor{grey41}{rgb}{0.411765,0.411765,0.411765}
  \definecolor{gray42}{rgb}{0.419608,0.419608,0.419608}
  \definecolor{grey42}{rgb}{0.419608,0.419608,0.419608}
  \definecolor{gray43}{rgb}{0.431373,0.431373,0.431373}
  \definecolor{grey43}{rgb}{0.431373,0.431373,0.431373}
  \definecolor{gray44}{rgb}{0.439216,0.439216,0.439216}
  \definecolor{grey44}{rgb}{0.439216,0.439216,0.439216}
  \definecolor{gray45}{rgb}{0.450980,0.450980,0.450980}
  \definecolor{grey45}{rgb}{0.450980,0.450980,0.450980}
  \definecolor{gray46}{rgb}{0.458824,0.458824,0.458824}
  \definecolor{grey46}{rgb}{0.458824,0.458824,0.458824}
  \definecolor{gray47}{rgb}{0.470588,0.470588,0.470588}
  \definecolor{grey47}{rgb}{0.470588,0.470588,0.470588}
  \definecolor{gray48}{rgb}{0.478431,0.478431,0.478431}
  \definecolor{grey48}{rgb}{0.478431,0.478431,0.478431}
  \definecolor{gray49}{rgb}{0.490196,0.490196,0.490196}
  \definecolor{grey49}{rgb}{0.490196,0.490196,0.490196}
  \definecolor{gray50}{rgb}{0.498039,0.498039,0.498039}
  \definecolor{grey50}{rgb}{0.498039,0.498039,0.498039}
  \definecolor{gray51}{rgb}{0.509804,0.509804,0.509804}
  \definecolor{grey51}{rgb}{0.509804,0.509804,0.509804}
  \definecolor{gray52}{rgb}{0.521569,0.521569,0.521569}
  \definecolor{grey52}{rgb}{0.521569,0.521569,0.521569}
  \definecolor{gray53}{rgb}{0.529412,0.529412,0.529412}
  \definecolor{grey53}{rgb}{0.529412,0.529412,0.529412}
  \definecolor{gray54}{rgb}{0.541176,0.541176,0.541176}
  \definecolor{grey54}{rgb}{0.541176,0.541176,0.541176}
  \definecolor{gray55}{rgb}{0.549020,0.549020,0.549020}
  \definecolor{grey55}{rgb}{0.549020,0.549020,0.549020}
  \definecolor{gray56}{rgb}{0.560784,0.560784,0.560784}
  \definecolor{grey56}{rgb}{0.560784,0.560784,0.560784}
  \definecolor{gray57}{rgb}{0.568627,0.568627,0.568627}
  \definecolor{grey57}{rgb}{0.568627,0.568627,0.568627}
  \definecolor{gray58}{rgb}{0.580392,0.580392,0.580392}
  \definecolor{grey58}{rgb}{0.580392,0.580392,0.580392}
  \definecolor{gray59}{rgb}{0.588235,0.588235,0.588235}
  \definecolor{grey59}{rgb}{0.588235,0.588235,0.588235}
  \definecolor{gray60}{rgb}{0.600000,0.600000,0.600000}
  \definecolor{grey60}{rgb}{0.600000,0.600000,0.600000}
  \definecolor{gray61}{rgb}{0.611765,0.611765,0.611765}
  \definecolor{grey61}{rgb}{0.611765,0.611765,0.611765}
  \definecolor{gray62}{rgb}{0.619608,0.619608,0.619608}
  \definecolor{grey62}{rgb}{0.619608,0.619608,0.619608}
  \definecolor{gray63}{rgb}{0.631373,0.631373,0.631373}
  \definecolor{grey63}{rgb}{0.631373,0.631373,0.631373}
  \definecolor{gray64}{rgb}{0.639216,0.639216,0.639216}
  \definecolor{grey64}{rgb}{0.639216,0.639216,0.639216}
  \definecolor{gray65}{rgb}{0.650980,0.650980,0.650980}
  \definecolor{grey65}{rgb}{0.650980,0.650980,0.650980}
  \definecolor{gray66}{rgb}{0.658824,0.658824,0.658824}
  \definecolor{grey66}{rgb}{0.658824,0.658824,0.658824}
  \definecolor{gray67}{rgb}{0.670588,0.670588,0.670588}
  \definecolor{grey67}{rgb}{0.670588,0.670588,0.670588}
  \definecolor{gray68}{rgb}{0.678431,0.678431,0.678431}
  \definecolor{grey68}{rgb}{0.678431,0.678431,0.678431}
  \definecolor{gray69}{rgb}{0.690196,0.690196,0.690196}
  \definecolor{grey69}{rgb}{0.690196,0.690196,0.690196}
  \definecolor{gray70}{rgb}{0.701961,0.701961,0.701961}
  \definecolor{grey70}{rgb}{0.701961,0.701961,0.701961}
  \definecolor{gray71}{rgb}{0.709804,0.709804,0.709804}
  \definecolor{grey71}{rgb}{0.709804,0.709804,0.709804}
  \definecolor{gray72}{rgb}{0.721569,0.721569,0.721569}
  \definecolor{grey72}{rgb}{0.721569,0.721569,0.721569}
  \definecolor{gray73}{rgb}{0.729412,0.729412,0.729412}
  \definecolor{grey73}{rgb}{0.729412,0.729412,0.729412}
  \definecolor{gray74}{rgb}{0.741176,0.741176,0.741176}
  \definecolor{grey74}{rgb}{0.741176,0.741176,0.741176}
  \definecolor{gray75}{rgb}{0.749020,0.749020,0.749020}
  \definecolor{grey75}{rgb}{0.749020,0.749020,0.749020}
  \definecolor{gray76}{rgb}{0.760784,0.760784,0.760784}
  \definecolor{grey76}{rgb}{0.760784,0.760784,0.760784}
  \definecolor{gray77}{rgb}{0.768627,0.768627,0.768627}
  \definecolor{grey77}{rgb}{0.768627,0.768627,0.768627}
  \definecolor{gray78}{rgb}{0.780392,0.780392,0.780392}
  \definecolor{grey78}{rgb}{0.780392,0.780392,0.780392}
  \definecolor{gray79}{rgb}{0.788235,0.788235,0.788235}
  \definecolor{grey79}{rgb}{0.788235,0.788235,0.788235}
  \definecolor{gray80}{rgb}{0.800000,0.800000,0.800000}
  \definecolor{grey80}{rgb}{0.800000,0.800000,0.800000}
  \definecolor{gray81}{rgb}{0.811765,0.811765,0.811765}
  \definecolor{grey81}{rgb}{0.811765,0.811765,0.811765}
  \definecolor{gray82}{rgb}{0.819608,0.819608,0.819608}
  \definecolor{grey82}{rgb}{0.819608,0.819608,0.819608}
  \definecolor{gray83}{rgb}{0.831373,0.831373,0.831373}
  \definecolor{grey83}{rgb}{0.831373,0.831373,0.831373}
  \definecolor{gray84}{rgb}{0.839216,0.839216,0.839216}
  \definecolor{grey84}{rgb}{0.839216,0.839216,0.839216}
  \definecolor{gray85}{rgb}{0.850980,0.850980,0.850980}
  \definecolor{grey85}{rgb}{0.850980,0.850980,0.850980}
  \definecolor{gray86}{rgb}{0.858824,0.858824,0.858824}
  \definecolor{grey86}{rgb}{0.858824,0.858824,0.858824}
  \definecolor{gray87}{rgb}{0.870588,0.870588,0.870588}
  \definecolor{grey87}{rgb}{0.870588,0.870588,0.870588}
  \definecolor{gray88}{rgb}{0.878431,0.878431,0.878431}
  \definecolor{grey88}{rgb}{0.878431,0.878431,0.878431}
  \definecolor{gray89}{rgb}{0.890196,0.890196,0.890196}
  \definecolor{grey89}{rgb}{0.890196,0.890196,0.890196}
  \definecolor{gray90}{rgb}{0.898039,0.898039,0.898039}
  \definecolor{grey90}{rgb}{0.898039,0.898039,0.898039}
  \definecolor{gray91}{rgb}{0.909804,0.909804,0.909804}
  \definecolor{grey91}{rgb}{0.909804,0.909804,0.909804}
  \definecolor{gray92}{rgb}{0.921569,0.921569,0.921569}
  \definecolor{grey92}{rgb}{0.921569,0.921569,0.921569}
  \definecolor{gray93}{rgb}{0.929412,0.929412,0.929412}
  \definecolor{grey93}{rgb}{0.929412,0.929412,0.929412}
  \definecolor{gray94}{rgb}{0.941176,0.941176,0.941176}
  \definecolor{grey94}{rgb}{0.941176,0.941176,0.941176}
  \definecolor{gray95}{rgb}{0.949020,0.949020,0.949020}
  \definecolor{grey95}{rgb}{0.949020,0.949020,0.949020}
  \definecolor{gray96}{rgb}{0.960784,0.960784,0.960784}
  \definecolor{grey96}{rgb}{0.960784,0.960784,0.960784}
  \definecolor{gray97}{rgb}{0.968627,0.968627,0.968627}
  \definecolor{grey97}{rgb}{0.968627,0.968627,0.968627}
  \definecolor{gray98}{rgb}{0.980392,0.980392,0.980392}
  \definecolor{grey98}{rgb}{0.980392,0.980392,0.980392}
  \definecolor{gray99}{rgb}{0.988235,0.988235,0.988235}
  \definecolor{grey99}{rgb}{0.988235,0.988235,0.988235}
  \definecolor{gray100}{rgb}{1.000000,1.000000,1.000000}
  \definecolor{grey100}{rgb}{1.000000,1.000000,1.000000}
  \definecolor{dark grey}{rgb}{0.662745,0.662745,0.662745}
  \definecolor{DarkGrey}{rgb}{0.662745,0.662745,0.662745}
  \definecolor{dark gray}{rgb}{0.662745,0.662745,0.662745}
  \definecolor{DarkGray}{rgb}{0.662745,0.662745,0.662745}
  \definecolor{dark blue}{rgb}{0.000000,0.000000,0.545098}
  \definecolor{DarkBlue}{rgb}{0.000000,0.000000,0.545098}
  \definecolor{dark cyan}{rgb}{0.000000,0.545098,0.545098}
  \definecolor{DarkCyan}{rgb}{0.000000,0.545098,0.545098}
  \definecolor{dark magenta}{rgb}{0.545098,0.000000,0.545098}
  \definecolor{DarkMagenta}{rgb}{0.545098,0.000000,0.545098}
  \definecolor{dark red}{rgb}{0.545098,0.000000,0.000000}
  \definecolor{DarkRed}{rgb}{0.545098,0.000000,0.000000}
  \definecolor{light green}{rgb}{0.564706,0.933333,0.564706}
  \definecolor{LightGreen}{rgb}{0.564706,0.933333,0.564706}